\documentclass[twocolumn]{aastex61}
\usepackage{natbib, color}
\usepackage[normalem]{ulem}
\usepackage{epstopdf}
\usepackage{graphicx}
\definecolor{red}{rgb}{1.0,0.0,0.0}

\newcommand{\Mj}[1]{$M_\mathrm{Jup}$}

\usepackage[utf8]{inputenc}

\begin{document}
\title{Evidence that the Directly-Imaged Planet HD 131399 A\lowercase{b} is a Background Star}

\correspondingauthor{Eric L. Nielsen}
\email{enielsen@seti.org}

\author[0000-0001-6975-9056]{Eric L. Nielsen}
\affiliation{SETI Institute, Carl Sagan Center, 189 Bernardo Ave.,  Mountain View CA 94043, USA}
\affiliation{Kavli Institute for Particle Astrophysics and Cosmology, Stanford University, Stanford, CA 94305, USA}

\author[0000-0002-4918-0247]{Robert J. De Rosa}
\affiliation{Department of Astronomy, University of California, Berkeley, CA 94720, USA}

\author[0000-0003-0029-0258]{Julien Rameau}
\affiliation{Institut de Recherche sur les Exoplan{\`e}tes, D{\'e}partement de Physique, Universit{\'e} de Montr{\'e}al, Montr{\'e}al QC, H3C 3J7, Canada}

\author[0000-0003-0774-6502]{Jason J. Wang}
\affiliation{Department of Astronomy, University of California, Berkeley, CA 94720, USA}

\author[0000-0002-0792-3719]{Thomas M. Esposito}
\affiliation{Department of Astronomy, University of California, Berkeley, CA 94720, USA}

\author[0000-0001-6205-9233]{Maxwell A. Millar-Blanchaer}
\affiliation{Jet Propulsion Laboratory, California Institute of Technology, Pasadena, CA 91125, USA}
\affiliation{NASA Hubble Fellow}

\author[0000-0002-4164-4182]{Christian Marois}
\affiliation{National Research Council of Canada Herzberg, 5071 West Saanich Rd, Victoria, BC, V9E 2E7, Canada}
\affiliation{University of Victoria, 3800 Finnerty Rd, Victoria, BC, V8P 5C2, Canada}

\author[0000-0002-5902-7828]{Arthur Vigan}
\affiliation{Aix Marseille Univ, CNRS, LAM, Laboratoire d'Astrophysique de Marseille, Marseille, France}

\author[0000-0001-5172-7902]{S. Mark Ammons}
\affiliation{Lawrence Livermore National Laboratory, Livermore, CA 94551, USA}

\author[0000-0003-3506-5667]{Etienne Artigau}
\affiliation{Institut de Recherche sur les Exoplan{\`e}tes, D{\'e}partement de Physique, Universit{\'e} de Montr{\'e}al, Montr{\'e}al QC, H3C 3J7, Canada}

\author[0000-0002-5407-2806]{Vanessa P. Bailey}
\affiliation{Kavli Institute for Particle Astrophysics and Cosmology, Stanford University, Stanford, CA 94305, USA}

\author{Sarah Blunt}
\affiliation{Department of Physics, Brown University, Providence, RI 02912, USA}
\affiliation{SETI Institute, Carl Sagan Center, 189 Bernardo Ave.,  Mountain View CA 94043, USA}
\affiliation{Kavli Institute for Particle Astrophysics and Cosmology, Stanford University, Stanford, CA 94305, USA}

\author{Joanna Bulger}
\affiliation{Subaru Telescope, NAOJ, 650 North A{'o}hoku Place, Hilo, HI 96720, USA}

\author[0000-0001-6305-7272]{Jeffrey Chilcote}
\affiliation{Dunlap Institute for Astronomy \& Astrophysics, University of Toronto, Toronto, ON M5S 3H4, Canada}

\author[0000-0003-0156-3019]{Tara Cotten}
\affiliation{Department of Physics and Astronomy, University of Georgia, Athens, GA 30602, USA}

\author{Ren{\'e} Doyon}
\affiliation{Institut de Recherche sur les Exoplan{\`e}tes, D{\'e}partement de Physique, Universit{\'e} de Montr{\'e}al, Montr{\'e}al QC, H3C 3J7, Canada}

\author[0000-0002-5092-6464]{Gaspard Duch\^ene}
\affiliation{Department of Astronomy, University of California, Berkeley, CA 94720, USA}
\affiliation{Univ. Grenoble Alpes/CNRS, IPAG, F-38000 Grenoble, France}

\author[0000-0003-3750-0183]{Daniel Fabrycky}
\affiliation{Department of Astronomy and Astrophysics, University of Chicago, 5640 South Ellis Avenue, Chicago, IL 60637, USA}

\author[0000-0002-0176-8973]{Michael P. Fitzgerald}
\affiliation{Department of Physics \& Astronomy, University of California, Los Angeles, CA 90095, USA}

\author[0000-0002-7821-0695]{Katherine B. Follette}
\affiliation{Amherst College Department of Physics and Astronomy, Merrill Science Center, 15 Mead Drive, Amherst, MA 01002, USA}

\author{Benjamin L. Gerard}
\affiliation{University of Victoria, 3800 Finnerty Rd, Victoria, BC, V8P 5C2, Canada}
\affiliation{National Research Council of Canada Herzberg, 5071 West Saanich Rd, Victoria, BC, V9E 2E7, Canada}

\author[0000-0002-4144-5116]{Stephen J. Goodsell}
\affiliation{Gemini Observatory, 670 N. A'ohoku Place, Hilo, HI 96720, USA}

\author{James R. Graham}
\affiliation{Department of Astronomy, University of California, Berkeley, CA 94720, USA}

\author[0000-0002-7162-8036]{Alexandra Z. Greenbaum}
\affiliation{Department of Astronomy, University of Michigan, Ann Arbor, MI 48109, USA}

\author[0000-0003-3726-5494]{Pascale Hibon}
\affiliation{Gemini Observatory, Casilla 603, La Serena, Chile}

\author{Sasha Hinkley}
\affiliation{School of Physics, College of Engineering, Mathematics and Physical Sciences, University of Exeter, Exeter EX4 4QL, UK}

\author[0000-0003-1498-6088]{Li-Wei Hung}
\affiliation{Department of Physics \& Astronomy, University of California, Los Angeles, CA 90095, USA}

\author{Patrick Ingraham}
\affiliation{Large Synoptic Survey Telescope, 950N Cherry Ave., Tucson, AZ 85719, USA}

\author{Rebecca Jensen-Clem}
\affiliation{Department of Astro physics, California Institute of Technology, 1200 E. California Blvd., Pasadena, CA 91101, USA}

\author{Paul Kalas}
\affiliation{Department of Astronomy, University of California, Berkeley, CA 94720, USA}
\affiliation{SETI Institute, Carl Sagan Center, 189 Bernardo Ave.,  Mountain View CA 94043, USA}

\author[0000-0002-9936-6285]{Quinn Konopacky}
\affiliation{Center for Astrophysics and Space Science, University of California San Diego, La Jolla, CA 92093, USA}

\author{James E. Larkin}
\affiliation{Department of Physics \& Astronomy, University of California, Los Angeles, CA 90095, USA}

\author[0000-0003-1212-7538]{Bruce Macintosh}
\affiliation{Kavli Institute for Particle Astrophysics and Cosmology, Stanford University, Stanford, CA 94305, USA}

\author{J\'er\^ome Maire}
\affiliation{Center for Astrophysics and Space Science, University of California San Diego, La Jolla, CA 92093, USA}

\author[0000-0001-7016-7277]{Franck Marchis}
\affiliation{SETI Institute, Carl Sagan Center, 189 Bernardo Ave.,  Mountain View CA 94043, USA}

\author[0000-0003-3050-8203]{Stanimir Metchev}
\affiliation{Department of Physics and Astronomy, Centre for Planetary Science and Exploration, The University of Western Ontario, London, ON N6A 3K7, Canada}
\affiliation{Department of Physics and Astronomy, Stony Brook University, Stony Brook, NY 11794-3800, USA}

\author[0000-0002-1384-0063]{Katie M. Morzinski}
\affiliation{Steward Observatory, University of Arizona, Tucson, AZ 85721, USA}

\author{Ruth A. Murray-Clay}
\affiliation{Department of Astronomy, UC Santa Cruz, 1156 High St., Santa Cruz, CA 95064, USA }

\author[0000-0001-7130-7681]{Rebecca Oppenheimer}
\affiliation{Department of Astrophysics, American Museum of Natural History, New York, NY 10024, USA}

\author{David Palmer}
\affiliation{Lawrence Livermore National Laboratory, Livermore, CA 94551, USA}

\author{Jennifer Patience}
\affiliation{School of Earth and Space Exploration, Arizona State University, PO Box 871404, Tempe, AZ 85287, USA}

\author[0000-0002-3191-8151]{Marshall Perrin}
\affiliation{Space Telescope Science Institute, Baltimore, MD 21218, USA}

\author{Lisa Poyneer}
\affiliation{Lawrence Livermore National Laboratory, Livermore, CA 94551, USA}

\author{Laurent Pueyo}
\affiliation{Space Telescope Science Institute, Baltimore, MD 21218, USA}

\author{Roman R. Rafikov}
\affiliation{Department of Applied Mathematics and Theoretical Physics, Centre for Mathematical Sciences, University of Cambridge, Wilberforce Road, Cambridge CB3 0WA, UK}
\affiliation{Institute for Advanced Study, Einstein Drive, Princeton, NJ 08540, USA}

\author[0000-0002-9246-5467]{Abhijith Rajan}
\affiliation{School of Earth and Space Exploration, Arizona State University, PO Box 871404, Tempe, AZ 85287, USA}

\author[0000-0002-9667-2244]{Fredrik T. Rantakyr\"o}
\affiliation{Gemini Observatory, Casilla 603, La Serena, Chile}

\author[0000-0003-2233-4821]{Jean-Baptiste Ruffio}
\affiliation{Kavli Institute for Particle Astrophysics and Cosmology, Stanford University, Stanford, CA 94305, USA}

\author[0000-0002-8711-7206]{Dmitry Savransky}
\affiliation{Sibley School of Mechanical and Aerospace Engineering, Cornell University, Ithaca, NY 14853, USA}

\author{Adam C. Schneider}
\affiliation{School of Earth and Space Exploration, Arizona State University, PO Box 871404, Tempe, AZ 85287, USA}

\author[0000-0003-1251-4124]{Anand Sivaramakrishnan}
\affiliation{Space Telescope Science Institute, Baltimore, MD 21218, USA}

\author[0000-0002-5815-7372]{Inseok Song}
\affiliation{Department of Physics and Astronomy, University of Georgia, Athens, GA 30602, USA}

\author[0000-0003-2753-2819]{Remi Soummer}
\affiliation{Space Telescope Science Institute, Baltimore, MD 21218, USA}

\author{Sandrine Thomas}
\affiliation{Large Synoptic Survey Telescope, 950N Cherry Ave., Tucson, AZ 85719, USA}

\author{J. Kent Wallace}
\affiliation{Jet Propulsion Laboratory, California Institute of Technology, Pasadena, CA 91125, USA}

\author[0000-0002-4479-8291]{Kimberly Ward-Duong}
\affiliation{School of Earth and Space Exploration, Arizona State University, PO Box 871404, Tempe, AZ 85287, USA}

\author{Sloane Wiktorowicz}
\affiliation{Department of Astronomy, UC Santa Cruz, 1156 High St., Santa Cruz, CA 95064, USA }

\author[0000-0002-9977-8255]{Schuyler Wolff}
\affiliation{Department of Physics and Astronomy, Johns Hopkins University, Baltimore, MD 21218, USA}


\begin{abstract}
We present evidence that the recently discovered, directly-imaged planet HD~131399~Ab is a background star with non-zero proper motion. From new {\it JHK1L}$^{\prime}$ photometry and spectroscopy obtained with the Gemini Planet Imager, VLT/SPHERE, and  Keck/NIRC2, and a reanalysis of the discovery data obtained with VLT/SPHERE, we derive colors, spectra, and astrometry for HD~131399~Ab. The broader wavelength coverage and higher data quality allow us to re-investigate its status. Its near-infrared spectral energy distribution excludes spectral types later than L0 and is consistent with a K or M dwarf, which are the most likely candidates for a background object in this direction at the apparent magnitude observed. If it were a physically associated object, the projected velocity of HD~131399~Ab would exceed escape velocity given the mass and distance to HD~131399~A.  We show that HD~131399~Ab is also not following the expected track for a stationary background star at infinite distance. Solving for the proper motion and parallax required to explain the relative motion of HD~131399~Ab, we find a proper motion of 12.3\,mas\,yr$^{-1}$.  When compared to predicted background objects drawn from a galactic model, we find this proper motion to be high, but consistent with the top 4\% fastest-moving background stars. From our analysis we conclude that HD~131399~Ab is a background K or M dwarf.

\end{abstract}

\section{Introduction}

Since 2005, multiple planets have been detected by direct imaging \citep[e.g.,][]{Chauvin:2004cy,Chauvin:2005dh,Kalas:2008,Marois:2008ei,Marois:2010b,Lagrange:2010,Carson:2013fw, Quanz:2013ii,Rameau:2013dr,Macintosh:2015ew}. Following the submission of this work, \citet{Chauvin:2017} announced the discovery of a planet orbiting the star HIP~65426.  For planets at wide separation ($>5$\,au), it is particularly interesting to consider the dynamics of the system that could influence the formation and migration of the planets \citep[e.g.,][]{Rodet:2017}.  Indeed, several of the stars that host directly-imaged planets are components of a multiple system, including 51~Eridani, which is orbited at $\sim$2000\,au by GJ 3305, a 6\,au binary M dwarf pair \citep{Macintosh:2015ew, DeRosa:2015jl, montet:2015}, and Fomalhaut \citep{Kalas:2008}, with TW~Piscis~Austrini and LP~876-10 at $\sim$54000 and $\sim$160000\,au projected separation \citep{Mamajek:2013a}.  Both these cases have a planet much closer to its parent star than the stellar companions, and so locating planets at more intermediate distance between primary star and stellar companions will help guide our understanding of how planets in binaries form and evolve.

HD~131399 is a young ($16\pm7$\,Myr) triple star system in the Upper Centaurus Lupus (UCL) association, a sub-group of the Scorpius-Centaurus (ScoCen) association \citep{deZeeuw:1999fe,Rizzuto:2011gs,Pecaut:2016} located at a distance of $98.0\pm6.9$\,pc \citep{vanLeeuwen:2007dc}. The hierarchical system comprises the central A-type star with a spectral type of A1V \citep{Houk:1988wv} and a tight pair composed of a G and a K star at a projected separation more than $3\arcsec$ ($\sim300$ au) from A \citep{Dommanget:2002ud}. During a survey carried out with the Spectro-Polarimetric High-contrast Exoplanet REsearch instrument \citep[SPHERE,][]{Beuzit:2008} at the VLT, a candidate planet was recently discovered in the system at a projected separation of $0\farcs 83$ ($\simeq82$\,au) \citep[hereafter W16]{wagner2016}. To assess the status of the source, astrometric follow-up was carried out eleven months later. The stationary background hypothesis was ruled out since both the star and the source share common proper motion. The co-moving scenario was also supported by a probability of $6.6\times10^{-6}$ to detect a cold ($<1500~$K) but unbound object along the line of sight at this stage of their survey. Moreover, the follow-up showed a motion consistent with an orbit around HD~131399~A. W16 reported a luminosity-based model-dependent mass of $4\pm1$\,$M_{\rm Jup}$, effective temperature of $850\pm50$\,K, and a spectral type of T2--T4 with the detection of methane in the {\it H} and {\it K} bands. The importance of HD~131399~Ab in the field is threefold: wide-orbit giant planets can be formed in hierarchical systems; the system is a good example to test dynamical evolution; and the planet is one of the few known at a low temperature ($<1000$\,K) to test atmospheric models.

Given the significance of this discovery, HD~131399~Ab was observed in 2017 with the Gemini Planet Imager \citep[GPI,][]{Macintosh:2014js} at the Gemini South observatory, with SPHERE at the VLT, and with the Near-Infrared Camera and Coronagraph (NIRC2) and the facility adaptive optics system \citep{Wizinowich:2006cp} at Keck observatory. The analysis of the data reveals unexpected spectroscopic and astrometric results that motivated the reanalysis of some of the already published data obtained with VLT/SPHERE. In Section \ref{sec:obs}, we discuss the observations, data reduction, astrometric and spectral extraction. The spectral energy distribution (SED) of HD~131399~A and HD~131399~Ab are presented and analyzed in Section \ref{sec:analysis}, and the astrometric measurements and analysis are presented in Section \ref{sec:astrometry}. The status of HD~131399~Ab is discussed in Section \ref{sec:discuss} and conclusions are drawn in Section \ref{sec:conc}.

\section{Observations and Data Reduction}\label{sec:obs}

\begin{deluxetable*}{ccccccccccc}
\tablecaption{\label{tab:obs}Observing log}
\tablewidth{0pt}
\tablehead{
\colhead{UT Date} & \colhead{Instrument} & \colhead{Mode} & \colhead{Filter(s)} & \colhead{Resolution} & \colhead{$t_{\rm int}$} & \colhead{$N_{\rm coadd}$} & \colhead{$N_{\rm exp}$} & \colhead{Field of view} & \colhead{DIMM seeing} & \colhead{\% time with Ab}   \\
&  & & &  & (s) & & & rotation (deg)  & ($\arcsec$)  & on chip \\
}
\startdata
2015 Jun 12 & SPH-IFS    & Spectroscopy & {\it YJH} & 30 & 32 & 1  &  50  & 38.0 & 1.0 & 46  \\
             & SPH-IRDIS & Imaging & $K1K2$ & \nodata & 16 & 1  & 96 & 37.2 &  1.0 & 100 \\
2016 Mar 06 &  SPH-IFS    & Spectroscopy & {\it YJH} & 30 & 32 & 1  & 84  & 41.1 &  1.1 & 67 \\
             & SPH-IRDIS & Imaging & $K1K2$ & \nodata & 32 & 1  & 63 & 34.0 & 1.1 & 100  \\
2016 Mar 17 &  SPH-IFS    &  Spectroscopy &{\it YJH} & 30 & 32 & 1  & 56  & 37.8 & 1.2 & 100   \\
             & SPH-IRDIS & Imaging & $K1K2$ & \nodata & 32 & 1  & 56 & 37.3 & 1.2 & 100  \\
2016 May 07 &  SPH-IFS    &  Spectroscopy & {\it YJH} & 30 & 32 & \nodata  & 56  & 41.3 & 1.0 & 30   \\
             & SPH-IRDIS & Imaging & $K1K2$ & \nodata & 32 & \nodata  & 56 & 40.4 & 1.0 & 100  \\
2017 Feb 08 & NIRC2 & Imaging & {\it L}$^\prime$ & \nodata & 0.9 & 30 & 166 & 37.0 & \nodata & 100 \\
2017 Feb 14 & GPI & Spectroscopy & $K1$ & 66 & 60 & 1 & 112 & 93.5 & 0.9 & 100 \\
2017 Feb 15 & GPI & Spectroscopy & {\it H} & 46 & 60 & 1 & 83 & 107.9 & 1.0 & 100 \\
2017 Feb 16 & GPI & Spectroscopy & {\it J} & 37 & 60 & 1 & 96 & 110.4 & 0.7 & 100  \\
2017 Mar 15 & SPH-IRDIS & Polarimetry & {\it J} & \nodata & 64 & 1 & 20 & 5.3 & 0.6 & 100 \\
2017 Apr 20 & GPI & Spectroscopy & {\it H} & 46 & 60 & 1 & 62 & 133.3 & \nodata & 100 \\
\enddata
\end{deluxetable*}

This paper uses ten datasets that were obtained with three different adaptive optics instruments, mounted on three different telescopes, all making use of the angular differential imaging technique \citep[ADI,][]{Marois:2006df}. Six out of the ten datasets are new from GPI, SPHERE, and NIRC2. The remaining data come from SPHERE and were previously published in W16, but reanalyzed as part of this work. The date, instrument, filter and resolution, exposure times, parallactic angle extent, and DIMM seeing of the observations are detailed in Table~\ref{tab:obs}. We also computed the fraction of time HD\,131399\,Ab (over one full width at half maximun, FWHM) was effectively on the detector for each dataset, because of its particular orientation with respect to the SPHERE-IFS detector. We provide more details on the observing sequence and data reduction below.

\subsection{New Gemini-South/GPI Observations}
\label{sec:GPI}
\begin{figure}
\centering
\includegraphics[width=0.5\columnwidth]{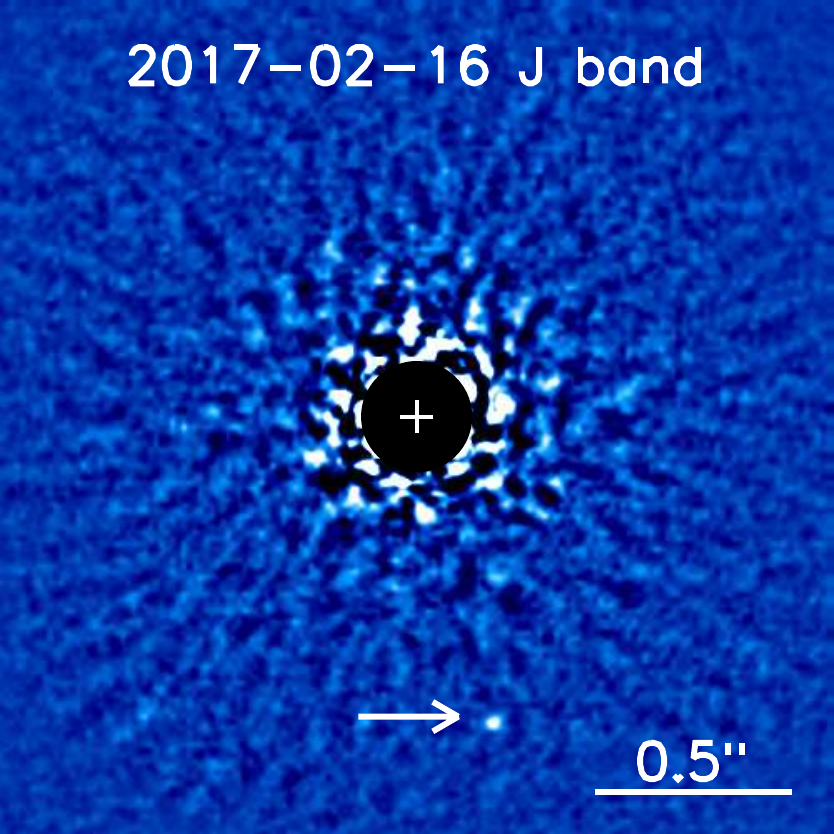}\includegraphics[width=0.5\columnwidth]{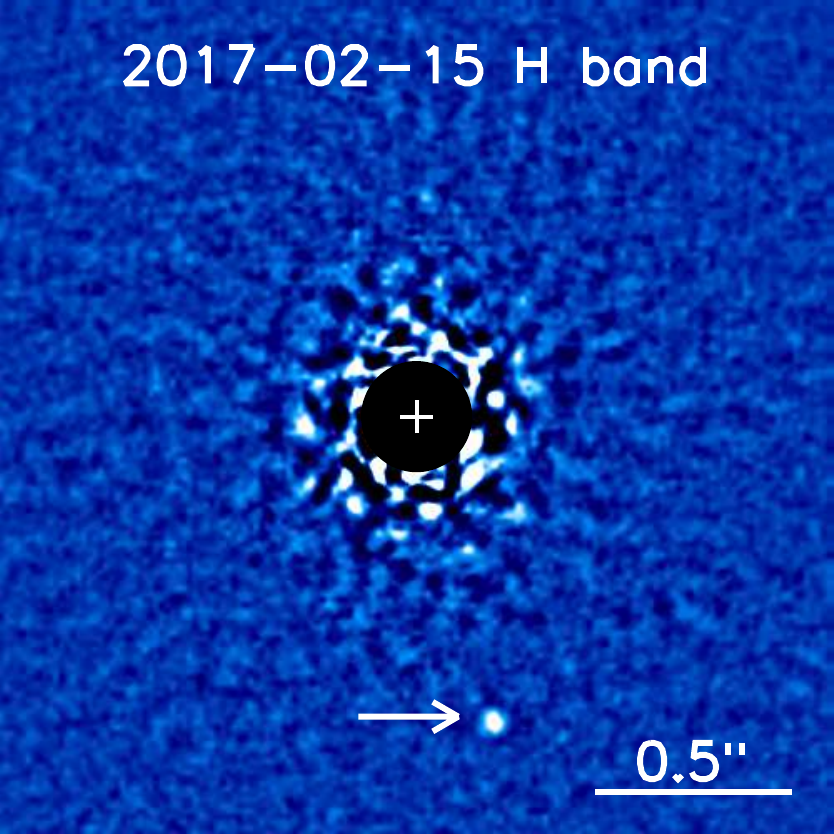}\\
\includegraphics[width=0.5\columnwidth]{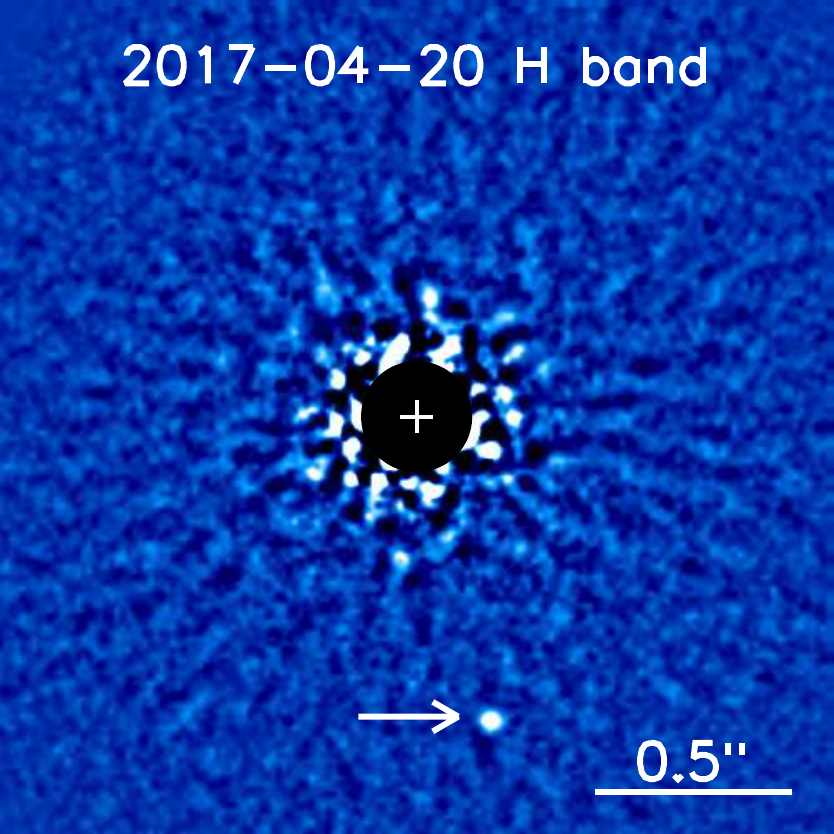}\includegraphics[width=0.5\columnwidth]{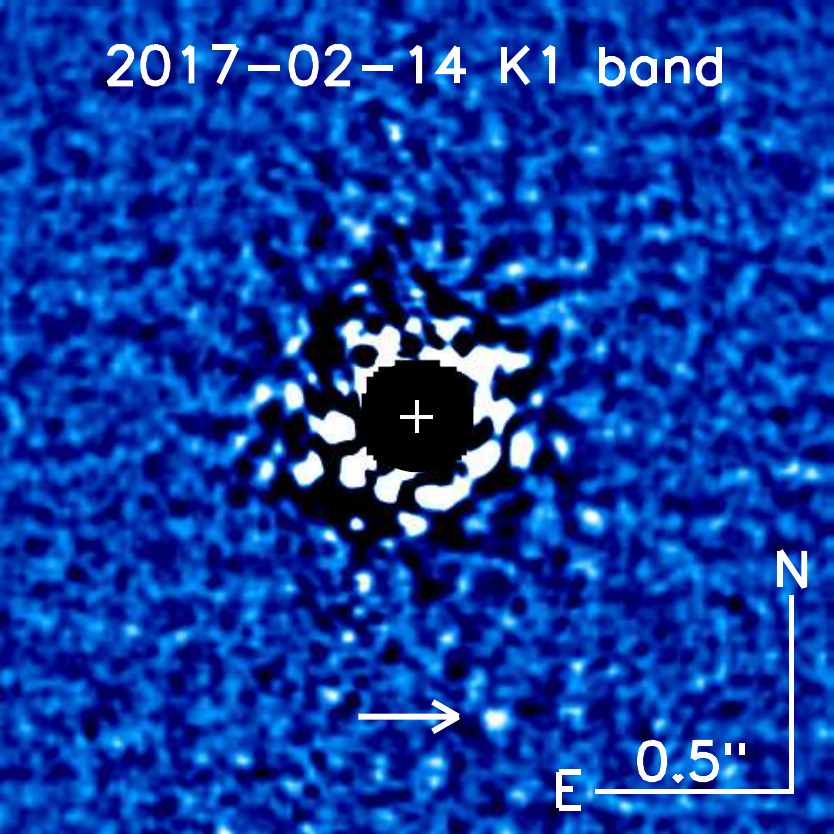}
\caption{\label{fig:imGPI}cADI PSF-subtracted images of HD~131399~Ab obtained with GPI in 2017 in {\it J} (top left), {\it H} (top right and bottom left), and $K1$ (bottom right) bands. A two-pixel low-pass filter was applied on the images to suppress shot noise. Intensity scales are linear, different in each image, and chosen to saturate the PSF of HD~131399~Ab. The central star is masked numerically, and its position is marked by the white cross.}
\end{figure}

HD~131399~A was observed with GPI at two epochs, 2017 February and 2017 April, as part of the GPI Exoplanet Survey (GS-2015B-Q-501). Three datasets were obtained on consecutive nights in 2017 February, with a total on-source integration time of 1.87\,hr at $K1_{\rm GPI}$ ($\lambda_{\rm eff} = 2.06$\,\micron), 1.38\,hr at {\it H} ($\lambda_{\rm eff} = 1.64$\,\micron) and 1.60\,hr at {\it J} ($\lambda_{\rm eff} = 1.23$\,\micron). An additional dataset was obtained on 2017 April 20 at {\it H} with an on-source integration time of 1.03\,hr. Each dataset was obtained in the spectral coronagraphic mode of the instrument.

To create spectral datacubes, the raw data were reduced with the GPI Data Reduction Pipeline v1.4.0 \citep[DRP,][]{Perrin:2014,Perrin:2016gm}, which subtracts the dark current, removes the microphonics noise \citep{Chilcote:2012hd, Ingraham:2014ef}, and identifies and removes bad pixels. Instrument flexure is compensated using observations of an argon arc lamp taken immediately prior to each sequence at the target elevation \citep{Wolff:2014}. Microspectra are then extracted to create 37-channel datacubes \citep{Maire:2014gs}, which are corrected for any remaining bad pixels and finally for distortion \citep{Konopacky:2014}. The last step consists of measuring in each image the location of the four satellite spots---attenuated replicas of the central point spread function (PSF) created by a diffraction grating in the pupil-plane---to accurately measure the position and flux of the central star during the sequence \citep{Wang:2014}. The position of each satellite spot flux is written in the header so that they can be used for calibration.

Further processing to remove the stellar PSF and extract the astrometry and spectrophotometry of HD\,131399\,Ab was performed using two different pipelines to mitigate biases and systematics introduced by the data processing.

In the first pipeline, a Fourier high-pass filter with a smooth cutoff frequency of four spatial cycles was applied to each image. The speckle field was then estimated and subtracted using the classical ADI algorithm \citep[cADI][]{Marois:2006df} (following the definition of \citealt{Lagrange:2010} as a median-combination) for each sequence in each wavelength slice, which was then rotated to align north with the vertical axis and averaged over the sequence. Broad-band images were further created from the stack of the individual slices, examples of which are shown in Figure \ref{fig:imGPI}. The astrometry and broad-band contrasts of HD~131399~Ab were extracted in each dataset from the broad-band images using the negative simulated planet technique \citep{Marois:2010b,Lagrange:2010}. A template PSF of HD\,131399\,Ab was created from the temporal and spectral average of the four satellite spots. The template was injected in the raw datacubes at a trial position but opposite flux of HD\,131399\,Ab and the same reduction as for the original set was executed. The process was iterated over these three parameters (separation, position angle, flux) to minimize the integrated squared pixel noise in a wedge of $3\times3$ FWHM centered at the trial position. The minimization was performed with the amoeba-simplex optimization algorithm \citep{Nelder:1965in} and provided the best fit broad-band contrast and position. Uncertainties on HD\,131399\,Ab location and contrast were calculated by injecting independently twenty positive templates at the same separation and contrast as HD\,131399\,Ab but different position angles. The fitting procedure was repeated for each simulated source and the measurement errors obtained from the statistical dispersion on the three parameters. Finally, the contrasts---and associated measurement errors---in individual slices in each set were then extracted following the same procedure at the best fit position, which is fixed, and varying only the flux of the template, which is built for each wavelength from the corresponding satellite spots.

The second pipeline used \texttt{pyKLIP} \citep{Wang:2015th}, an open-source Python implementation of the Karhunen-Lo\`eve Image Projection algorithm \citep[KLIP][]{Soummer:2012ig}. Before PSF subtraction, the images were high-pass filtered using a seven-pixel FWHM Gaussian filter in Fourier space to remove the smooth background. KLIP was run on a 22-pixel wide annulus centered on the location of the source. To build the model of the stellar PSF, we used the 150 most-correlated reference images in which HD\,131399\,Ab moved at least a certain number of pixels due to ADI and SDI observing methods (the exclusion criteria). Since we will forward model the PSF of the planet, including the effects of self-subtraction, we use an aggressive exclusion criteria of 1.5\,pixels for all wavelengths except {\it J}-band where we found using images very close in time most accurately modeled the speckles and thus a 0.2\,pixel exclusion criteria worked best. As the source is far from the star and thus from the majority of the speckle noise, we used only the first five KL basis vectors to reconstruct the stellar PSF. All images were then rotated to align north up, and collapsed in time and wavelength, resulting in one 2-D image per epoch. The astrometry and broad-band photometry were measured from these images using the Bayesian KLIP-FM Astrometry (BKA) technique \citep{Wang:2016gl} that is implemented in \texttt{pyKLIP}. In BKA, we concurrently forward model the PSF of HD\,131399\,Ab during KLIP. To do this, we used the average of the satellite spots to model the instrumental PSF at each wavelength, and assumed HD\,131399\,Ab had a spectral shape that was the same as HD\,131399\,A.  As noted in \citet{Wang:2016gl}, spectra differing by even 20\% did not affect the astrometry, so we did not require a precise input spectral template for our forward model. After generating the forward model, we used the affine-invariant sampler implemented in \texttt{emcee} \citep{ForemanMackey:2013io} to compute the posterior distribution of the location and flux of HD\,131399\,Ab. Our MCMC sampler used 100 walkers, each iterating for 800 steps after 300 steps were discarded as the ``burn in". To obtain accurate uncertainties, the residual speckle noise in the image was modelled as a Gaussian process with a spatial correlation described by the Mat\'{e}rn covariance function. We adopt the 50th percentile values as the position of HD\,131399\,Ab and the 16th and 84th percentile values as the 1$\sigma$ uncertainty range. To obtain the spectrum of HD\,131399\,Ab in each filter, we performed a PSF subtraction with KLIP that only used ADI to model the stellar PSF, allowing us to forward model the PSF of HD\,131399\,Ab without any spectral dependencies. Then, we modified BKA to run independently on each spectral channel to obtain the flux and uncertainty on the flux at each wavelength. As the planet has significantly lower signal-to-noise ratio (SNR) in each spectral channel than in a collapsed broad-band image, we restricted the position of the planet to be within 0.1 pixels of the position we measured in the broad-band data. 

Astrometric calibration was obtained with observations of the $\theta^1$~Ori~B field and other calibration binaries following the procedure described in \citet{Konopacky:2014} and used to convert the detector positions into on-sky astrometry. The astrometric error budget consists of the following added in quadrature: the measurement errors described previously; a star registration error of $0.7$\,mas from \citet{Wang:2014}; a plate scale error of $0.007$\,mas\,lenslet$^{-1}$; and position angle offset error of $0.13$\,deg, the last two from \citet{Konopacky:2014}.

The raw astrometric and photometric measurements from the two pipelines ($i=1,2$) agreed very well to better than 1$\sigma$ at each epoch. The pairs ($x_i,\sigma_i$) from the two pipelines for each dataset were combined with a weighted average $x_\mathrm{tot}=\sum_iw_ix_i/\sum_iw_i$, where $w_i=1/\sigma_i^2$. The measurement errors were computed as $\sigma_\mathrm{tot}=\sqrt{\sum_i\sigma_i^2w_i/\sum_iw_i}$ since they are not independent. The systematic errors (registration, calibration) were then added in quadrature to calculate the final astrometric uncertainties. 

Photometric measurements from different epochs ($j=1,2$) were also combined with the same weighted mean but the errors were computed as $\sigma_\mathrm{tot}=\sqrt{1/\sum_jw_j}$ since they are independent. Finally, the systematic uncertainties of the star-to-satellite-spot ratios ($0.03$\,mag in {\it J} band, $0.06$\,mag in {\it H} band, and $0.07$\,mag in $K1$ band, \citealt{Maire:2014gs}) were added in quadrature to the final contrast errors. The spectrum was then obtained by multiplying the contrasts with the spectrum of the central star (see Section \ref{sec:sedA}).

SNRs for each dataset were computed using the \texttt{pyKLIP} implementation of the Forward Model Matched Filter (FMMF) algorithm \citep{Ruffio2017}, using the stellar spectrum of HD~131399~A as the spectral template in the matched filter. Like the two pipelines to extract astrometric and photometric data, FMMF similarly utilizes forward modelling of point sources through the PSF subtraction process for the data analysis, but is better optimized for planet detection. Thus, FMMF produces SNRs that are comparable or slightly better than the SNRs inferred from the astrometric for photometric errors.

All measurements are discussed in Sections \ref{sec:SEDb} and \ref{sec:astrometry}.

\subsection{Public VLT/SPHERE data and new observations}
\label{sec:obs_SPH}

\subsubsection{Reanalysis of public data}
\label{subsec:public_sphere}

\begin{figure*}
\centering
\includegraphics[width=0.3\textwidth]{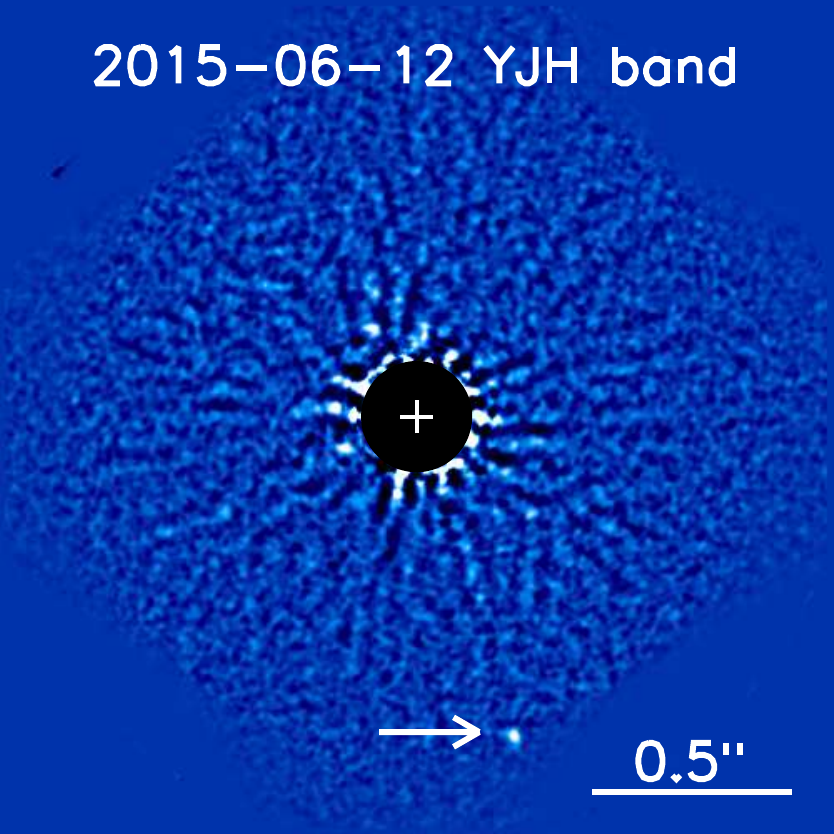}\includegraphics[width=0.3\textwidth]{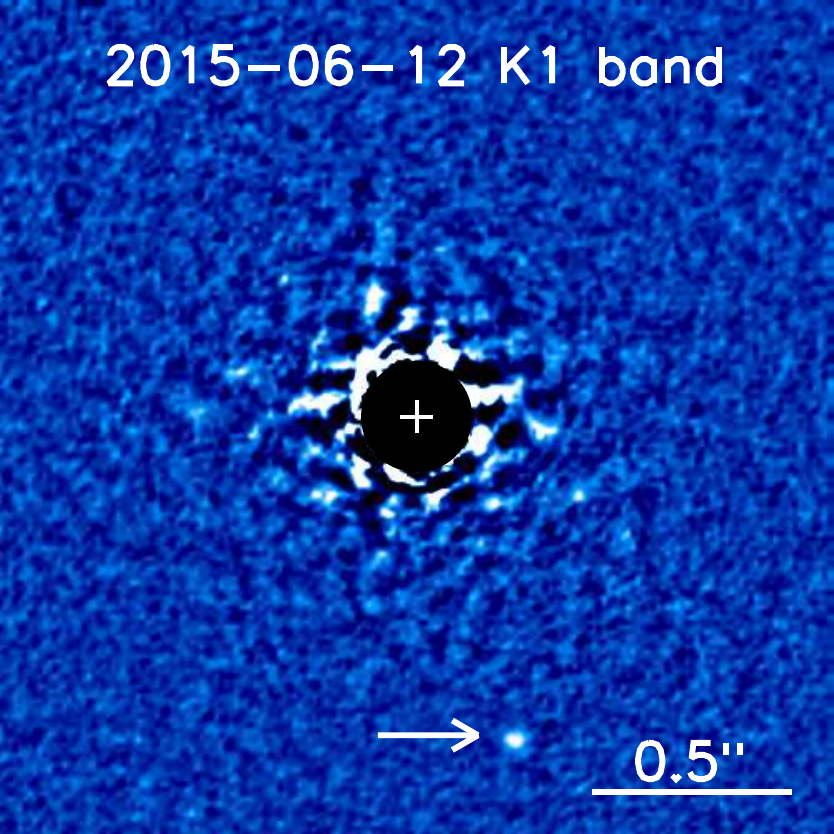}\includegraphics[width=0.3\textwidth]{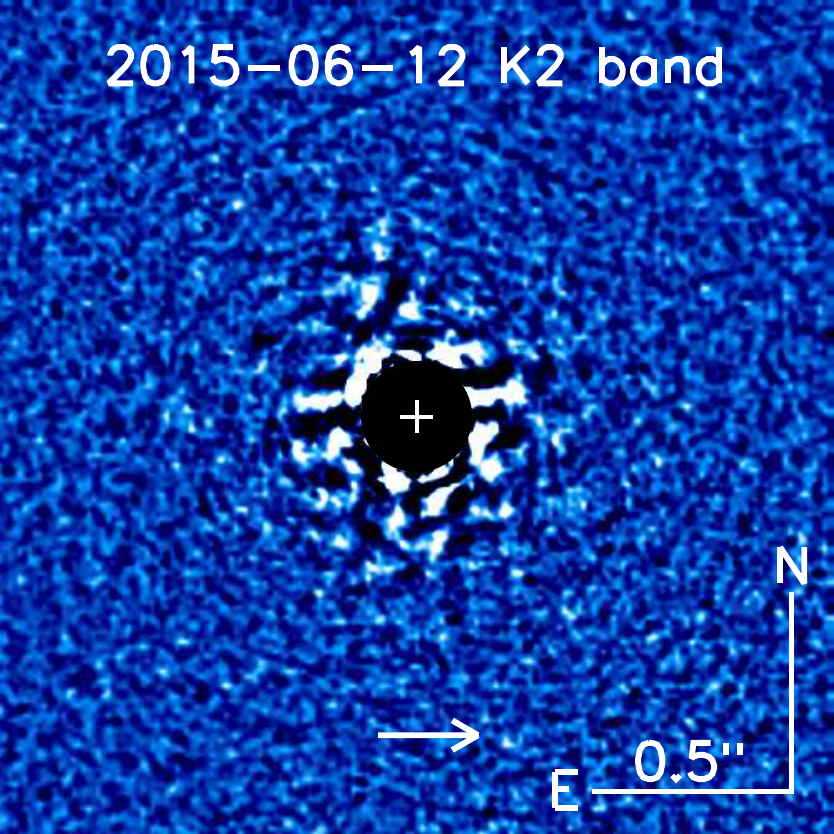}
\includegraphics[width=0.3\textwidth]{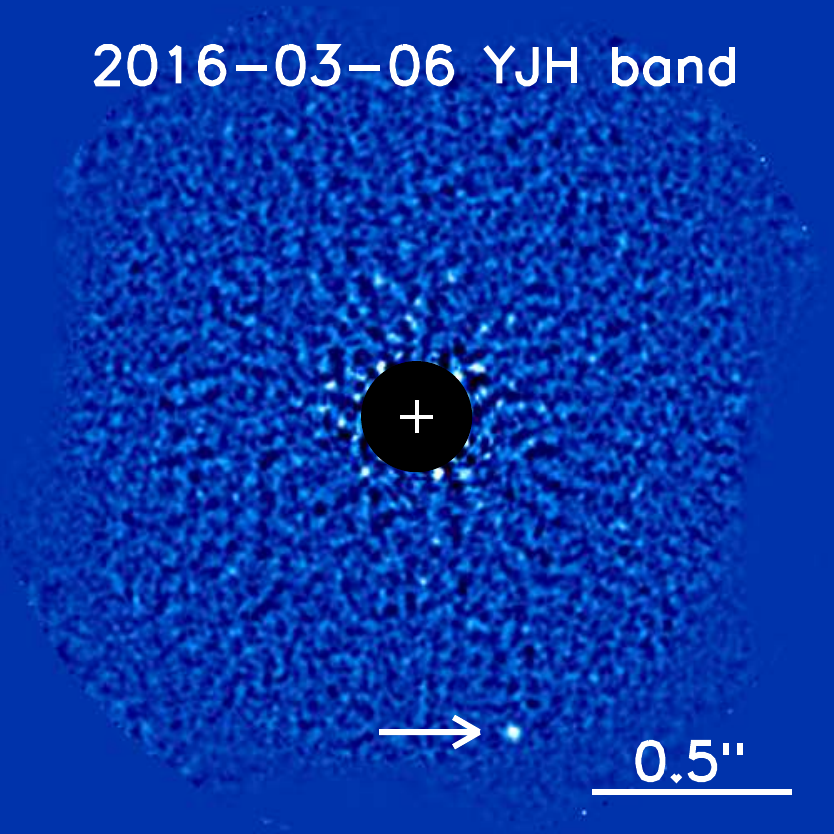}\includegraphics[width=0.3\textwidth]{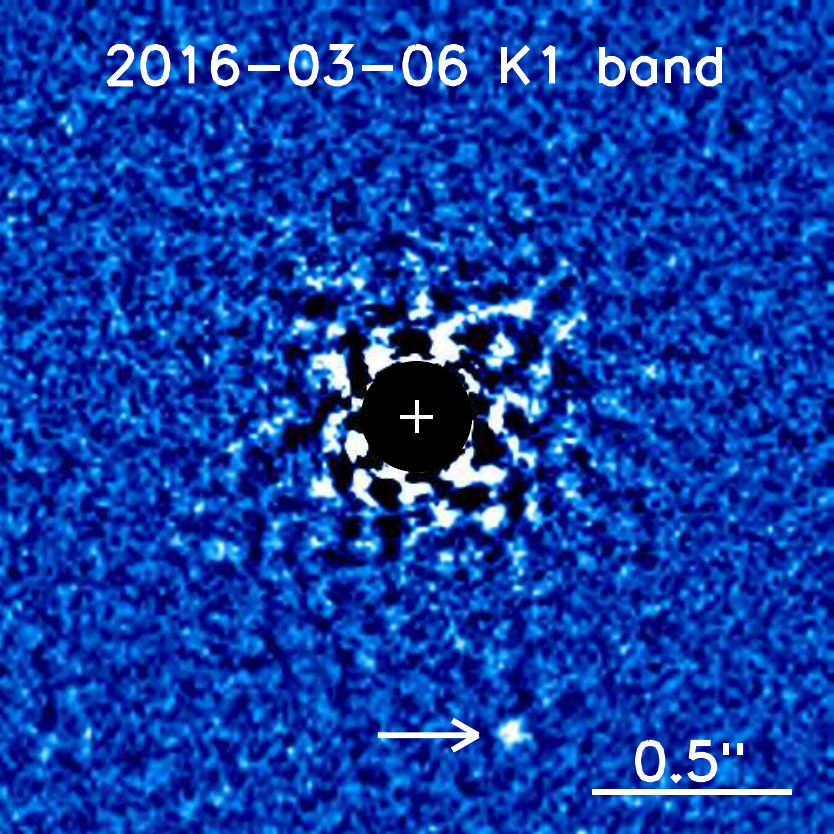}\includegraphics[width=0.3\textwidth]{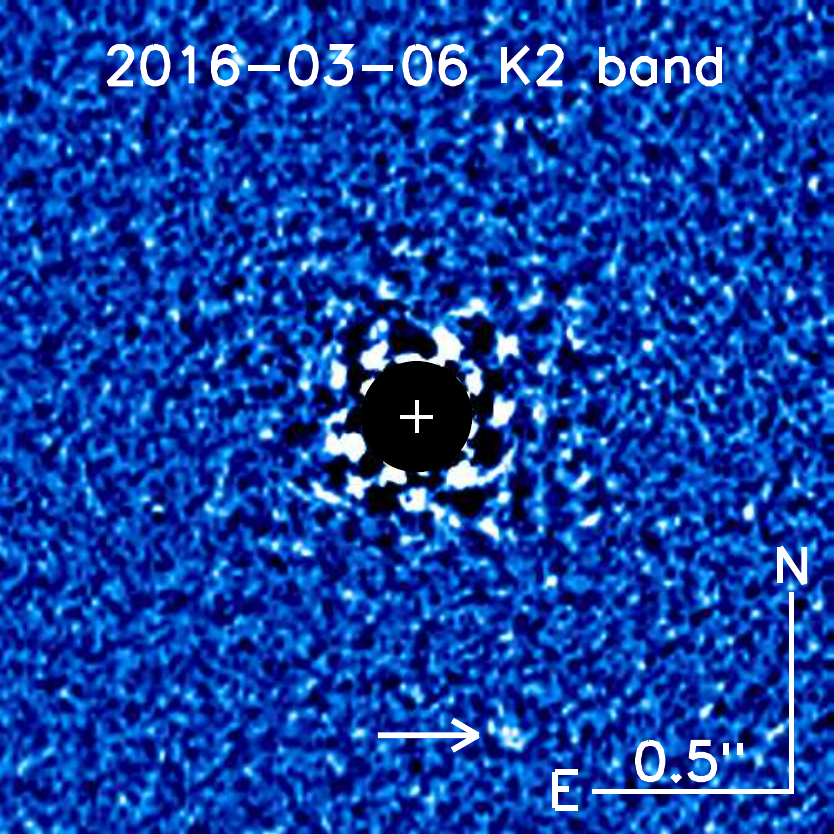}
\includegraphics[width=0.3\textwidth]{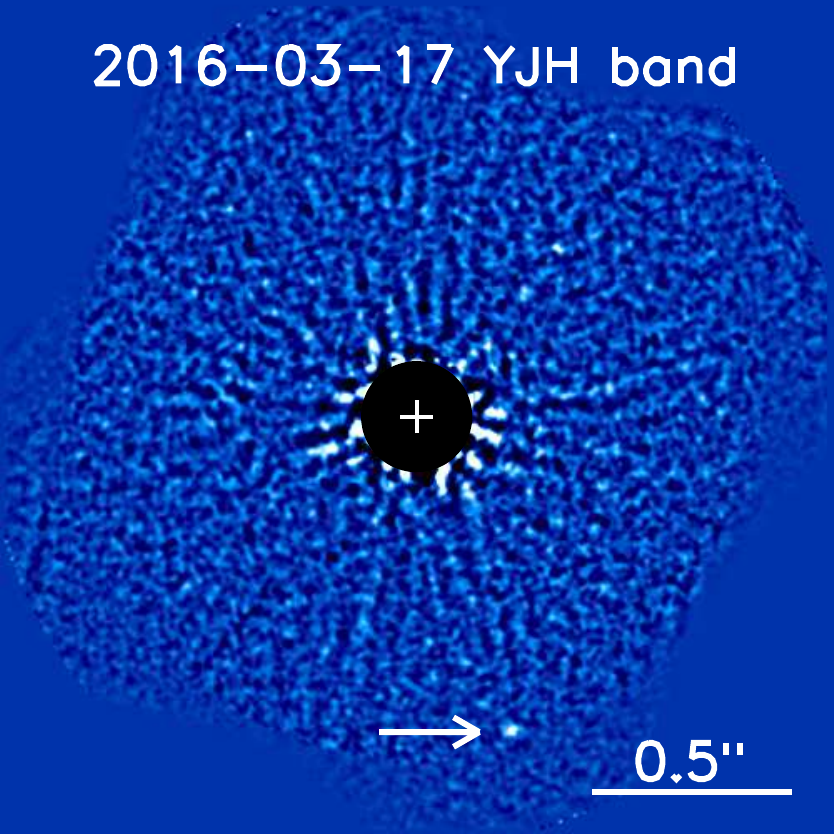}\includegraphics[width=0.3\textwidth]{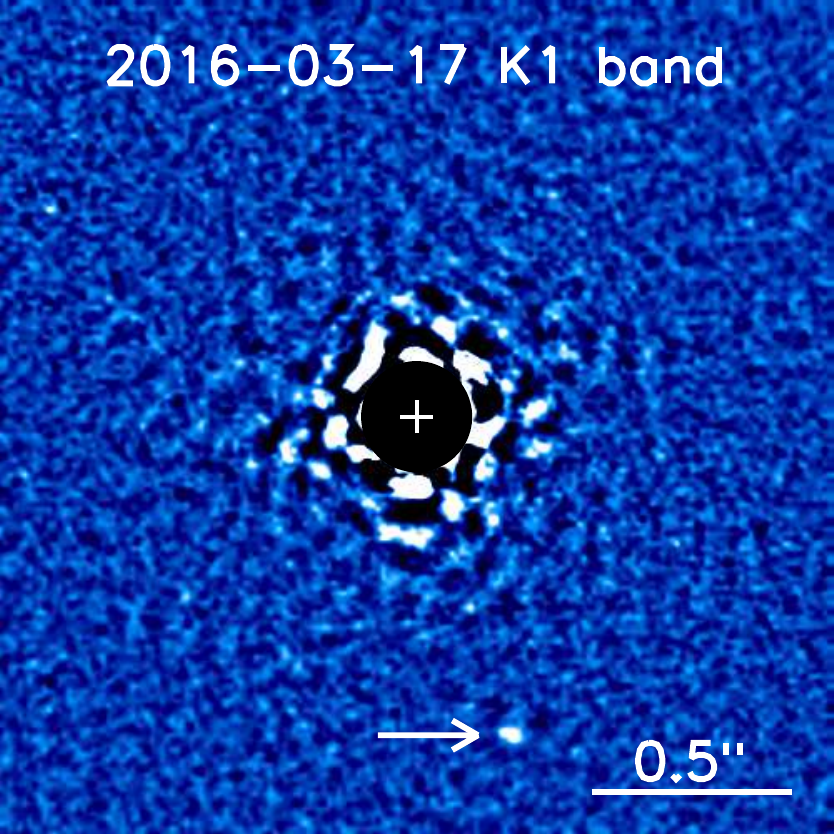}\includegraphics[width=0.3\textwidth]{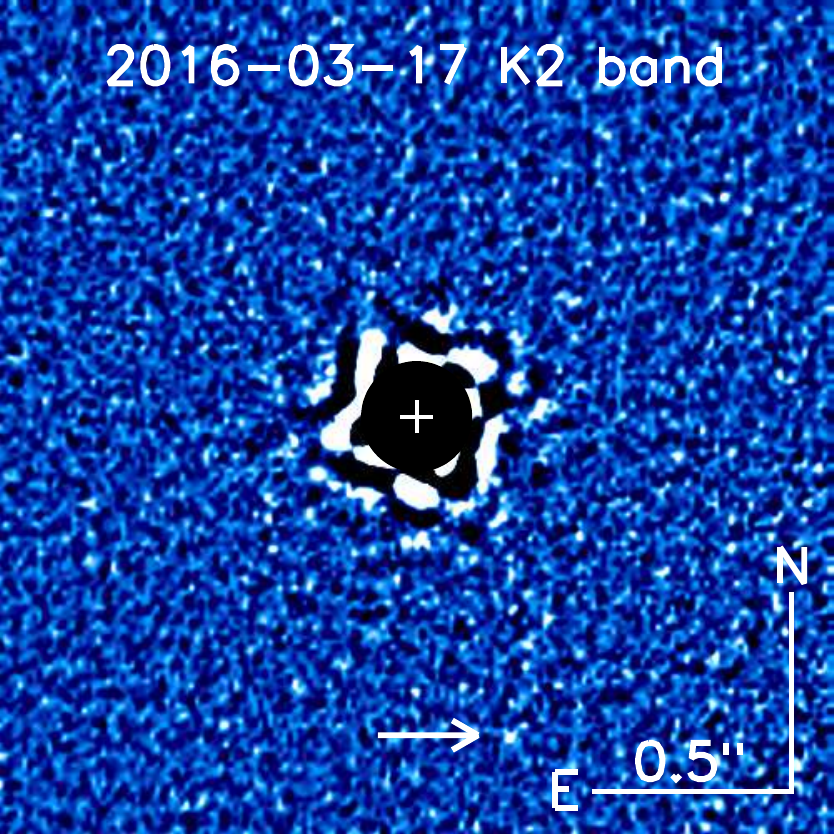}
\includegraphics[width=0.3\textwidth]{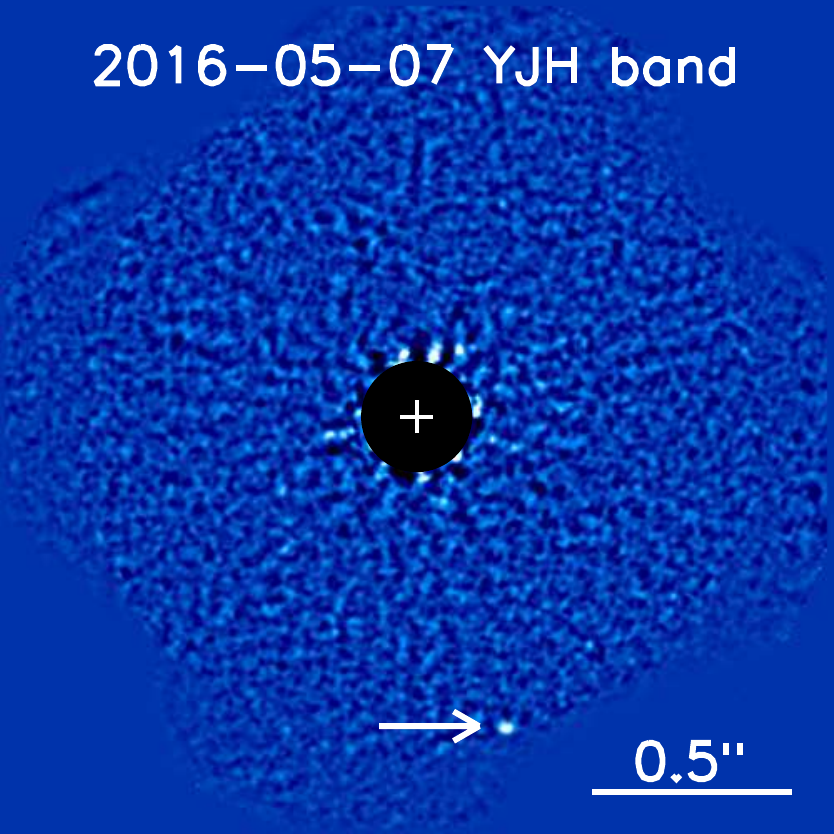}\includegraphics[width=0.3\textwidth]{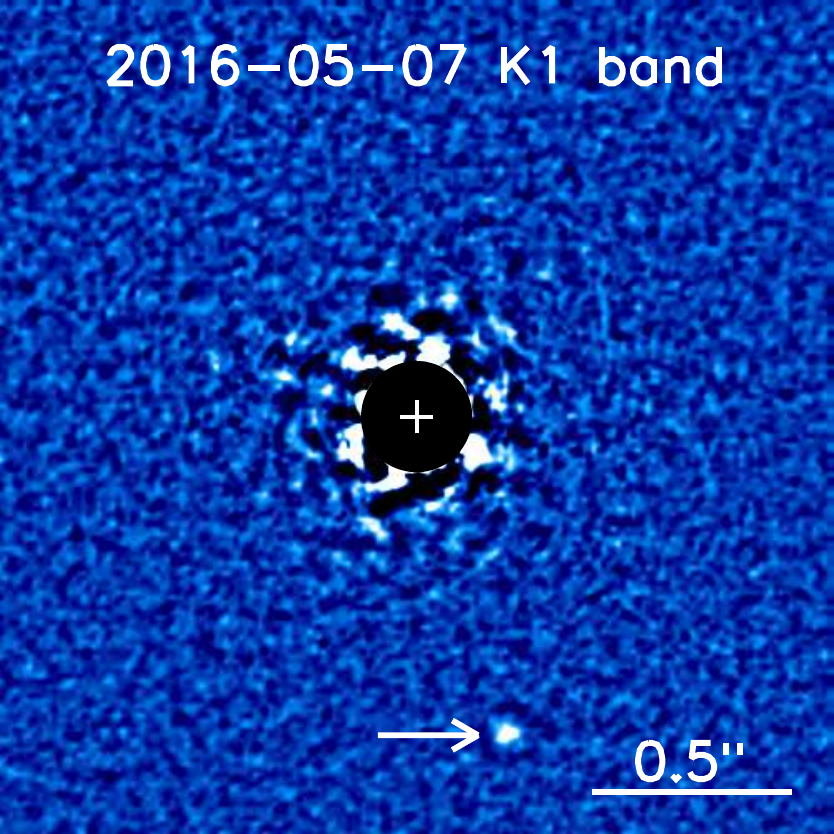}\includegraphics[width=0.3\textwidth]{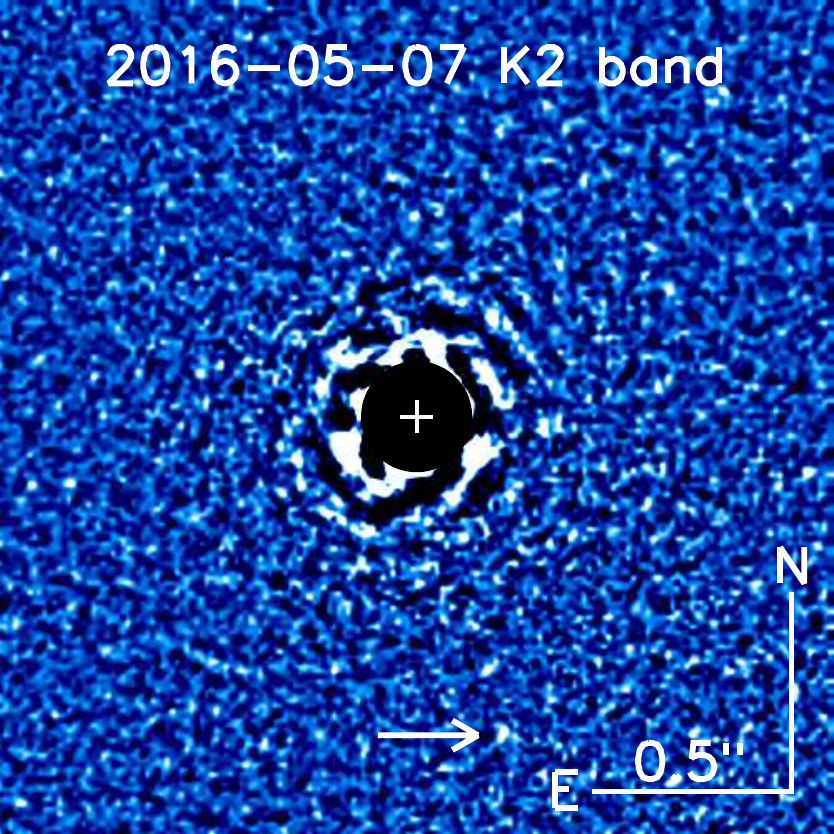}
\caption{\label{fig:imSPHERE}Reanalyzed cADI PSF-subtracted images of the publicly-available datasets of HD~131399~Ab obtained with SPHERE on 2015 June 12 (top row), 2016 March 06 (second row), 2016 March 17 (third row), and 2016 May 07 (bottom row). In each row, the left column contains IFS 39-channel cubes stacked into single {\it YJH} image. The source lies at the very edge of the field-of-view. The middle and right columns contain IRDIS images at $K1$ and $K2$ respectively. The source is barely detected in $K2$. The image design is similar to Figure \ref{fig:imGPI}.}
\end{figure*}

Four epochs of observations were obtained with SPHERE by W16 between 2015 June and 2016 May, all of which are publicly available on the ESO archive \footnote{\url{http://archive.eso.org/}}. We downloaded the data as well as the associated raw calibration files. Briefly, the HD~131399 system was observed with the \texttt{IRDIFS\_EXT} mode using simultaneously the Integral Field Spectrograph (IFS, \citealt{Claudi:2008}) instrument in spectroscopic mode from $0.95$--$1.65$\,\micron\ ({\it YJH}) and the Infra-Red Dual-beam Imaging and Spectroscopy (IRDIS, \citealt{Dohlen:2008}) instrument in dual-band imaging mode (DBI, \citealt{Vigan:2010}) at $K1_{\rm SPH}$ ($\lambda_{\rm eff}=2.10$\,\micron) and $K2_{\rm SPH}$ ($\lambda_{\rm eff}=2.25$\,\micron), with all SPHERE filter profiles being different from those of GPI (see Section~\ref{sec:sedA} and Figure \ref{fig:filters}). The IRDIS detector was dithered on a $4\times4$ pattern. A total of 0.44\,hr, 0.75\,hr, 0.50\,hr, and 0.50\,hr were obtained on the IFS on 2015 June 12, 2016 March 06, 2016 March 17, and 2016 May 07 respectively, and 0.43\,hr, 0.56\,hr, 0.50\,hr and 0.50\,hr on IRDIS, the difference between the two detectors being due to readout overheads. Each observing sequence started and finished with a brief ``star-center'' coronagraphic sequence in which four satellite spots are created from a periodic modulation introduced on the deformable mirror, the barycenter of these spots being used to measure the position of the star behind the focal plane mask during the sequence. In practice, the star position is very stable \citep{Zurlo:2016, Vigan:2015}. A brief off-axis ($\simeq0\farcs4$) ``flux'' sequence with the neutral density filter ND3.5 (attenuation factor from $4\times10^2$ to $2\times10^4$ dependant on wavelength) was then executed to obtain a template and the flux of the target PSF. The on-axis coronagraphic sequence was then carried out. Calibration data were obtained during the following days: darks, detector flat fields, integral field unit flat (broad-bang lamp image to register the IFS microspectra), and a wavelength calibration frame.

\paragraph{IFS data processing}
The raw data and calibration files were reduced using the SPHERE IFS pre-processing tools v1.2\footnote{\url{http://astro.vigan.fr/tools.html}} \citep{Vigan:2015}, which make use of custom IDL routines and the ESO Data Reduction and Handling (DRH) package v22.0 \citep{Pavlov:2008}. These tools were updated with the latest calibration values provided by \citet{Maire:2016} and the ESO SPHERE user manual 7th edition\footnote{\url{https://www.eso.org/sci/facilities/paranal/instruments/sphere/doc/VLT-MAN-SPH-14690-0430_v100.pdf}}: instrument angle updates (pupil offset of $135.99$\,deg, and IFS angle offset of $-100.48$\,deg), the IFS anamorphism correction ($1.0059$ along the horizontal direction, $1.0011$ along the vertical direction), and the parallactic angle correction $\epsilon$, a small factor to correct the parallactic angle calculation from a mis-synchronisation between the VLT and SPHERE internal clock that affect data taken before 2016 July 13. Additionally, the tools were updated to process the entire field-of-view (it was originally cropped by five pixels on the edges). The pre-processing tools used the DRH package to create the master darks, bad pixel maps, the microspectra position map, the IFU flat-field, and the wavelength calibration file. Detector flats were created with a custom IDL routine. The data pre-processing were then executed by a custom IDL routine, which subtracts the dark current, removes the bad pixels and corrects for cross-talk. This was followed by processing through the DRH, which corrects for flat-fielding and extracts the microspectra to create 39-channel datacubes. The 3-D datacubes were then digested by a custom IDL routine to remove the remaining bad pixels, to correct from the anamorphism, to register the spot locations in the star-center frames, to align the coronagraphic and the off-axis PSF frame at the center, and to recalibrate the wavelengths.

\paragraph{IRDIS data processing}
A custom set of tools to reduce IRDIS-DBI data was developed following the IFS philosophy, combining both DRH and IDL routines. IRDIS DBI raw data are made from images in two side-by-side quadrants, being associated to the $K1$ (left) and $K2$ (right) filters. The DRH first created the master darks, flat-fields, and associated bad pixel maps. Our IDL routine then performed the dark current subtraction, flat-field division, bad pixel removal, vertical anamorphism correction by a factor of $1.006$ \citep{Maire:2016}, and parallactic angle calculation and correction by the $\epsilon$ factor. For each image, the two quadrants were separated at the end to create a master datacube for each filter. The locations of the satellite spots and frame registration, taking into account the dithering offset from the header keywords, were performed as for the IFS data as final processing steps.

\begin{figure}
\includegraphics[width=\columnwidth]{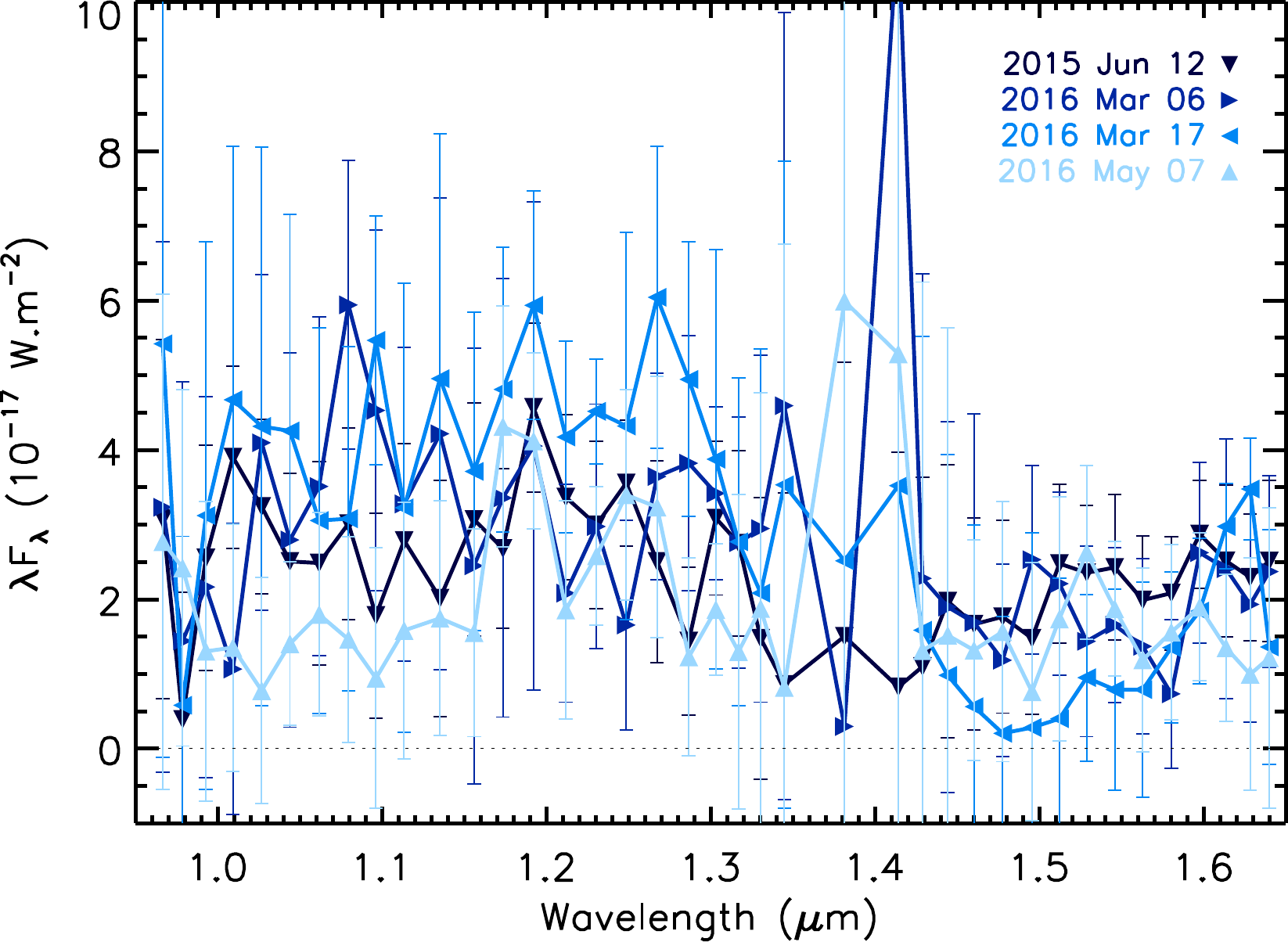}
\caption{\label{fig:SPH-spec}
{\it YJH} SPHERE-IFS spectra of HD~131399~Ab extracted from 2015 June 12 (dark blue downward triangles), 2016 March 06 (blue rightward triangles), 2016 March 17 (sky blue leftward triangles) and 2016 May 07 (light blue upward triangles). As HD~131399~Ab is barely detected in individual channels, all epochs are noisy, but the spectra are  consistent in the {\it YJ} band. Only the third epoch exhibits a steep slope in the {\it H} band, with a peak near $1.62$\,\micron. Discrepancies around $1.35-1.40$\,\micron\ can be explained by the significantly lower atmospheric transmission at these wavelengths.}
\end{figure}
\begin{figure}
\centering
\includegraphics[width=0.6\columnwidth]{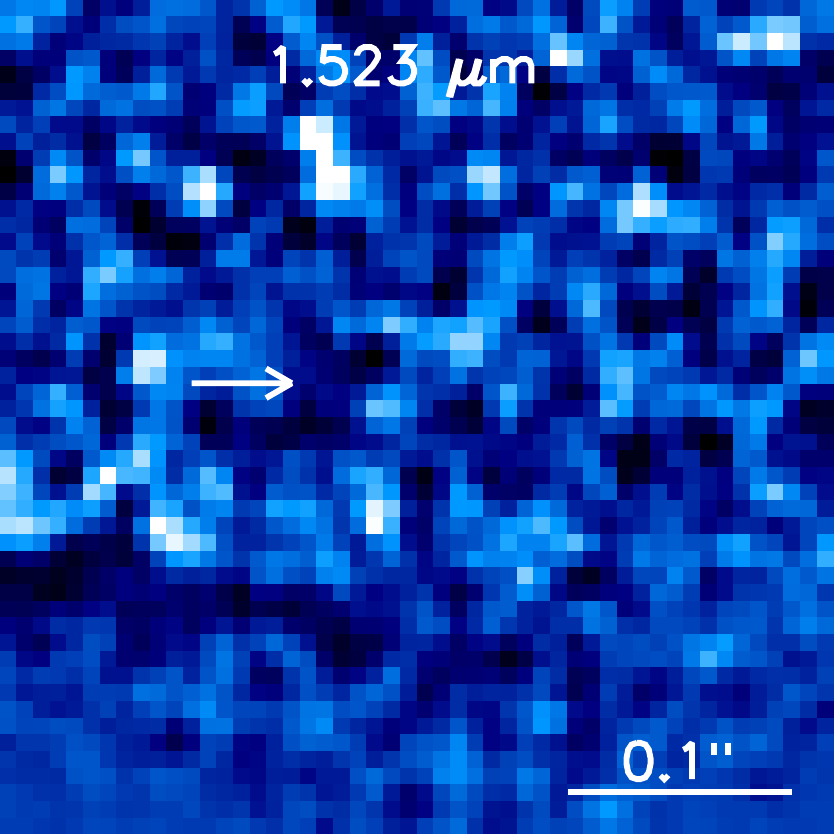}
\includegraphics[width=0.6\columnwidth]{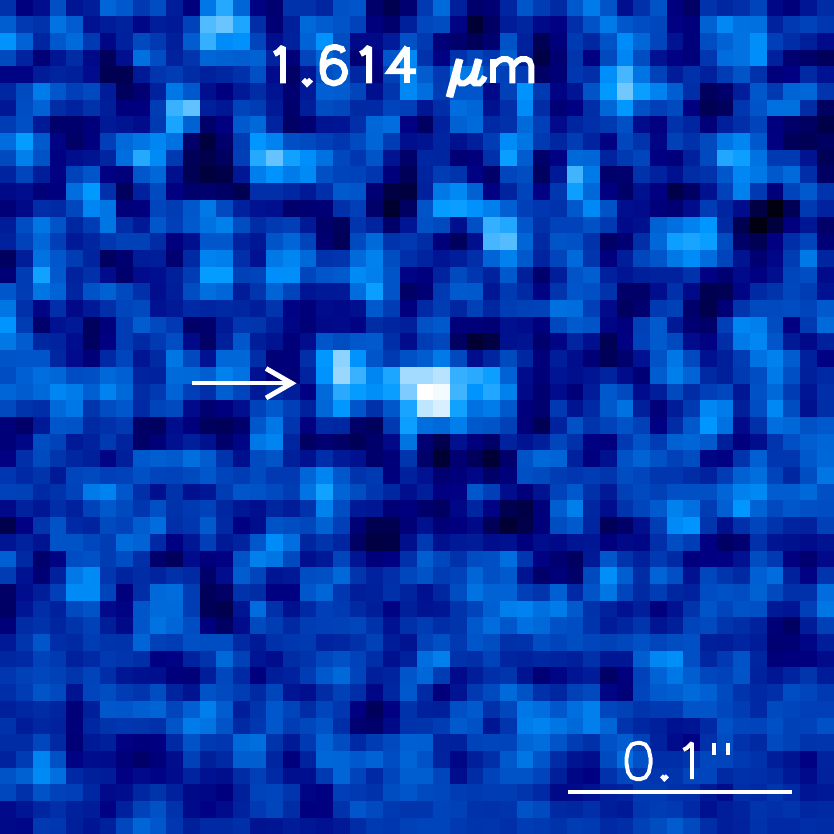}
\includegraphics[width=0.6\columnwidth]{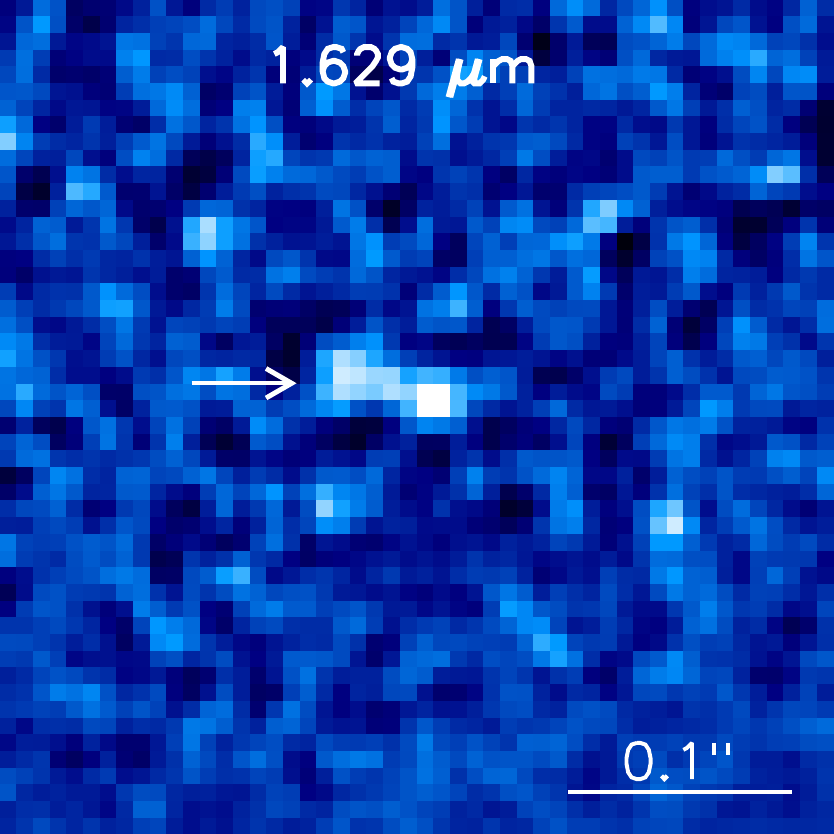}
\caption{\label{fig:SPH-speckle}Stamps ($220\times220$\,mas) of HD~131399~Ab from the SPHERE-IFS cADI-reduced data of 2016 Mar 17, at $1.523$\,\micron\ (top), $1.614$\,\micron\ (middle) and $1.629$\,\micron\ (bottom). The PSF is affected by a nearby speckle (indicated by the arrow) that becomes the most prominent in {\it H} band at $1.629$\,\micron. This speckle is present in all ADI reductions and might bias the spectrum of HD~131399~Ab to create a spurious peak at {\it H} band. Scales are linear and identical between the three panels. North is up, and east is to the left.}
\end{figure}

Similarly to the GPI data, the speckle field in both IRDIS and IFS datacubes was removed using the two post-processing pipelines as described in Section \ref{sec:GPI}. Final broadband images at $K1$, $K2$, and {\it YJH}, created from the stack of the 39-channel IFS datacubes, are shown for each epoch in Figure \ref{fig:imSPHERE}. The position, contrast, and measurement uncertainties of HD\,131399\,Ab were also obtained using the same techniques as for GPI, the PSF templates for the IRDIS and IFS data were built from the unsaturated off-axis images of the star. The astrometric calibrations of the platescale and position angle for both instruments are given by \citet{Maire:2016} and the ESO SPHERE user manual 7th edition to convert the on-chip measurements into on-sky positions. These calibration values have been stable since the commissioning of the instrument, when taking into account the mis-synchronisation correction between the SPHERE and VLT clocks. The final astrometric error budget consists of the following added in quadrature: the measurement errors described in Section~\ref{sec:GPI}; a star registration error of 0.1~px \citep{Vigan:2015, Zurlo:2016}; a plate scale error of $0.02$\,mas\,lenslet$^{-1}$ (IFS) and $0.021$\,mas\,px$^{-1}$ (IRDIS); a pupil angle offset error of $0.11$\,deg; a position angle offset error of $0.08$\,deg; and an IFS angle offer error of $0.13$\,deg.

The spectro-photometric and astrometric measurements from the two pipelines agreed very well to better than $1\sigma$ at each epoch and were combined following the procedure used for the GPI data (see Section \ref{sec:GPI}). The SNRs for all of the datasets, except the 2016 May 7 IFS data, were also computed using the same FMMF algorithm as the GPI data. Due to the short amount of time HD~131399~Ab stays on the chip and some artifacts on the edge of the images, the 2016 May 7 IFS data seemed to be a pathological dataset for the FMMF algorithm. Instead, for this dataset, we computed the SNRs by cross-correlating each broadband-collapsed image with a Gaussian PSF, and comparing the peak of the cross correlation of Ab with the standard deviation of the cross correlation of the noise at the same separation.

Figure \ref{fig:SPH-spec} shows the spectrum of HD~131399~Ab extracted from each epoch of IFS data. The spectra are very noisy because HD\,131399\,Ab is barely detected in individual slices, especially in {\it Y} and {\it J} bands. As reported in Table \ref{tab:obs}, the source lies on the detector a small fraction of the total time in three datasets (as low as $30\%$), lies very close to the edge of the detector another significant portion, particularly in the May 2016 dataset, and falls off the chip up to $39\%$ of the time in our reduced IFS images. Ultimately, this reduced effective observing time strongly affects the data quality. The continuum and flux are nevertheless consistent between the different epochs, except between the {\it J} and {\it H} bands where the atmospheric transmission is low. However, the third epoch strongly differs from the other three in {\it H} band, exhibiting a steep slope with a peak at $1.61-1.63$\,\micron. To assess this feature, the 2016 March 17 data were reduced using LOCI and the spectrum was extracted following the same procedure as for the cADI/\texttt{pyKLIP} analysis. In both cases, the slope and peak were both recovered. A visual inspection of the reduced datacubes reveals the presence of a speckle very close to HD~131399~Ab, which becomes more prominent in the $1.61$ and $1.63$\,\micron\ channels (see Figure \ref{fig:SPH-speckle}). More aggressive high-pass filters and algorithm parameters are not able to suppress this speckle. We therefore propose that the peak of the spectrum of HD~131399~Ab in the 2016 March 17 may be biased by this speckle, particularly in less aggressive reductions.

To mitigate this effect and also to improve the SNR of the spectrum, we followed the strategy of W16 and combined the four datasets using both pipelines. 

In the first pipeline, the cADI flux-loss ($\simeq 5\%$) was compensated in each dataset by injecting and reducing simulated sources at the same separation as HD~131399~Ab, but at twenty other position angles. A stamp of $30\times30$\,pixels centered at the measured position of HD~131399~Ab was then extracted in the cADI-reduced image at each epoch. The stamps of the four epochs were averaged for each wavelength slice. To extract the flux at each wavelength from the combined data, we created a forward model of the PSF of HD~131399~Ab. At each epoch and wavelength slice, the off-axis PSF was injected in a noise-free datacube at the separation and position angle of HD~131399~Ab and reduced using the parallactic angle exploration of each epoch with cADI. Stamps of the model were then extracted and combined similarly. The combined model was used to fit the flux of HD~131399~Ab using the amoeba-simplex minimization procedure. To estimate the uncertainties, the exercise (injection of simulated sources in the raw data and forward model computation) was repeated at the same separation but at twenty different position angles. The statistical dispersion of the extracted fluxes was used as the uncertainty in the spectrum at each wavelength. 

In the second pipeline, we extracted from the \texttt{pyKLIP}-reduced data and forward-modelled PSF a $11\times11$\,pixel stamp centered at the location of HD~131399~Ab at each epoch. The stamps of both the data and forward model were averaged over the four epochs at each wavelength slice resulting in one stamp of both the data and forward model at each wavelength. Then, we follow the same BKA technique as before to measure the flux and quantify the uncertainties in each wavelength channel. 

The spectra were then combined in the same way as discussed previously. The results are discussed in Section \ref{sec:SEDb}.

Astrometry and photometry of HD~131399~B in the IRDIS $K1$ and $K2$ unsaturated off-axis images were obtained using the same technique used for the NIRC2 data and are discussed in sections \ref{sec:analysis} and \ref{sec:astrometry}.

\subsubsection{New observations}
\begin{figure}
\centering
\includegraphics[width=0.7\columnwidth]{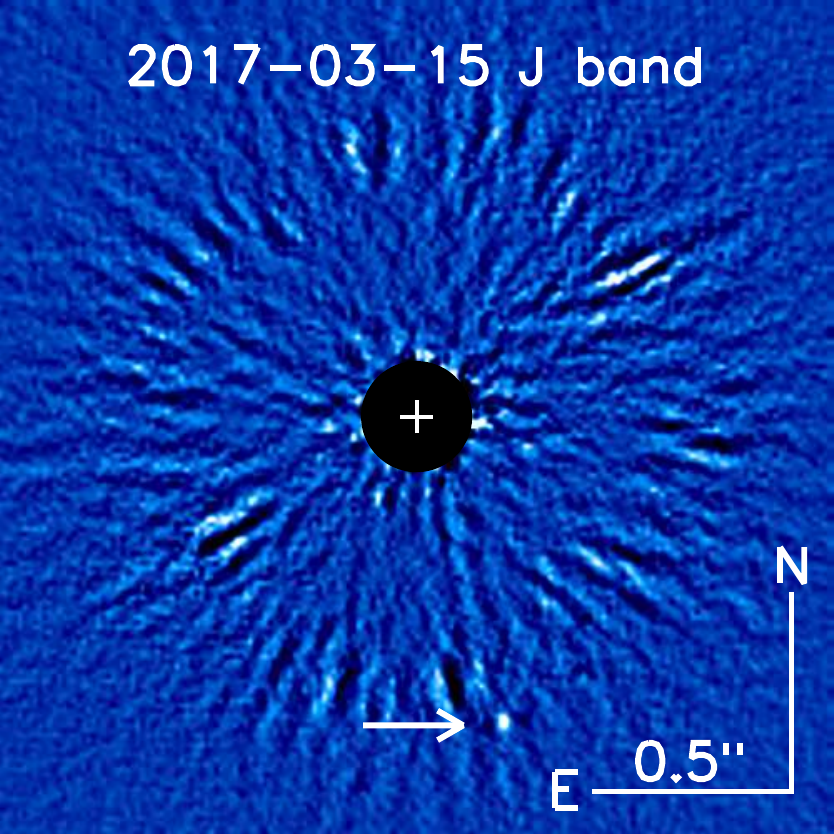}
\caption{\label{fig:SPHERE_J}PSF-subtracted image of HD~131399~Ab obtained with SPHERE-IRDIS in 2017 in {\it J} band. The image design is similar to Figure \ref{fig:imGPI}.}
\end{figure}

HD~131399 was observed on 2017 March 15 (098.C-0864(A), PI: Hinkley) with SPHERE-IRDIS in dual polarimetric imaging (DPI) mode at $J$ ($\lambda_\mathrm{eff}=1.23$\,\micron) as part of a program to measure the polarization of directly-imaged planets. The same ``star-center'', ``flux'', and ``coronagraphic'' sequences, as were executed for the public DBI observations described in Section~\ref{subsec:public_sphere}, were carried out in this program, for a total on-source integration time of $0.36$\,hr. Calibrations data were obtained on subsequent days, following the standard calibration plan for the instrument.

The raw data were reduced following the same procedure as the public IRDIS DBI data. However, since the data were taken in DPI mode, images in the two quadrants, corresponding to two orthogonal polarization states, were summed to create total-intensity images. PSF-subtracted (see Figure \ref{fig:SPHERE_J}) photometric and astrometric measurements were also obtained using the two post-processing pipelines and same parameters as described previously. Finally, the measurements from these two pipelines were combined with a weighted mean as for the other datasets and are reported in Table \ref{tab:b_phot} and in Table \ref{tab:astr}.

\subsection{New Keck/NIRC2 Observations}
\begin{figure}
\centering
\includegraphics[width=0.5\columnwidth]{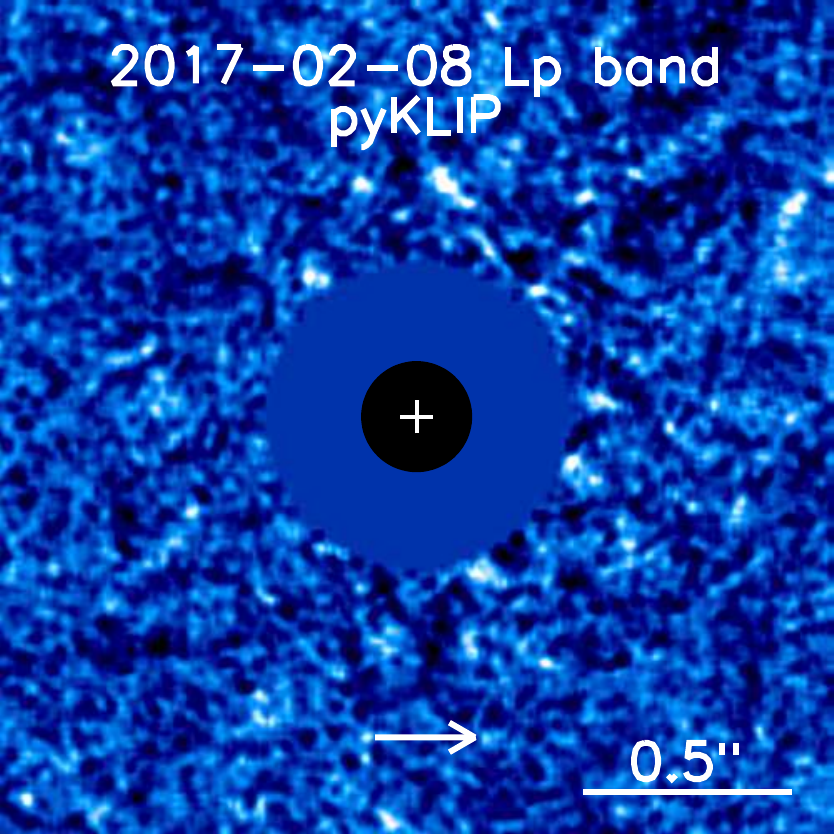}\includegraphics[width=0.5\columnwidth]{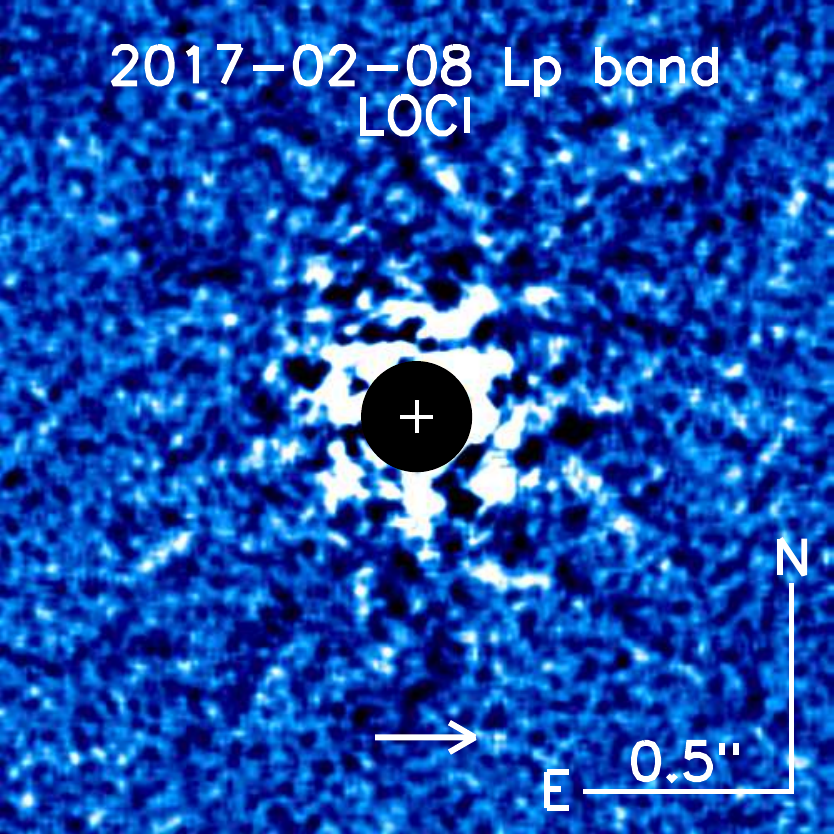}
\caption{\label{fig:imNIRC2}PSF subtracted images of HD~131399~Ab obtained with NIRC2 in 2017 at {\it L}$^{\prime}$ using \texttt{pyKLIP} (left) and LOCI (right). No source is significantly detected at the location of HD~131399~Ab (arrow). The image design is similar to Figure \ref{fig:imGPI}.}
\end{figure}

HD~131399~A was observed with the narrow camera of Keck/NIRC2 in the {\it L}$^\prime$ filter ($\lambda_{\rm eff} = 3.72$\,\micron) serving as its own natural guide star on consecutive nights 2017 February 7 and 8. We used only the Feb 8 data in our final analysis because high winds and poor seeing degraded the quality of the Feb 7 data. This resulted in 166 exposures of 0.9\,s and 30 coadds each for a total integration time of 1.25\,hr. The 400\,mas diameter coronagraph mask occulted the star in all exposures and the instrument was in vertical angle mode to enable ADI. The raw data were reduced with a custom set of tools that subtracts dark current and thermal background and then aligns all frames to a common star position.

To recover HD 131399 Ab, we subtracted the stellar halo and speckle pattern using a customized LOCI algorithm (``locally optimized combination of images''; \citealt{Lafreniere:2007}). We tested various levels of algorithm aggressiveness and present here a compromise between noise suppression and astrophysical source throughput, with LOCI parameter values of $N_{\delta}=0.3$, $W=10$\,px, $dr=10$\,px, $g=0.9$, and $N_a=10$ following the conventional definitions in \citet{Lafreniere:2007}. Speckle suppression in this data set particularly benefited from temporal proximity of reference images (i.e., small $N_{\delta}$), possibly due to high airmass and varying seeing conditions diminishing PSF stability. The PSF-subtracted frames were rotated to place North up and collapsed into a final median image (see Figure \ref{fig:imNIRC2}, left). We also performed a separate reduction using \texttt{pyKLIP} on the same aligned frames. The algorithm divided images into annuli that were 20 pixels wide radially and further divided into 10 azimuthal subsections each. To build the model of the stellar PSF, we used the first 50 KL basis vectors of the 200 most-correlated reference images where HD\,131399\,Ab moved at least 3 pixels due to the ADI observing method (\ref{fig:imNIRC2}, right).

In neither reduction was a source detected at the location of Ab with greater than 3$\sigma$ confidence over the background noise levels (see Figure \ref{fig:imNIRC2}). Therefore, we report only a lower limit of $11.10$\,mag for its {\it L}$^\prime$ contrast.

HD~131399~B and C are detected in individual images in which HD~131399~A is unocculted and unsaturated, so we performed astrometry on brighter component B as an independent confirmation of our SPHERE astrometry. To locate A, we fitted it with a bivariate Gaussian function using a least-squares minimization. We then jointly fitted B and C using the PSF of A as a template for a least-squares minimization. We repeated this process for six images divided between two dither positions, and report in Section \ref{sec:astrometry} the mean separation and PA of B from those fits. The measurement errors were estimated as the standard deviation of the separation and PA across the six images. The final astrometric uncertainties were calculated as the quadrature sum of these measurement errors, the star registration error estimated at 5\,mas, and the plate scale error of 0.004\,mas\,pixel$^{-1}$ and position angle offset error of 0.02\,deg \citep{Service:2016}.

\section{Spectro-photometric Analysis}\label{sec:analysis}

\begin{deluxetable}{lccc}
\tablecaption{Properties of the HD~131399 system} 
\tablewidth{0pt}
\tablehead{
\colhead{Property} & \multicolumn{2}{c}{Value} & \colhead{Unit}}
\startdata
$\pi$                & \multicolumn{2}{c}{$10.20\pm0.70$\tablenotemark{\it a}} & mas\\
$d$                & \multicolumn{2}{c}{$98.0^{+7.2}_{-6.3}$\tablenotemark{\it a}} & pc\\
$\mu_{\alpha}$          & \multicolumn{2}{c}{$-29.69\pm0.59$\tablenotemark{\it a}} & mas yr$^{-1}$\\
$\mu_{\delta}$          & \multicolumn{2}{c}{$-31.52\pm0.55$\tablenotemark{\it a}} & mas yr$^{-1}$\\
Age                     & \multicolumn{2}{c}{$16\pm7$\tablenotemark{\it b}} & Myr \\
\tableline
& A & Ab\\
\tableline
$\Delta Y_{\rm SPH-IFS}$     & \multicolumn{2}{c}{$13.73 \pm 0.23$\tablenotemark{\it c}} & mag\\
$\Delta J_{\rm SPH-IFS}$     & \multicolumn{2}{c}{$13.32 \pm 0.14$\tablenotemark{\it c}} & mag\\
$\Delta H_{\rm SPH-IFS}$     & \multicolumn{2}{c}{$13.04 \pm 0.16$\tablenotemark{\it c}} & mag\\
$\Delta K1_{\rm SPH}$    & \multicolumn{2}{c}{$12.70 \pm 0.05$\tablenotemark{\it c}} & mag\\
$\Delta K2_{\rm SPH}$    & \multicolumn{2}{c}{$12.50 \pm 0.13$\tablenotemark{\it c}} & mag\\
$\Delta J_{\rm GPI}$     & \multicolumn{2}{c}{$13.37 \pm 0.17$} & mag\\
$\Delta H_{\rm GPI}$     & \multicolumn{2}{c}{$12.84 \pm 0.06$\tablenotemark{\it c}} & mag\\
$\Delta K1_{\rm GPI}$    & \multicolumn{2}{c}{$12.61 \pm 0.17$} & mag\\
$\Delta L^{\prime}$      & \multicolumn{2}{c}{$>11.10$}                      & mag\\
$Y_{\rm SPH-IFS}$        & $6.928\pm0.015$\tablenotemark{\it d} & $20.64 \pm 0.16$ & mag\\
$J_{\rm GPI}$            & $6.904\pm0.016$\tablenotemark{\it d} & $20.27 \pm 0.17$ & mag\\
$H_{\rm GPI}$            & $6.895\pm0.017$\tablenotemark{\it d} & $19.73 \pm 0.07$ & mag\\
$K1_{\rm GPI}$           & $6.872\pm0.018$\tablenotemark{\it d} & $19.48 \pm 0.17$ & mag\\
$K1_{\rm SPH}$           & $6.869\pm0.018$\tablenotemark{\it d} & $19.56 \pm 0.06$ & mag\\
$K2_{\rm SPH}$           & $6.865\pm0.019$\tablenotemark{\it d} & $19.36 \pm 0.13$ & mag\\
$L^{\prime}$             & $6.862\pm0.020$\tablenotemark{\it d} & $>17.96$ & mag\\
\enddata
\tablenotetext{a}{\citet{vanLeeuwen:2007dc}}
\tablenotetext{b}{Combining median age and uncertainty with intrinsic age spread from \citet{Pecaut:2016}}
\tablenotetext{c}{Obtained from a weighted mean of the different epochs presented in Table \ref{tab:b_phot}}
\tablenotetext{d}{Synthetic photometry derived from SED fit described in Section~\ref{sec:sedA}}
\label{tab:properties}
\end{deluxetable}

\subsection{SED and Mass of HD~131399~A}\label{sec:sedA}

A flux-calibrated spectrum of the primary was required to convert the measured contrast between HD~131399~A and Ab within the SPHERE and GPI datasets. As no near-IR spectrum of HD~131399~A was available within the literature, we used a stellar evolutionary model and a grid of synthetic stellar spectra to fit the observed SED of HD~131399~A. From this fit we estimated both the spectrum of the star and synthetic photometry within the GPI and SPHERE passbands.

Optical and near-infrared photometry were found in the literature for a number of systems: Tycho ($B_TV_T$; \citealp{Hog:2000wk}), {\it Hipparcos} ($H_p$; \citealp{ESA:1997ws}), and 2MASS ($JHK_s$; \citealp{skrutskie:2006}). Optical color indices in the Str\"{o}mgren $uvby$ \citep{Hauck:1986uf} and Geneva\footnote{\url{http://obswww.unige.ch/gcpd/}} systems were also found \citep{Mermilliod:1997ch}. An uncertainty of 0.1\,mag was assumed for these color indices as none were presented within the literature. As the angular separation between HD~131399~A and BC is comparable to the angular resolution of the telescopes used to obtain these photometric measurements, the measures reported within these catalogs are of the blended system rather than of HD~131399~A. At shorter wavelengths the contrast between HD~131399~A and the BC pair is large enough that the faint pair has a negligible impact on the optical photometry of the system. At longer wavelengths this effect becomes significant, approximately 10\,\% at $K$. To account for this we simultaneously fit the combined flux of the three stars using the photometric measurements of the system described previously, and apparent magnitudes of the BC pair obtained from the literature.

We used the {\tt emcee} parallel-tempered affine-invariant MCMC sampler  to fully explore parameter space and estimate uncertainties on the near-IR spectrum of HD~131399~A. At each step within a chain an age $t$, parallax $\pi$, mass for each component $M_{\rm A}$, $M_{\rm B}$, $M_{\rm C}$, and extinction $A_V$ were selected. We used a Gaussian prior for age ($16\pm7$\,Myr), and a Gaussian ($10.20\pm0.70$\,mas) multiplied by a $\pi^{-4}$ power law---to account for a uniform space density of stars as expected at the distance to HD 131399~Ab---as the prior for parallax. The prior on the three masses was based on the \citet{Kroupa:2001ki} initial mass function. Age and mass were converted into an effective temperature ($T_{\rm eff}$) and surface gravity ($\log g$) using the MIST evolutionary models \citep{Dotter:2016fa, Choi:2016kf}. Given the rapid rotation seen for young early-type stars \citep[e.g.,][]{Strom:2005}, we used the evolutionary models that incorporated stellar rotation ($v/v_{\rm crit} = 0.4$). A solar metallicity was assumed, consistent with the observed metallicity of other stars within the ScoCen association \citep{Mamajek:2013}.

Synthetic photometry and color indices were computed from a {\sc BT-NextGen} model atmosphere \citep{Allard:2012fp}\footnote{\url{https://phoenix.ens-lyon.fr/Grids/BT-NextGen/AGSS2009/SPECTRA/}} of the appropriate $T_{\rm eff}$ and $\log g$, scaled by the $R^2/d^2$ dilution factor, where $R$ is the radius of the star computed from $M$ and $\log g$, and $d = 1/\pi$ is the distance to the star. Model atmospheres at temperatures and surface gravities between grid points were estimated using a linear interpolation of the logarithm of the flux. These synthetic spectra were first reddened using the selected $A_V$ value and the \citet{Cardelli:1989dp} extinction law, and then convolved with the throughput of each filter to obtain synthetic photometry. Filter transmission profiles and zero points were obtained from \citet{Mann:2015gp} for the optical filters, and from \citet{Cohen:2003gg} for the 2MASS filters. A probability ($\ln p = -\chi^2/2$) was calculated at each step by comparing the synthetic magnitudes and color indices for the blended system to the observed values, the synthetic magnitudes of the B and C components to the $K1_{\rm SPH}$ contrasts (the SPHERE/IRDIS filters are described later in this section) given in W16 ($\Delta K1_{\rm SPH} = 1.86\pm0.10$\,mag and $3.86\pm0.10$\,mag for B and C, respectively), and the apparent $H_p$ magnitude for the blended BC pair of $11.161\pm0.187$\,mag reported in the Catalogue of the Components of Double and Multiple Stars (CCDM; \citealp{Dommanget:2002ud}).

\begin{figure}
\includegraphics[width=\columnwidth]{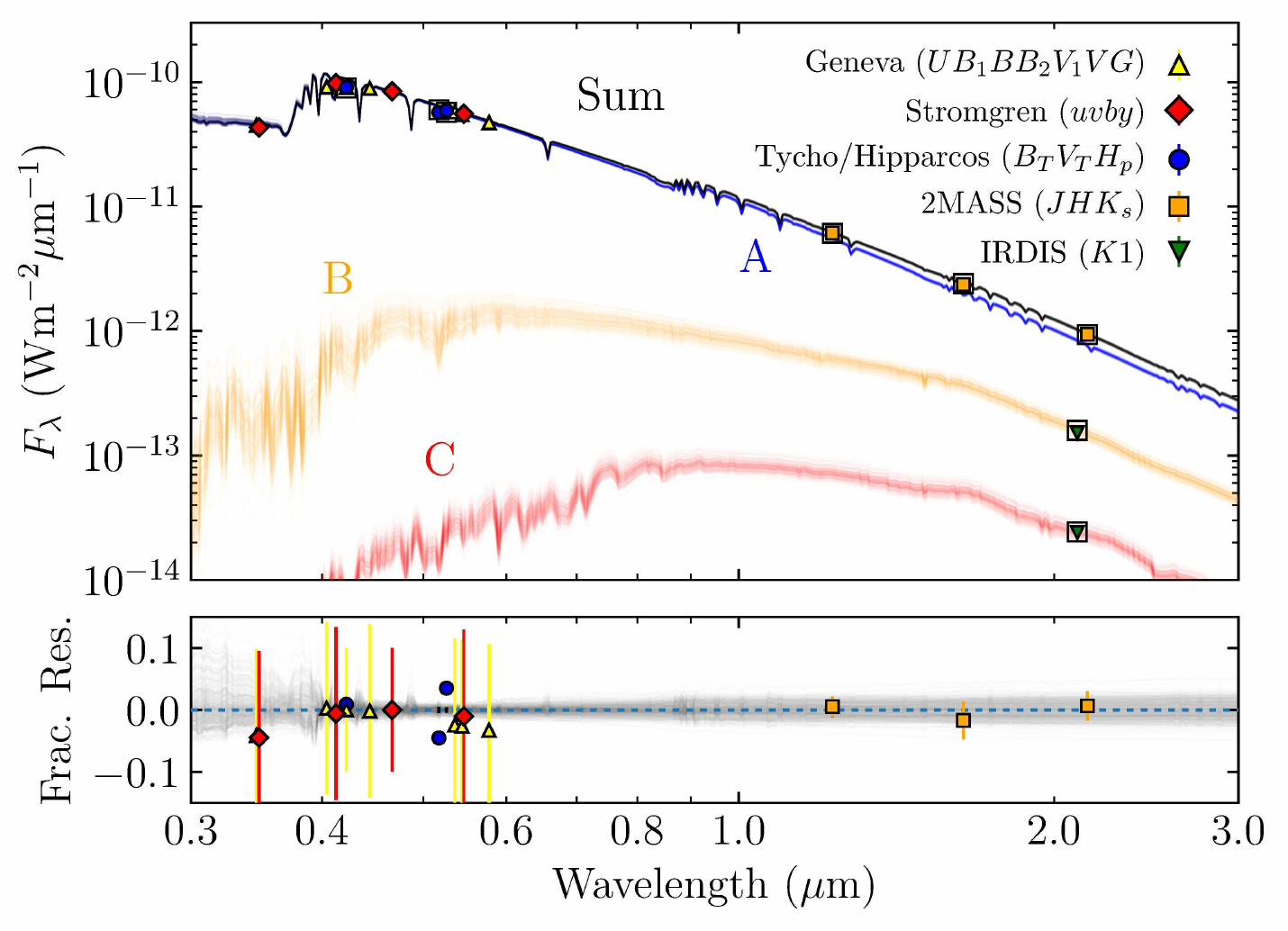}
\caption{\label{fig:sedA}(Top panel): One hundred realizations of the spectral energy distribution of HD~131399~A (blue), B (orange), and C (red) drawn randomly from the MCMC posterior distributions described in Section~\ref{sec:sedA}. The SED of the blended system is also shown (black). Photometric measurements of the system, and of the B and C components, are plotted as filled symbols; Tycho/{\it Hipparcos} (blue circle), 2MASS (orange square), IRDIS (green downward triangle). Predicted fluxes in these systems are shown as open squares. The Geneva (yellow upward triangle) and $uvby$ (red diamond) fluxes are tied to the predicted flux in the $B$/$b$ filter. (Bottom panel): Fractional residuals for each of the one hundred SEDs of the blended system (grey curves) and for the photometric measurements (symbols as before).}
\end{figure}

\begin{deluxetable}{ccccc}
\tablecaption{\label{tab:sed_mcmc}Stellar parameters derived from SED fit}
\tablewidth{0pt}
\tablehead{
\colhead{Property} & \colhead{Unit} & \multicolumn{3}{c}{HD 131399 system}  
}
\startdata
$t$ & Myr & \multicolumn{3}{c}{$21.9^{+4.1}_{-3.8}$}\\
$\pi$ & mas & \multicolumn{3}{c}{$9.27^{+0.33}_{-0.37}$}\\ 
$d$ & pc & \multicolumn{3}{c}{$107.9^{+4.5}_{-3.7}$}\\
$A_V$ & mag & \multicolumn{3}{c}{$0.22\pm0.09$}  \\
\hline
& & A & B & C \\
\hline
$M$ & $M_{\odot}$ & $2.08^{+0.12}_{-0.11}$ & $0.95\pm0.04$ & $0.35\pm0.04$ \\
$T_{\rm eff}$ & K & $9480^{+420}_{-410}$ & $4890^{+190}_{-170}$ & $3460\pm60$ \\
$\log g$ & [dex] & $4.32\pm0.01$ & $4.40\pm0.03$ & $4.45\pm0.05$ \\
\enddata
\end{deluxetable}

We initialized 512 walkers at each of 16 different temperatures to ensure the parameter space was fully explored; lower temperatures sample the posterior distribution, while higher temperatures fully explore the prior distributions. Each walker was advanced for 1,000 steps as an initial burn in stage, and then advanced for a further 9,000 steps to fully sample the posterior distribution for each parameter. The median and 1$\sigma$ range calculated from the posterior distribution of the six fitted parameters ($t$, $\pi$, $M_{\rm A}$, $M_{\rm B}$, $M_{\rm C}$, $A_V$), and that of the derived $T_{\rm eff}$ and $\log g$ for each component, are given in Table~\ref{tab:sed_mcmc}.

We find a mass of $2.08^{+0.12}_{-0.11}$\,$M_{\odot}$, a temperature of $9480^{+420}_{-410}$\,K, and a surface gravity of $\log g = 4.32\pm0 .01$\,[dex] for HD~131399~A. These parameters are consistent with an A1V spectral type \citep{Houk:1982vl} at an age of 16\,Myr. The extinction towards HD~131399 of $A_V = 0.22\pm0.09$\,mag estimated from the SED fit is consistent with literature estimates that range from 0.14--0.28\,mag \citep{deGeus:1989vn,Sartori:2003ej,Chen:2012ki}. The photometric distance of $107.9^{+4.5}_{-3.7}$\,pc is 1.2$\sigma$ discrepant from the trigonometric distance of $98.0^{+7.2}_{-6.3}$\,pc from the {\it Hipparcos} parallax \citep{vanLeeuwen:2007dc}. Repeating the SED fit using only a $p(\pi) \propto \pi^{-4}$ prior, corresponding to an assumed uniform space density of stars, results in a similar photometric distance of $112.2^{+5.2}_{-5.1}$\,pc. The stated uncertainties on the fitted parameters do not incorporate any model uncertainty, and are therefore likely underestimated.  

The SED of each component, and that of the blended system, are shown in Figure~\ref{fig:sedA}. Uncertainties on the near-IR portion of the SED of HD~131399~A, estimated by sampling randomly from the posterior distributions ($t$, $\pi$, $M_{\rm A}$, and $A_V$), ranged between 1.5--2.0\,\%. The SED of A was degraded to the spectral resolving power of the GPI and SPHERE IFS observations to convert the contrasts between HD~1313199~A and Ab measured in Section~\ref{sec:obs} into apparent fluxes for Ab.

\begin{deluxetable}{cccccc}
\tablecaption{\label{tab:filters}Atmosphere throughput-corrected filter properties}
\tablewidth{0pt}
\tablehead{
\colhead{Filter} & \colhead{$\lambda_{\rm eff}$} & \colhead{$W_{\rm eff}$} & \colhead{Zero point}\\
 & (\micron ) & (\micron ) & ($10^{-9}$\,Wm$^{-2}\micron^{-1}$)}
\startdata
$J_{\rm GPI}$ & 1.23 & 0.19 & 3.12 \\
$H_{\rm GPI}$ & 1.64 & 0.27 & 1.15 \\
$K1_{\rm GPI}$ & 2.06 & 0.20 &  0.50 \\
\hline
$Y_{\rm SPH-IFS}$ & 1.03 & 0.16 & 5.65 \\
$J_{\rm SPH-IFS}$ & 1.24 & 0.24 & 2.99 \\
$H_{\rm SPH-IFS}$ & 1.54 & 0.19 & 1.41 \\
\hline
{\it J}$_{\rm SPH}$ & 1.23 & 0.20 & 3.11 \\
$K1_{\rm SPH}$ & 2.10 & 0.09 & 0.47 \\
$K2_{\rm SPH}$ & 2.25 & 0.11 & 0.36\\
\hline
$L^{\prime}$ & 3.72 & 0.59 & 0.054\\
\enddata
 \tablecomments{The subscript SPH refers to the SPHERE IRDIS filters and SPH-IFS to the derived SPHERE IFS filters, to differentiate between them.}
\end{deluxetable}

\begin{figure}
\includegraphics[width=\columnwidth]{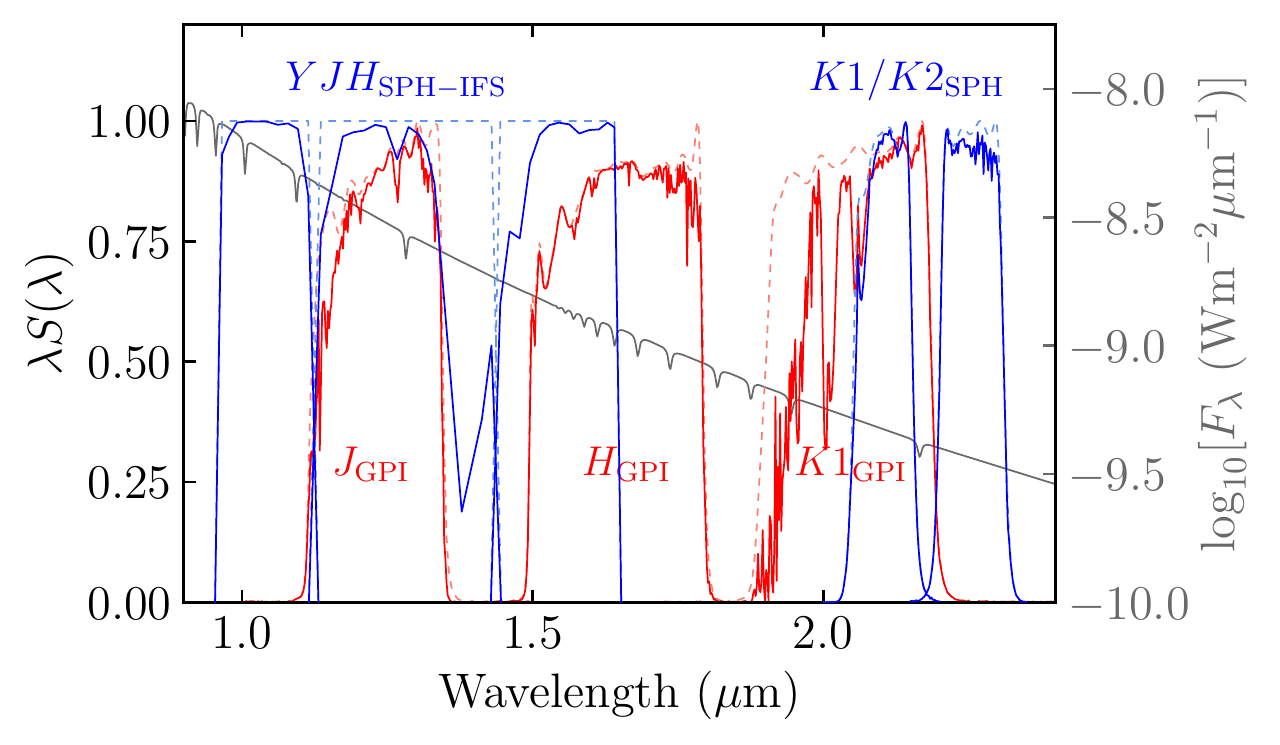}
\caption{\label{fig:filters}Energy response functions for the GPI (red curves) and SPHERE (blue curves) filters, following the definitions of \citet{Bessell:2012bq}. The response functions are shown before (dashed curves) and after (solid curves) multiplication by either a median Cerro Pach\'{o}n or Paranal atmosphere. The $J_{\rm SPH}$ filter is not plotted as it is very similar to the $J_{\rm GPI}$ filter. Plotted in gray is the CALSPEC spectrum of Vega used to compute the zero points given in Table~\ref{tab:filters}.}
\end{figure}

Synthetic photometry of HD~131399~A was also computed for the GPI, SPHERE, and NIRC2 filters to convert the measured broad-band contrasts between A and Ab into apparent magnitudes for Ab. Filter transmission profiles for the GPI filters were obtained from the GPI DRP, and were combined with a median Cerro Pach\'{o}n atmosphere (4.3\,mm precipitable water vapor) at one airmass \citep{Lord:1992to}. The SPHERE IRDIS filter curves were obtained from the ESO website\footnote{\url{ https://www.eso.org/sci/facilities/paranal/instruments/sphere/inst/filters.html}}, while the IFS throughput was assumed to be uniform between 0.96--1.11\,\micron\ at {\it Y}, 1.13--1.42\,\micron\ at {\it J}, and 1.44--1.64\,\micron\ at {\it H}. These filter curves were combined with a median Paranal atmosphere (2.5\,mm precipitable water vapor) at one airmass \citep{Moehler:2014ba}. The NIRC2 $L^{\prime}$ filter curve was obtained from the Keck website\footnote{\url{https://www2.keck.hawaii.edu/inst/nirc2/filters.html}}, and was combined with a Mauna Kea atmosphere \citep{Lord:1992to} with 1.6\,mm of precipitable water vapor at two airmasses (chosen to match the observing conditions on 2107 Feb 08). The throughput of the GPI and SPHERE filters are plotted in Figure~\ref{fig:filters}. Zero points and effective wavelengths for all of the filters were estimated using the CALSPEC Vega spectrum\footnote{\url{ftp://ftp.stsci.edu/cdbs/current_calspec/alpha_lyr_stis_008.fits}} \citep{Bohlin:2014bs}, and are given in Table~\ref{tab:filters}. The properties of the $H_{\rm GPI}$ and $K1_{\rm GPI}$ filters match those derived from observations of the white dwarf HD~8049~B presented in \citet{DeRosa:2016kh}.

\subsection{SED and Spectral Type of HD~131399~Ab}\label{sec:SEDb}
\begin{figure*}
\centering
\includegraphics{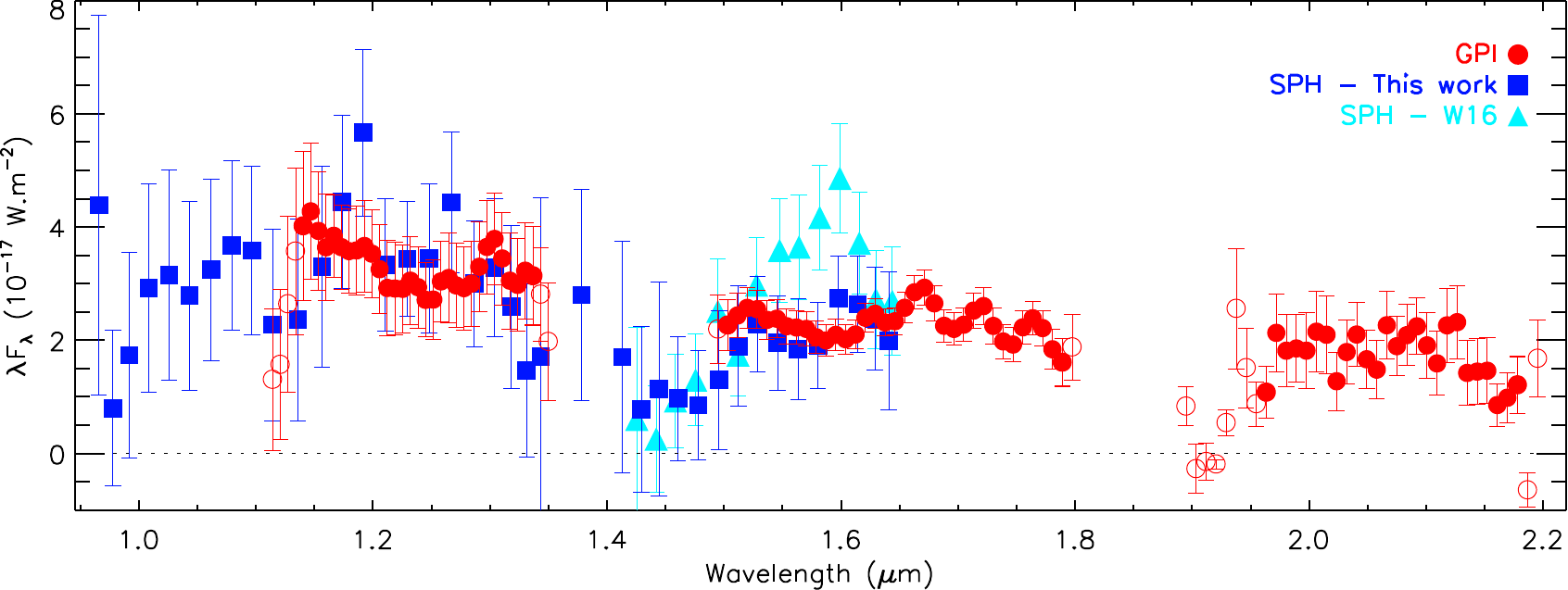}
\caption{\label{fig:b_spec}
$YJHK1$ spectra of HD~131399~Ab extracted from GPI data (red circles, resolving power of 45) and all SPHERE data (blue squares, resolving power of 30). The {\it H} band spectrum published by W16 is also shown for comparison (cyan triangles). The {\it H}-band flux is consistent between SPHERE and GPI but we argue that the shape of the published SPHERE spectrum may be biased by the speckle discussed in Figure \ref{fig:SPH-speckle}. Open circles correspond to GPI wavelength channels where the atmospheric+instrument+filter throughput is lower than $50\%$ (see Figure \ref{fig:filters}). }
\end{figure*}

\begin{deluxetable}{ccccc}
\tabletypesize{\scriptsize}
\tablecaption{\label{tab:b_phot}Contrast measurements of HD 131399 Ab}
\tablewidth{0pt}
\tablehead{
\colhead{UT Date} & \colhead{Instrument} & \colhead{Filter} & \colhead{Contrast (mag.)} & \colhead{SNR}}
\startdata
 & (SPH W16) & ({\it J}) & ($13.23\pm0.20$\tablenotemark{\it a}) & (13.2)  \\
  & (SPH W16) & ({\it H}) & ($12.99\pm0.20$\tablenotemark{\it a}) & (15.5) \\
 & (SPH W16) & ($K1$) & ($12.45\pm0.10$\tablenotemark{\it a}) & (23.5) \\
  & (SPH W16) & ($K2$) & ($12.64\pm0.16$\tablenotemark{\it a}) & (11.9) \\
 2015 Jun 12       & SPH-IFS    & {\it Y} &        $13.73\pm0.33$\tablenotemark{\it b} & 4.0\tablenotemark{\it b} \\
                &                    & {\it J} &  $13.19\pm0.23$\tablenotemark{\it c}  & 6.3\tablenotemark{\it c} \\
                &                      & {\it H}  & $12.94\pm0.24$\tablenotemark{\it d} & 5.3\tablenotemark{\it d} \\
        &  SPH-IRDIS & $K1$  &  $12.75\pm0.11$  & 11.4 \\
                    &  SPH-IRDIS & $K2$  &  $12.76\pm0.41$ & 6.1 \\
2016 Mar 06          & SPH-IFS   & {\it Y}  &  $13.67\pm0.44$\tablenotemark{\it b} & 3.3\tablenotemark{\it b}  \\
                   &             & {\it J} &  $13.32\pm0.38$\tablenotemark{\it c} & 4.6\tablenotemark{\it c} \\
                &              & {\it H} &   $13.09\pm0.29$\tablenotemark{\it d} & 3.4\tablenotemark{\it d} \\
      &  SPH-IRDIS & $K1$  &  $12.69\pm0.11$ & 9.6 \\
            &  SPH-IRDIS & $K2$  &  $12.32\pm0.36$ & 5.7 \\
2016 Mar 17          & SPH-IFS   & {\it Y}  &  $13.82\pm0.45$\tablenotemark{\it b} & 2.7\tablenotemark{\it b} \\
                   &             & {\it J} &  $13.32\pm0.39$\tablenotemark{\it c} & 5.5\tablenotemark{\it c} \\
                &              & {\it H} &   $13.19\pm0.36$\tablenotemark{\it d} & 3.2\tablenotemark{\it d} \\
      &  SPH-IRDIS & $K1$  &  $12.58\pm0.11$ & 12.3 \\
            &  SPH-IRDIS & $K2$  &  $12.31\pm0.31$ & 5.7 \\
2016 May 07          & SPH-IFS   & {\it Y}  &  $>13.74$\tablenotemark{\it b} & 1.2\tablenotemark{\it b,e}  \\
                   &             & {\it J} &  $13.47\pm0.25$\tablenotemark{\it c} & 4.4\tablenotemark{\it c,e} \\
                &              & {\it H} &   $>13.11$\tablenotemark{\it d} & 2.0\tablenotemark{\it d,e} \\
      &  SPH-IRDIS & $K1$  &  $12.79\pm0.10$ & 14.0 \\
            &  SPH-IRDIS & $K2$  &  $12.54\pm0.17$ & 7.2 \\
2017 Feb 08 & NIRC2 & {\it L}$^\prime$ & $>11.10$ & - \\
2017 Feb 14 & GPI & $K1$ & $12.61\pm0.17$ & 6.2 \\
2017 Feb 15 & GPI & {\it H} & $12.81\pm0.09$ & 10.6 \\
2017 Feb 16 & GPI & {\it J} & $13.37\pm0.17$ & 7.7 \\
2017 Mar 15 &  SPH-IRDIS & {\it J}  &  $13.50\pm0.14$ & 7.6 \\
2017 Apr 20 & GPI & {\it H} & $12.86\pm0.09$ & 12.0 \\
\enddata
\tablenotetext{a}{\citet{wagner2016} reports only one apparent magnitude measurement for all four epochs.}
\tablenotetext{b}{Obtained by averaging channels between $0.96-1.11$\,\micron}
\tablenotetext{c}{Obtained by averaging channels between $1.13-1.42$\,\micron}
\tablenotetext{d}{Obtained by averaging channels between $1.44-1.64$\,\micron}
\tablenotetext{e}{Gaussian cross-correlation instead of FMMF used for SNR}
\end{deluxetable}

Photometric measurements, SNR, and spectra obtained from GPI, NIRC2, from the new SPHERE data, from our reanalysis of the SPHERE data, and those published by W16 are given in Table \ref{tab:b_phot} and Figure \ref{fig:b_spec}. The measurements provide {\it YJH} contrasts consistent at the 1$\sigma$ level between the four SPHERE sets, between our average SPHERE contrasts and those published in W16, and between our average SPHERE and average GPI measurements, with the caveat that the GPI and SPHERE filters are different (especially {\it H}, see Figure~\ref{fig:filters}). However, the reanalyzed SPHERE contrast at $K1$ and $K2$ differ significantly (2$\sigma$ at $K1$ and 1$\sigma$ at $K2$) with that of W16. The origin of these discrepancies remains unclear since the $K1$ contrasts of HD~131399~B and C are in agreement between our reanalysis ($\Delta K1$~$=1.95\pm0.07$ and $3.84\pm0.10$\,mag) and that of W16 ($\Delta K1$~$=1.86\pm0.10$ and $3.86\pm0.10$\,mag, for B and C respectively).

The GPI spectrum is flat, except for some correlated noise, at a high confidence level, without any indication of the methane absorption beyond $1.6$\,\micron\ that is expected in the spectra of mid-T dwarfs. The GPI spectrum is also in agreement with that of the combined four SPHERE sets in both {\it J} and {\it H} bands. However, the published SPHERE {\it H} band spectrum (W16) peaks at $1.61$\,\micron, a peak that does not appear either in the GPI spectrum nor in our reanalysis. The peak flux is also nearly twice than the plateau of the other two spectra. These differences might be explained by: (1) a different technique used to combine the multiple datasets; and/or (2) the technique used to extract the photometry of Ab, with different techniques being biased by nearby speckles to varying degrees. The presence of a speckle close to HD~131399~Ab in the 2016 March 17 SPHERE dataset may be significantly biasing the spectrum at $\sim$1.6\,\micron\ (see Sec.~\ref{sec:obs_SPH}), with the spectrum being featureless in the three other sets.

\subsubsection{Color-magnitude and color-color diagrams}\label{sec:ccd}
\begin{figure}
\includegraphics[width=\columnwidth]{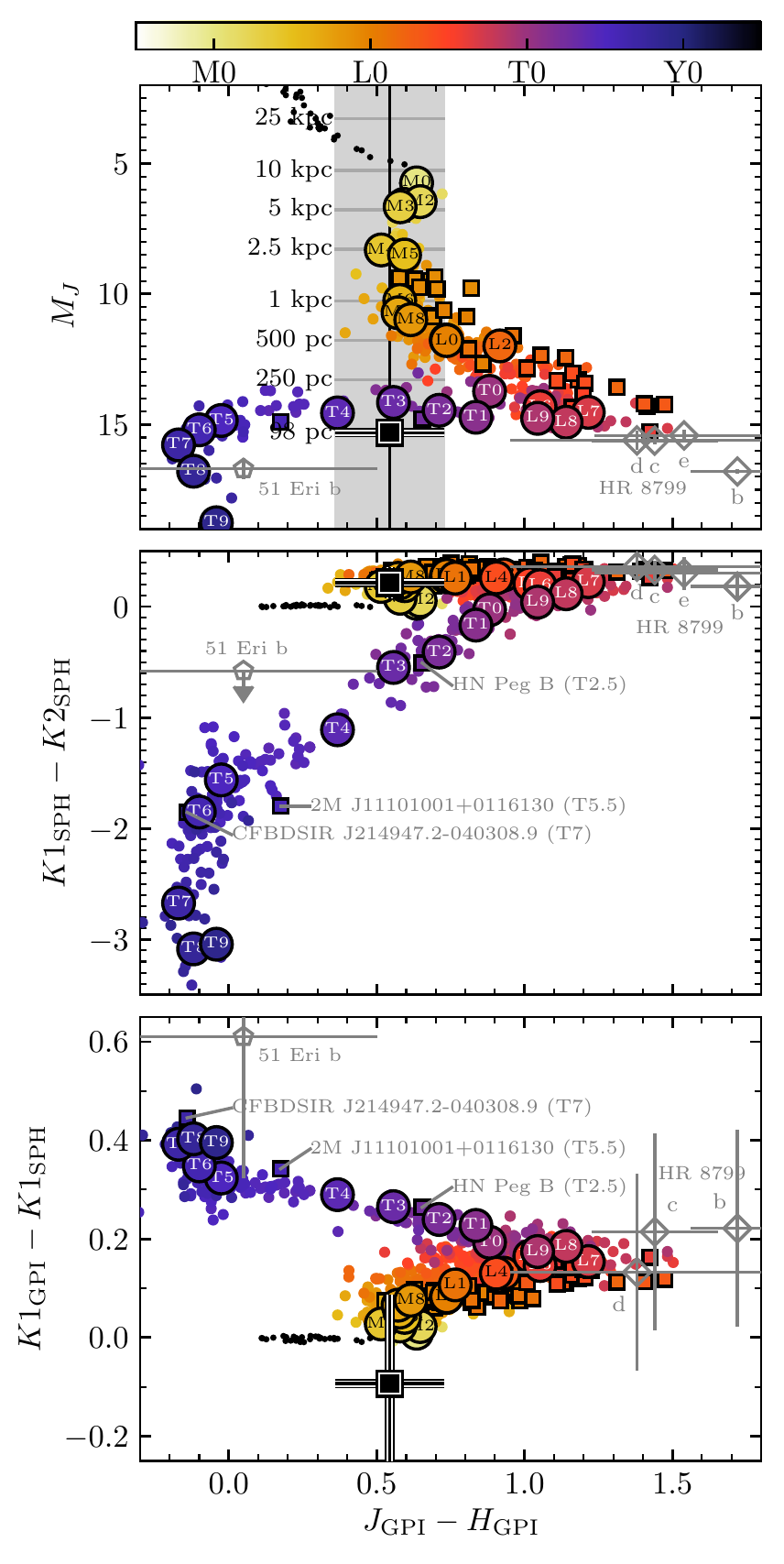}
\caption{\label{fig:CCD} CMD (top panel) and CCDs (middle and bottom panels) showing HD~131399~Ab (black square) relative to stars, brown dwarfs, and directly-imaged planets. Low-gravity (\textsc{vl-g}/$\gamma$) objects are plotted as squares, and field-gravity standards are highlighted \citep{Burgasser:2006cf,Kirkpatrick:2010dc}. Also highlighted are several young T-dwarfs, as well as 51~Eri~b (gray pentagon, \citealp{Samland:2017vi, Rajan:2017ur}) and the HR 8799 planets (gray diamonds, \citealp{Barman:2011fe, Skemer:2012gr, Currie:2014fm, Ingraham:2014gx, Zurlo:2016}). Stars with spectral type earlier than M0 are plotted as black points. In addition to the absolute magnitude assuming a distance of 98\,pc, the absolute magnitude if it is a background object is also shown for a range of distances (top panel).}
\end{figure}
\begin{figure}
\includegraphics[width=\columnwidth]{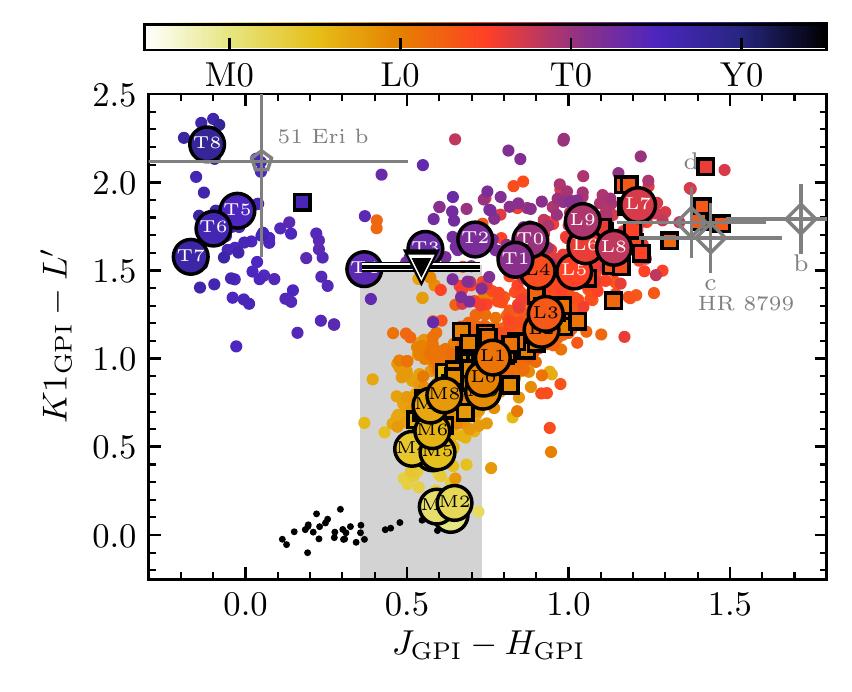}
\caption{\label{fig:CCD2}$K1_{\rm GPI}-L^{\prime}$ vs. $J_{\rm GPI}-H_{\rm GPI}$ CCD showing HD~131399~Ab (black triangle, upper limit on $K1_{\rm GPI}-L^{\prime}$ denoted by shaded gray region) relative to stars, brown dwarfs, and directly-imaged planets. Symbols and colors are as in Figure~\ref{fig:CCD}.}
\end{figure}
The physical nature of HD~131399~Ab can be assessed by placing it on a color-magnitude or color-color diagram (CMD/CCD) and comparing it to the location of other objects of known spectral types. A library of medium-resolution ($R\sim200$) near-IR spectra of stars and brown dwarfs was compiled from the SpeX Prism library\footnote{{\tt http://pono.ucsd.edu/\~{}adam/browndwarfs/spexprism}} \citep{Burgasser:2014tr}, the IRTF Spectral Library\footnote{{\tt http://irtfweb.ifa.hawaii.edu/\~{}spex/IRTF\_Spectral\_Library}} \citep{Cushing:2005ed}, and the Montreal Spectral Library\footnote{{\tt https://jgagneastro.wordpress.com/ \\ the-montreal-spectral-library/}} (e.g., \citealp{Gagne:2015dc,Robert:2016gh}). The spectra were normalized to literature 2MASS and/or MKO photometry. Parallax measurements were obtained from \citet{Dupuy:2012bp, Dupuy:2013ks, Liu:2016co} (and references therein) for the brown dwarfs, and from \citet{vanLeeuwen:2007dc} for the stars. Synthetic magnitudes in the GPI and SPHERE filters were calculated for each object using the filter curves shown in Figure~\ref{fig:filters}. We generated a $M_J$ vs. $J_{\rm GPI}-H_{\rm GPI}$ CMD, and $K1_{\rm SPH} - K2_{\rm SPH}$ vs. $J_{\rm GPI}-H_{\rm GPI}$, and $K1_{\rm GPI} - K1_{\rm SPH}$ vs. $J_{\rm GPI}-H_{\rm GPI}$ CCDs, all of which are plotted in Figure~\ref{fig:CCD}. A $K1_{\rm GPI}-L^{\prime}$ vs. $J_{\rm GPI}-H_{\rm GPI}$ CCD was also created, shown in Figure~\ref{fig:CCD2} using literature MKO $L^{\prime}$ photometry, or estimated from the WISE $W1$ to MKO $L^{\prime}$ color transformation given in \citet{DeRosa:2016kh}. No extinction correction was applied to the colors, although this is expected to be small ($A_V\sim0.22$, $A_J\sim 0.06$, $A_H\sim 0.04$\,mag) at the distance to HD~131399~A, increasing to $A_V\sim1$\,mag ($A_J\sim 0.29$, $A_H\sim0.18$\,mag) due to a combination of the extinction within the UCL region, and the predicted extinction from galactic dust. The location of field-gravity standards for spectral types later than M0 are highlighted in each diagram \citep{Burgasser:2006cf,Kirkpatrick:2010dc}.

Using the contrasts between A and Ab reported in Table~\ref{tab:b_phot}, and the synthetic magnitudes for A calculated in Section~\ref{sec:sedA}, we derive colors of $J_{\rm GPI}-H_{\rm GPI} = 0.54\pm0.18$\,mag, $K1_{\rm SPH} - K2_{\rm SPH} = 0.22\pm0.14$\,mag, and $K1_{\rm GPI} - K1_{\rm SPH} = -0.09\pm0.18$\,mag for HD~131399~Ab. We also derive an upper limit of $K1_{\rm GPI}-L^{\prime} > 1.52$\,mag, using the detection limit from the NIRC2 $L^{\prime}$ observations. On each of the CCDs in Figure~\ref{fig:CCD}, HD~131399~Ab is consistent with the colors of M-dwarfs, and is significantly different from the observed colors of early to mid-T dwarfs, a discrepancy that is most significant for the measured $K1_{\rm SPH}-K2_{\rm SPH}$ color. As a comparison, the upper limit on the color of 51~Eri~b of $K1_{\rm SPH}-K2_{\rm SPH}<-0.58\pm0.14$\,mag \citep{Samland:2017vi} is more than 3$\sigma$ discrepant. The position of HD~131399~Ab on the $K1_{\rm GPI}-L^{\prime}$ vs. $J_{\rm GPI}-H_{\rm GPI}$ CCD (Figure~\ref{fig:CCD2}) only excludes mid to late-Ls and late-Ts; M-dwarfs and mid-Ts are consistent with the measured $J_{\rm GPI}-H_{\rm GPI}$ color and the $K1_{\rm GPI}-L^{\prime}$ upper limit.

While the absolute $M_J$ magnitude is consistent with an early to mid-T dwarf, the $J-H$ color is far less diagnostic (Figure~\ref{fig:CCD}, top panel). If the distance to HD~131399~Ab was not known, the only constraint on the spectral type from the $J-H$ color would be that it is between mid-G and late-M, or between early to mid-T. Excluding the $J-H$ color, the only evidence in support of the bound T-dwarf companion hypothesis from these color-magnitude and color-color diagrams is the absolute $J$-band magnitude, which relies on the assumption that it is at the same distance as HD~131399~A ($98.0^{+7.2}_{-6.3}$\,pc), and the upper limit on the $K1_{\rm GPI}-L^{\prime}$ color, which is consistent with either a M-dwarf or a mid-T dwarf. The remaining color indices plotted in Figure~\ref{fig:CCD} except for $J-H$ are inconsistent with the observed colors of field T-dwarfs. Instead, they are consistent with those of field M-dwarfs, which would require HD~131399~Ab to be at a significantly greater distance of between 1--10\,kpc and not physically associated with HD~131399~A.

\subsubsection{Comparison to spectra of field objects}
\begin{figure*}
\includegraphics[width=\textwidth]{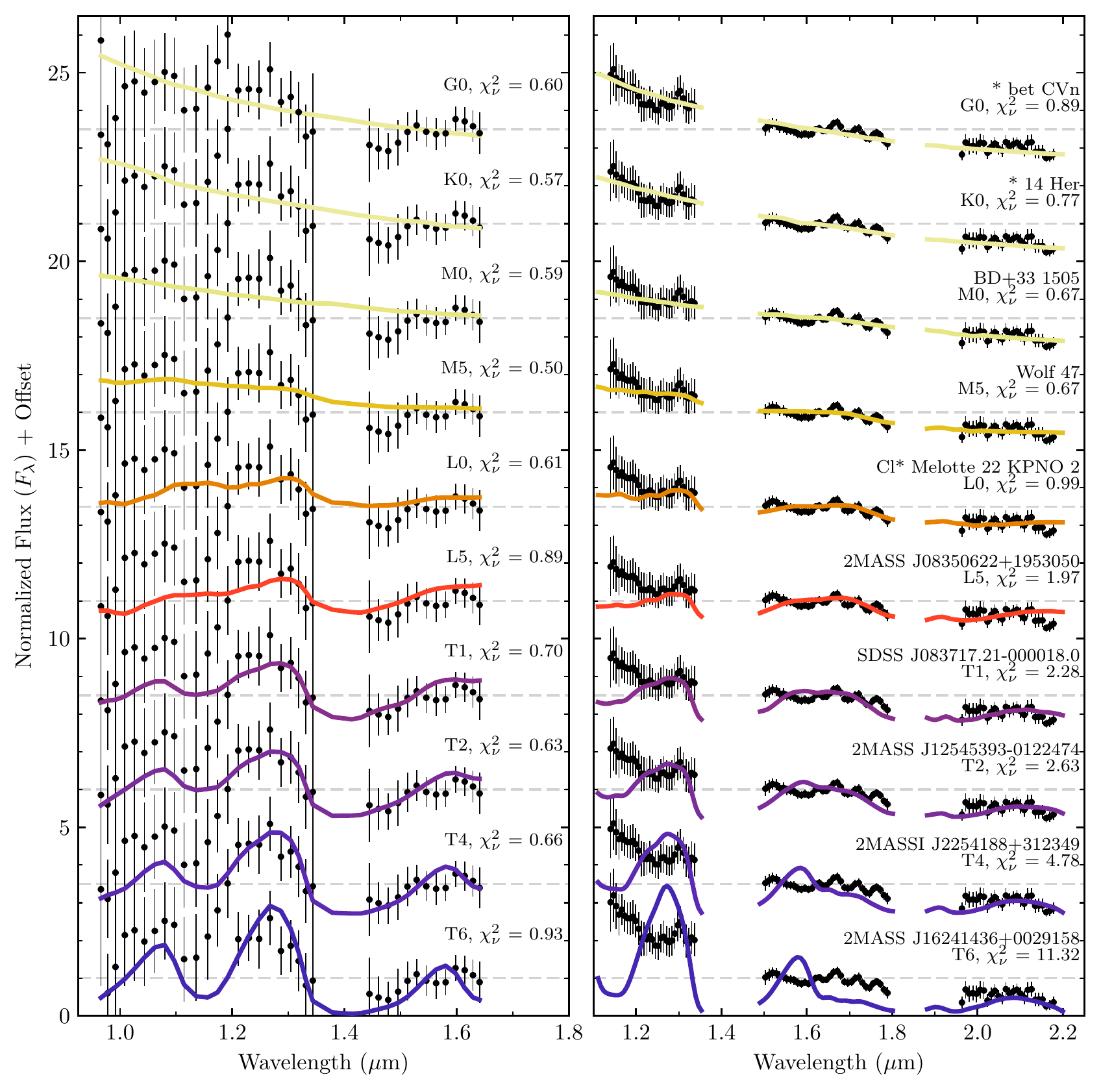}
\caption{\label{fig:spt_fit}Near-infrared spectra of representative objects of spectral types ranging from G0 to T6 compared to the measured spectrum of HD~131399~Ab obtained from our analysis of the SPHERE observations (left panel) and from the new GPI observations presented in this study (right panel). The same comparison object is plotted in both panels for each spectral type. Spectra were obtained from: \citealp{Rayner:2009ki} (G0, K0, M5), \citealp{Kirkpatrick:2010dc} (M0), \citealp{Burgasser:2006jj} (L0), \citealp{Chiu:2006jd} (L5), \citealp{Burgasser:2006cf} (T0), \citealp{Cushing:2005ed} (T2), \citealp{Burgasser:2004hg} (T4), and \citealp{Burgasser:2006el} (T6).}
\end{figure*}
\begin{figure}
\includegraphics[width=\columnwidth]{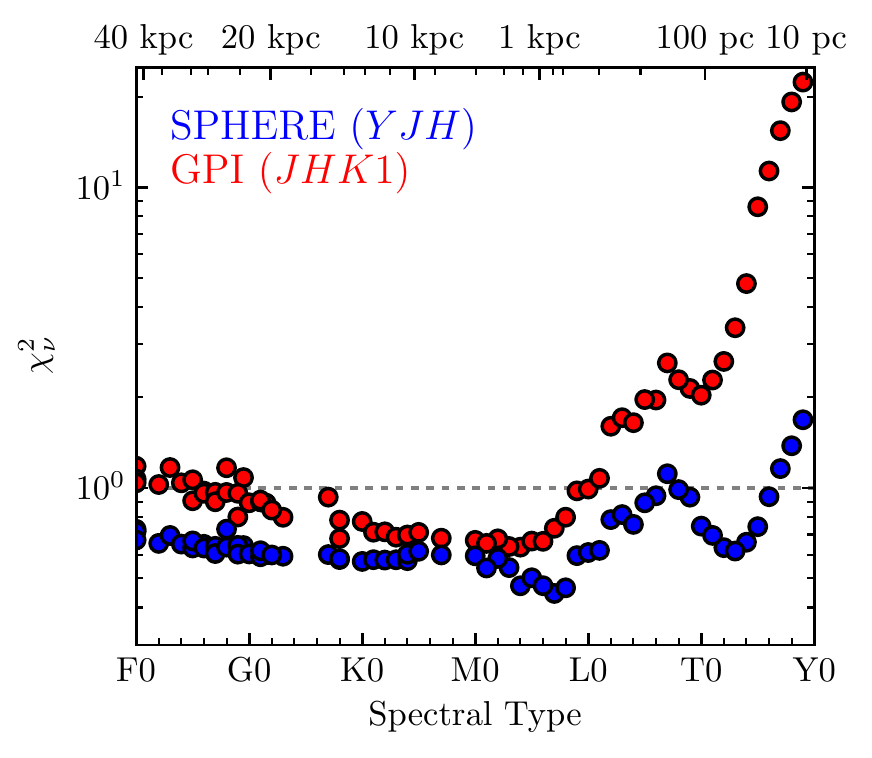}
\caption{\label{fig:spt_chi2}$\chi^2_{\nu}$ as a function of spectral type for the comparison between the spectral library and the SPHERE {\it YJH} spectrum (blue points) and the GPI {\it JHK1} spectrum (red points). Only field-gravity standards later than M0 are plotted for clarity. The SPHERE spectrum does not provide strong constraints on the spectral type earlier than $\sim$T8, while the GPI spectrum is only consistent with objects earlier than L0 ($\chi^2_{\nu} < 1$). The photometric distance required to match the apparent $J$-band magnitude of HD~131399~Ab is shown on the top axis.}
\end{figure}
One of the primary reasons why instruments such as GPI and SPHERE use an integral field spectrograph is the ability to immediately distinguish between background stars, which have relatively featureless spectra, and cool substellar companions with strong molecular absorption features. While the $J-H$ of HD~131399~Ab is consistent with both stars between mid-G and late-M and brown dwarfs between early-T and mid-T (Figure~\ref{fig:CCD}), the {\it JH} spectra of these two groups of objects are significantly different. With a high enough SNR spectrum, it should be possible to confirm or reject the presence of strong molecular absorption features that are seen in the spectra of cool brown dwarfs.

We compared the GPI and our SPHERE spectra of HD~131399~Ab to the library of near-IR spectra described in Section~\ref{sec:ccd}. The spectra of each object within the library was degraded to the resolution of the GPI/SPHERE spectra by convolving the spectrum with a Gaussian of appropriate width. The scaling factor that minimized $\chi^2$ was found analytically for the comparison to the SPHERE data, and numerically for the comparison to the GPI data where the separate bands were allowed to float independently to account for uncertainties in the satellite spot ratio \citep{Maire:2014gs}. Wavelengths with a throughput lower than 50\% (Figure~\ref{fig:filters}) were excluded from the fit.

The fits of the HD~131399~Ab SPHERE and GPI spectra to objects ranging from a spectral type of G0 to T6 are shown in Figure~\ref{fig:spt_fit}, with the minimum $\chi^2_{\nu}$ plotted as a function of spectral type in Figure~\ref{fig:spt_chi2}. The lower SNR of the SPHERE spectrum is apparent (Figure~\ref{fig:spt_fit}, left panel), with $\chi^2_{\nu}<1$ for all spectral types except for those between L5--L9 and T5--T9 (Figure~\ref{fig:spt_chi2}). The SPHERE spectrum is fit well ($\chi^2_{\nu} < 1$) by objects that have significantly different spectral morphologies: from an M5 dwarf (Wolf~47, $\chi^2_{\nu} = 0.50$), with a relatively featureless spectrum, to a T2 (2MASS~J12545393--0122474, $\chi^2_{\nu} = 0.63$) or a T4 brown dwarf (2MASSI~J2254188+312349, $\chi^2_{\nu} = 0.66$) which exhibit strong molecular absorption features. The {\it YJ} portion of the spectrum is consistent within the uncertainties with spectral types earlier than T6, providing little diagnostic power. The {\it H} band spectrum exhibits a rising slope towards longer wavelengths, similar to what is seen in the spectra of brown dwarfs later than L5, although this slope is not measured at a significant level given the low SNR.

The improved SNR and greater wavelength coverage of the GPI {\it JHK1} spectrum provide for better constraints on the spectral type of HD~131399~Ab (Figure~\ref{fig:spt_fit}). The spectrum appears relatively featureless, consistent with the near-IR SED of stars with a spectral type earlier than mid-M. The red end of the {\it H} spectrum appears to modulate on a characteristic length scale consistent with the intrinsic resolution of GPI at {\it H}. It is likely this is correlated noise due to the presence of speckles at those wavelengths rather than an astrophysical signal. The GPI spectrum is fit well by both an M0 (BD+33 1505, $\chi^2_{\nu}=0.67$) and an M5 (Wolf 47, $\chi^2_{\nu}=0.67$) dwarf. Earlier spectral types are also fit well ($\chi^2_{\nu} < 1$), although these would require HD~131399~Ab to be at a significantly greater distance, inconsistent with the predictions of Galactic population models described in Section~\ref{bg_sec2}. The minimum $\chi^2_{\nu}$ for the fit of the GPI spectrum as a function of spectral type plotted in Figure~\ref{fig:spt_chi2} displays a similar trend to that for the fit of the SPHERE data, with later spectral types being more strongly excluded. Objects earlier than a spectral type of L0 fit the spectrum relatively well ($\chi^2_{\nu} < 1$). One limitation of this analysis is the relative dearth of known young/low surface gravity T-dwarfs. The three within the library---HN~Peg~B (T2.5, $\chi^2_{\nu}=3.0$, \citealp{Luhman:2007}), 2MASS~J11101001+0116130 (T5.5, $\chi^2_{\nu} = 8.9$, \citealp{Burgasser:2006el}), and CFBDSIR~J214947.2-040308.9 (T7, $\chi^2_{\nu} = 16.0$, \citealp{Delorme:2013bo})---are all poor fits to the GPI spectrum of HD~1313199~Ab.

Using the color-magnitude and color-color diagrams in Figure~\ref{fig:CCD} and the fit of the SPHERE and GPI spectra to stars and brown dwarfs in Figures~\ref{fig:spt_fit} and \ref{fig:spt_chi2}, we find no strong evidence to suggest that HD~131399~Ab has a near-IR SED consistent with that of a cool planetary-mass companion of early to mid-T spectral type. We do not detect the characteristic H$_2$O and CH$_4$ absorption in the GPI spectrum at either {\it J} or, more significantly, at {\it H}, nor do we detect it based on the measured $K1_{\rm SPH}-K2_{\rm SPH}$ color, which is sensitive to methane absorption in the spectra of T-dwarfs (Figure~\ref{fig:CCD}, middle panel). Instead, our analysis of the near-IR SED suggests it has a relatively featureless spectrum, and has near-IR colors that are consistent with those of a low-mass star.

\section{Astrometric Analysis and Discussion}\label{sec:astrometry}

\begin{deluxetable*}{ccccccccc}
\tabletypesize{\scriptsize}
\tablecaption{Astrometry of HD~131399~Ab}
\tablewidth{0pt}
\tablehead{
\colhead{UT Date} & \colhead{Instrument} & \colhead{Filter} & \colhead{$\rho_{\rm raw}$} & \colhead{$\theta_{\rm raw}$}  &\colhead{Plate Scale} & \colhead{Position Angle} & \colhead{$\rho_{\rm true}$} & \colhead{$\theta_{\rm true}$}     \\
& & & (px) & (deg) & (mas~px$^{-1}$)\tablenotemark{\it a} & Offset (deg) & (mas) & (deg) 
}
\startdata
2015 Jun 12 & SPH-IFS    & {\it YJH} & $112.9\pm0.2$ & $196.81\pm0.16$ & $7.46\pm0.02$\tablenotemark{\it b} & $-1.75\pm0.19$\tablenotemark{\it b,c}  & $842.4\pm2.9$ & $195.06\pm0.25$  \\
              & SPH-IRDIS & $K1$ &  $69.1\pm0.3$ & $196.94\pm0.25$ & $12.267\pm0.021$\tablenotemark{\it b} & $-1.75\pm0.14$\tablenotemark{\it b,c} & $848.1\pm3.4$ & $195.19\pm0.29$  \\
              & (SPH W16\tablenotemark{\it d}) & (\nodata) & (\nodata) & (\nodata) & ($12.23\pm0.03$) & ($-1.56\pm0.2$) & ($839\pm4$) & ($194.2\pm0.3$) \\   
2016 Mar 06 &  SPH-IFS    & {\it YJH} & $111.6\pm0.2$ & $196.68\pm0.15$ & $7.46\pm0.02$\tablenotemark{\it b} & $-1.75\pm0.19$\tablenotemark{\it b,c} & $832.1\pm2.9$ & $194.94\pm0.24$ \\
              & SPH-IRDIS & $K1$ & $67.5\pm0.3$ & $196.58\pm0.27$ & $12.267\pm0.021$\tablenotemark{\it b} & $-1.75\pm0.14$\tablenotemark{\it b,c} & $828.2\pm4.1$ & $194.83\pm0.30$ \\
          & (SPH W16\tablenotemark{\it d}) & (\nodata) & (\nodata) & (\nodata) & ($12.24\pm0.03$) & ($-1.40\pm0.2$) & ($834\pm4$) & ($193.8\pm0.3$) \\   
2016 Mar 17 &  SPH-IFS    & {\it YJH} & $110.5\pm0.3$ & $196.69\pm0.16$ & $7.46\pm0.02$\tablenotemark{\it b} & $-1.75\pm0.19$\tablenotemark{\it b,c} & $824.3\pm3.2$ & $194.94\pm0.25$ \\
              & SPH-IRDIS & $K1$ & $67.3\pm0.3$ & $196.79\pm0.23$ & $12.267\pm0.021$\tablenotemark{\it b} & $-1.75\pm0.14$\tablenotemark{\it b,c} & $825.3\pm3.4$ & $195.03\pm0.27$ \\
     & (SPH W16\tablenotemark{\it d}) & (\nodata) & (\nodata) & (\nodata) & ($12.24\pm0.03$) & ($-1.40\pm0.2$) & ($834\pm4$) & ($193.8\pm0.3$) \\ 
2016 May 07   &  SPH-IFS    & {\it YJH} & $110.2\pm0.3$ & $196.11\pm0.17$ & $7.46\pm0.02$\tablenotemark{\it b} & $-1.75\pm0.19$\tablenotemark{\it b,c} & $822.4\pm3.4$ & $194.36\pm0.25$ \\
              & SPH-IRDIS & $K1$ & $67.2\pm0.3$ & $195.89\pm0.15$ & $12.267\pm0.021$\tablenotemark{\it b} & $-1.75\pm0.14$\tablenotemark{\it b,c} & $824.2\pm3.4$ & $194.14\pm0.20$ \\
            & (SPH W16\tablenotemark{\it d}) & (\nodata) & (\nodata) & (\nodata) & ($12.24\pm0.03$) & ($-1.47\pm0.2$) & ($830\pm4$) & ($193.5\pm0.3$) \\
2017 Feb 15 & GPI & {\it H}   & $56.7\pm0.2$ & $195.01\pm0.17$ &  $14.166\pm0.007$ & $-1.10\pm0.13$  & $802.9\pm2.4$ & $193.92\pm0.21$    \\
2017 Feb 16 & GPI & {\it J}   &  $56.8\pm0.2$ & $194.93\pm0.16$ &  $14.166\pm0.007$ & $-1.10\pm0.13$  & $804.6\pm2.4$ &$193.83\pm0.21$   \\
2017 Mar 15 & SPH-IRDIS & {\it J} & $65.4\pm0.2$ & $195.43\pm0.16$ & $12.263\pm0.021$\tablenotemark{\it b} & $-1.75\pm0.14$\tablenotemark{\it b,c} & $801.8\pm2.5$ & $193.68\pm0.21$ \\
2017 Apr 20 & GPI & {\it H}   & $56.4\pm0.1$ & $194.68\pm0.13$ &  $14.166\pm0.007$ & $-1.10\pm0.13$ & $799.0\pm1.8$ & $193.58\pm0.18$ \\ 
\enddata
\tablenotetext{a}{In reduced GPI/SPHERE-IFS datacubes, one pixel is equivalent to one lenslet.}
\tablenotetext{b}{\citet{Maire:2016} and ESO SPHERE user manual 7th edition.}
\tablenotetext{c}{Since the pupil offset ($135.99\pm0.11$\,deg) and the IFS offset ($-100.48\pm0.13$\,deg) are corrected by our pipeline, the position angle offset corresponds to the calibration of a reduced image with respect to true north ($-1.75\pm 0.08$\,deg). The uncertainty on the position angle offset is the quadratic sum of all three angles for the IFS data, and of the pupil offset and true north correction for the IRDIS data.}
\tablenotetext{d}{\citet{wagner2016} reports the astrometry without specifying the instrument and one point for the two epochs in March 2016.}
\label{tab:astr}
\end{deluxetable*}
\begin{deluxetable}{cccc}
\tabletypesize{\normalsize}
\tablecaption{\label{tab:astrometry_B}SPHERE Astrometry of HD~131399~B}
\tablewidth{0pt}
\tablehead{
\colhead{Epoch} & \colhead{Parameter} & \colhead{W16} & \colhead{This work}}
\startdata
2015 Jun 12 & $\rho$ (mas) & $3149\pm6$ & $3153\pm6$ \\
            & $\theta$ (deg) & $221.9\pm0.3$ & $222.2\pm0.2$ \\
2016 Mar 06 & $\rho$ (mas) & $3150\pm6$\tablenotemark{\it a} & $3151\pm6$ \\
            & $\theta$ (deg) & $221.5\pm0.3$\tablenotemark{\it a} & $221.9\pm0.2$ \\
2016 Mar 17 & $\rho$ (mas) & \nodata & $3152\pm6$ \\
            & $\theta$ (deg) & \nodata & $222.3\pm0.2$ \\
2016 May 07 & $\rho$ (mas) & $3149\pm6$ & $3154\pm6$ \\
            & $\theta$ (deg) & $221.8\pm0.3$ & $222.2\pm0.2$ \\
\enddata
\tablenotetext{a}{W16 report one point for the two epochs in March 2016.}
\end{deluxetable}

Measurements on the detector chip, calibration values, and calibrated astrometric positions, for each dataset are given in Table \ref{tab:astr}. At each epoch, both IFS and IRDIS measurements agree within the uncertainties. For reference, published calibration values and calibrated positions from W16 are also provided, although which SPHERE detector being used was not specified.  Our reanalysis of the SPHERE data shows a significant change in separation ($\sim$22 mas), much larger than that reported by W16 from an analysis of the same data ($\sim$9 mas).  Comparing the weighted mean of our IFS and IRDIS separation at each epoch to the separations reported by W16 (and using their March 2016 astrometry for both the 2016 March 06 and 2016 March 17 epochs), we find offsets of $+1.45$, $-0.80$, $-2.31$, and $-1.67\sigma$, and thus a much larger projected velocity.  In addition, our position angles are systematically offset by one degree (or $\sim 0.3 \lambda/D$) compared to W16.  

We investigated the origin of this one degree offset between our data reduction and that of W16. The pre-processing and reduction pipelines are similar but not exactly the same version, which mostly has a negligible impact except on the instrument angles. We find that the parallactic angle correction $\epsilon$ is insignificant (of the order of $0.05$\,deg). However, the calibration angles and the instrument angles used in W16 differ from the latest calibrated values \citep{Maire:2016} that were used in our analysis. These differences would make our discrepancies even higher by further lowering their position angles by 0.1--0.25\,deg for the IFS and 0.01-0.18\,deg for IRDIS. As a cross check of the astrometry, we looked at the separations and position angles of HD~131399~B. We find they are consistent at the $1\sigma$ level with that of W16 (see Table \ref{tab:astrometry_B}), though we find systematically higher (by more than the total errors reported by W16) position angles. The systematically larger separations are here due to the larger (by $0.2\%$) calibration platescale. Our astrometry is independently confirmed with the Keck/NIRC2 data at $3149\pm7$\,mas and $222.3\pm0.5$\,deg in February 2017, with the orbital motion being negligible at 400\,au over one year. Another plausible explanation for this offset is measurement biases on HD~131399~Ab. Our reanalysis leads to consistent astrometry using multiple PSF subtraction and astrometric extraction algorithms. Remaining biases due to differences in the way the astrometry was measured between this work and that of W16 however could exist.

When a candidate companion is detected next to a star by direct imaging, there are typically two scenarios that are considered: the candidate is a common proper motion companion orbiting the target star, or the candidate is at infinite distance with no proper motion. We investigate these possibilities with the new GPI and SPHERE astrometry as well as the revised SPHERE points in the following sections.

\subsection{Escape Velocity}\label{escape_sec}

Before fitting an orbit, we first consider whether the projected velocity of HD~131399~Ab is less than the escape velocity of the system, as should be true for a bound orbit.  The projected velocity (in RA and Dec) will in fact be a lower limit on the total velocity, since the total velocity will also include the unmeasured component along the line of sight.  We compute projected velocity by fitting straight lines to the astrometry in RA and Dec as a function of time.  This too represents a lower limit on the velocity, since any curvature not captured by the linear fit would represent a higher velocity.  This value is then converted to a physical velocity (km\,s$^{-1}$) using the distance to the system.

Escape velocity is given by $v_{\rm esc} = \sqrt{2GM/r}$, where $G$ is the gravitational constant, $M$ is the mass of the star, and $r$ is the total separation between star and planet.  In the direct imaging case this  corresponds to an upper limit on escape velocity, since we can only measure the projected separation in RA and Dec.  In fact, the presence of the binary BC would lower the effective escape velocity further beyond this upper limit, since the planet would not need sufficient velocity on its own to reach infinity, but only enough velocity to reach the gravitational sphere of influence of BC to eventually escape.  Separation is computed using the minimum value over the range of epochs of the astrometry (2015 Jun 12 through 2017 Apr 20) from the linear fit, with the minimum value chosen so we continue to define the upper limit of the escape velocity.

In order to compare the projected velocity to the escape velocity limit we use a Monte Carlo method to draw samples from both velocities given uncertainties in the astrometry, distance to the system, and mass of the star.  For each Monte Carlo trial, for both RA and Dec, we generate values of slope (projected velocity) and intercept (reference position) from the covariance matrix of the linear fit to the astrometry, as well as stellar mass and parallax from Gaussian distributions ($2.08 \pm 0.11$\,$M_\odot$ and $10.20 \pm 0.70$\,mas, respectively).  Finally, we compute the ratio of projected velocity to escape velocity, which should be less than unity for a bound orbit.  This method accounts for correlations in distance, since the same generated distance is used to calculate projected velocity and escape velocity, as well as between slope and intercept, since the same generated pair is used to compute the minimum separation for the escape velocity as well as the projected velocity.

\begin{figure}
\centering
\includegraphics{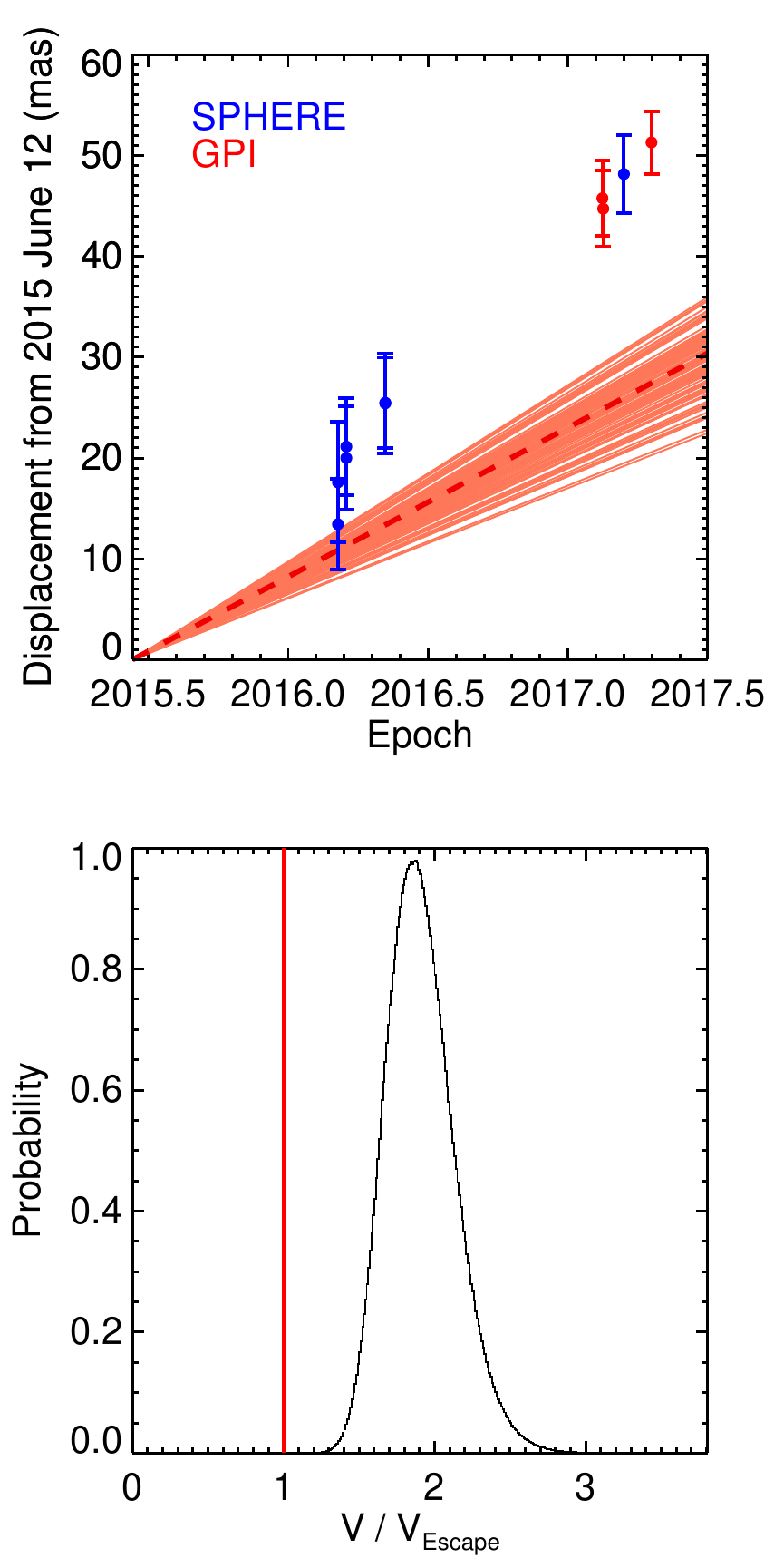}
\caption{(Top) Offset of HD~131399~Ab over time from the first epoch, 2015 June 12, as measured with VLT/SPHERE (blue) and Gemini/GPI (red). Red lines represent Monte Carlo draws of the escape velocity, with the dashed line showing the median escape velocity.  Later epochs, especially those in 2017, show the motion of HD~131399~Ab to be significantly above escape velocity.  The posterior probability distribution of the projected velocity of HD~131399~Ab divided by the escape velocity (bottom), illustrates that the system is not consistent with a bound orbit ($v / v_{\rm esc} = 1.89\pm0.23$).}
\label{fig:esc1}
\end{figure}

In Figure~\ref{fig:esc1} we plot the total projected displacement (total distance in both RA and Dec) between the first epoch and all subsequent epochs.  The reference location is taken as the average of the IRDIS and IFS astrometry at 2015 June 12.  We draw 100 values of the escape velocity from our Monte Carlo analysis, and plot these as red lines, normalized to pass through the reference location.  The astrometry clearly show a steeper slope (a faster velocity) than the escape velocity.  Computing the quotient of projected velocity and escape velocity (bottom panel of Figure~\ref{fig:esc1}) shows that the projected velocity is indeed always greater than the escape velocity, with the ratio having a value of $1.89\pm0.23$, which reached a minimum value of 1.07 out of $10^7$ trials. Thus the data are robustly inconsistent with the hypothesis that HD~131399~Ab is a bound planet.

The quantity $v / v_{\rm esc}$ is inversely proportional to the square root of stellar mass, and directly proportional to $d^{1.5}$, with a factor of $d$ coming from the projected velocity and another factor of $\sqrt{d}$ from the escape velocity.  The 2$\sigma$ lower limit on this quantity is 1.56 (95.45\% of the samples are larger that this number).  In order to bring this 2$\sigma$ limit to unity, it is therefore necessary to increase the mass of HD~131399~A by a factor of $1.56^2 = 2.43$, or decrease the distance by $1.56^{1/1.5} = 1.35$ (or else have a linear combination of these two changes).  Such a change would represent a 27$\sigma$ deviation in mass or a 5.1$\sigma$ deviation in distance.  Such a change in distance would likely exclude HD~131399~A from the UCL association, therefore the star and HD~131399~Ab would be much older, which ultimately affects the model-dependent mass estimate of the latter.
Of these two, the most susceptible to error is mass, since the mass of the primary comes from a SED fit, and lower-mass stellar companions, too close to be resolved with GPI, could add additional mass that would raise the escape velocity.  However, it is difficult to imagine there being 3\,$M_\odot$ of additional stars close to the 2.08\,$M_\odot$ HD~131399~A.  The most likely high-mass companion would be an equal-mass binary, which even then is not enough to make the orbital velocity equal escape velocity at the 2$\sigma$ level, and would be evident in the distance posterior from the SED fit.

This large upper limit on $v / v_{\rm esc}$ is not solely dependent on the astrometric calibration between SPHERE and GPI.  When we repeat the same analysis for only the SPHERE astrometry presented here, 5 epochs from 2015 to 2017, we find this factor has a value of $1.88\pm0.25$, with only 4 out of $10^7$ trials less than unity.  Using the original astrometry reported by W16, this factor becomes consistent with bound orbits, $1.02\pm0.42$, with 48\% of generated values below unity.  In contrast, when we use our astrometry for these same 4 epochs from 2015 to 2016, we find a value of $1.80\pm0.31$, with 0.09\% of trials less than unity, consistent with our finding of a significantly larger slope in our reduction of the 2015--2016 SPHERE data compared to W16.

\subsection{Exploring Orbital Phase Space}\label{orbit_sec}

\begin{figure}
\includegraphics[width=\columnwidth,trim={0 0cm 0 0},clip]{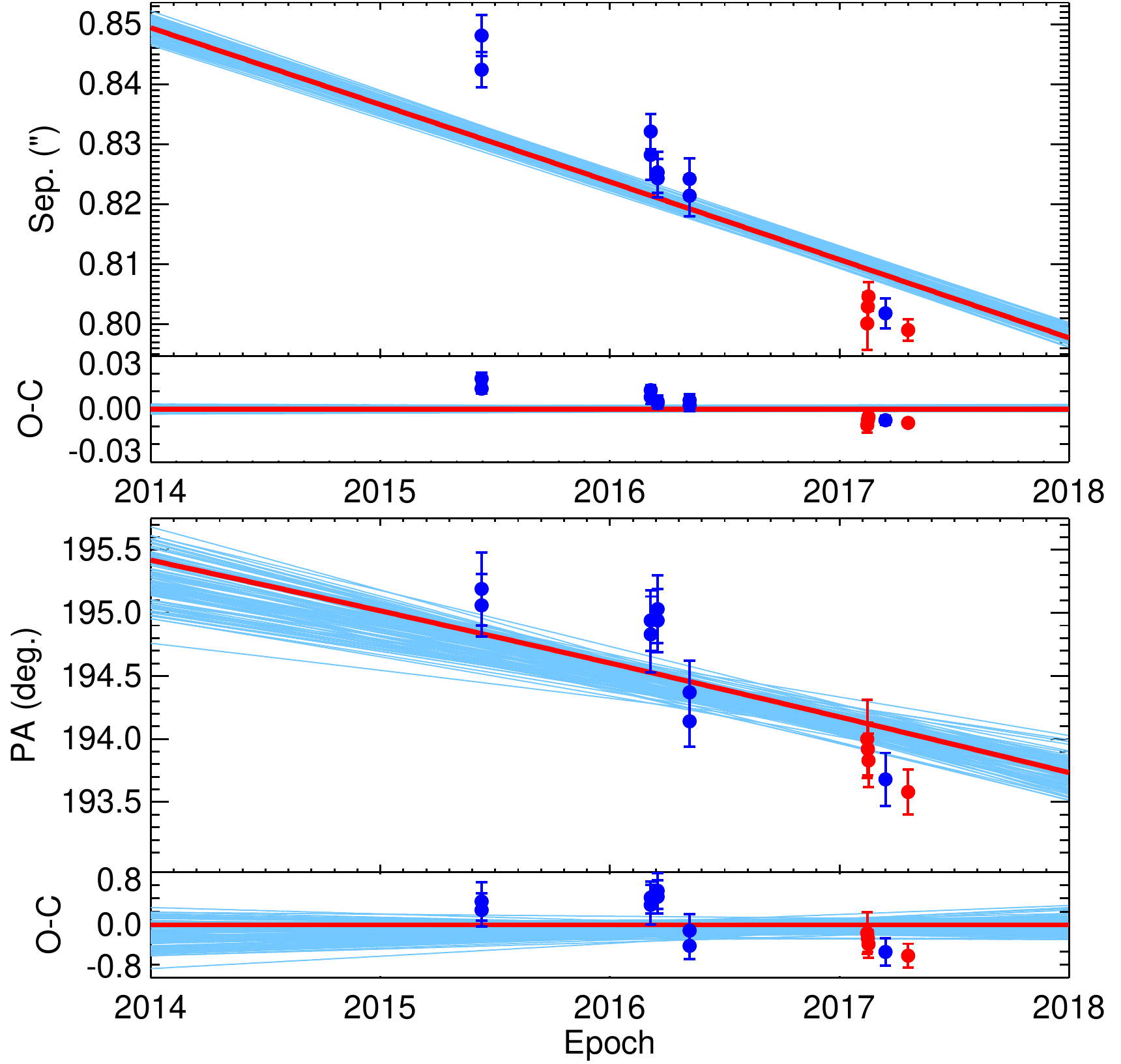}
\caption{Orbital fit to HD~131399~Ab when fixing the distance of the system to 98\,pc and mass to 2.08\,$M_{\odot}$.  As expected given our analysis of the escape velocity, no bound orbit has enough projected velocity to fit the data.  The plotted orbits are all highly eccentric, $e>0.95$, with minimum apastron of 597\,au, which is about twice the projected separation between HD~131399~A and HD~131399~B, so even this poorly-fitting orbit ($\chi_\nu^2=6.7$) is very unlikely given the system architecture.}
\label{fig:orbit1}
\end{figure}

In the previous section we have demonstrated that the projected velocity of HD~131399~Ab is significantly above the escape velocity.  We proceed to explore the magnitude of the offset required in the astrometry, mass, and distance in order to fit a bound orbit to the data.  We begin by assuming a fixed mass and distance of the star, 2.08\,M$_\odot$ and 98.0\,pc.  In order to investigate the orbital parameters required to fit the astrometry, we use the rejection sampling algorithm OFTI \citep{Blunt:2016, DeRosa:2015jl,rameau2016}.  Using OFTI, we generate 100 orbits drawn from the posterior probability distribution, and plot them in Figure~\ref{fig:orbit1}.

Unsurprisingly, since the projected motion is faster than escape velocity, the best-fitting orbit is a poor fit to the data, with a systematically steeper slope in the data than the fit.  With $\chi_\nu^2=6.7$, even the best-fitting orbit is clearly a bad fit to the data.  This high projected velocity is only possible with very high orbital eccentricity, $e>0.949$ for all generated orbits.  This results in a high value of apastron, with 68\% confidence between 1017 and 22433\,au, and a minimum value of 597\,au.  The projected separation of the closest of the BC pair, HD~131399~B, with respect to HD~131399~A, is 309\,au.  Thus a large semimajor axis for BC around A ($\gtrsim$1000 au) and a highly inclined orbit ($\approx 70-110^\circ$, so that the projected separation is only $\sim$300 au) would be required for these orbits to not cross each other.

When fitting orbits, it is more correct to incorporate errors on mass and distance, by varying these parameters in the orbit fit and imposing priors as Gaussians given the measurements.  We noted in Section~\ref{escape_sec} that the escape velocity problem can be ameliorated by increasing the mass of the star or decreasing the distance, and so this standard orbit method will have the result of balancing, in a Bayesian sense, the $\chi^2$ of the orbit fit, the mass, and parallax to find the most likely compromise between the three.

To investigate the effect of allowing the distance and mass of the star to vary within the orbit fit, we use the {\tt emcee} parallel-tempered affine-invariant MCMC sampler to estimate the orbital elements from the astrometry presented in Table~\ref{tab:astr}. We fit eight parameters: semi-major axis $a$, inclination $i$, eccentricity $e$, position angle of nodes and argument of periastron as $\Omega + \omega$ and $\Omega - \omega$, epoch of periastron $T_0$, parallax $\pi$, and $M$ (total mass, the mass of Ab being negligible if bound). Here we define $T_0$, in units of orbital period from the first epoch of the astrometric record (2015.44). We adopt uniform priors in $\log_{10} a$, $\cos i$ ($-1$ to 1), $e$ ($<1$), $\Omega + \omega$ and $\Omega - \omega$ (0--2$\pi$), and $T_0$ (0--1). The prior on $\pi$ was created by multiplying a Gaussian distribution centered at 10.20\,mas with a 1$\sigma$ width of 0.70\,mas, corresponding to the {\it Hipparcos} parallax of HD~131399~A, with a $\pi^{-4}$ power law distribution. The prior on the mass, $M$, was a Gaussian distribution centered at 2.08\,$M_{\odot}$ with a 1$\sigma$ width of 0.11\,$M_{\odot}$ (Table~\ref{tab:sed_mcmc}). We initialized 512 walkers at each of 32 temperatures. Each walker was advanced for $10^6$ steps, with the first half of each chain discarded as a ``burn in'' as they converged to their final value.

\begin{figure}
\includegraphics[width=\columnwidth,trim={0 0cm 0 0},clip]{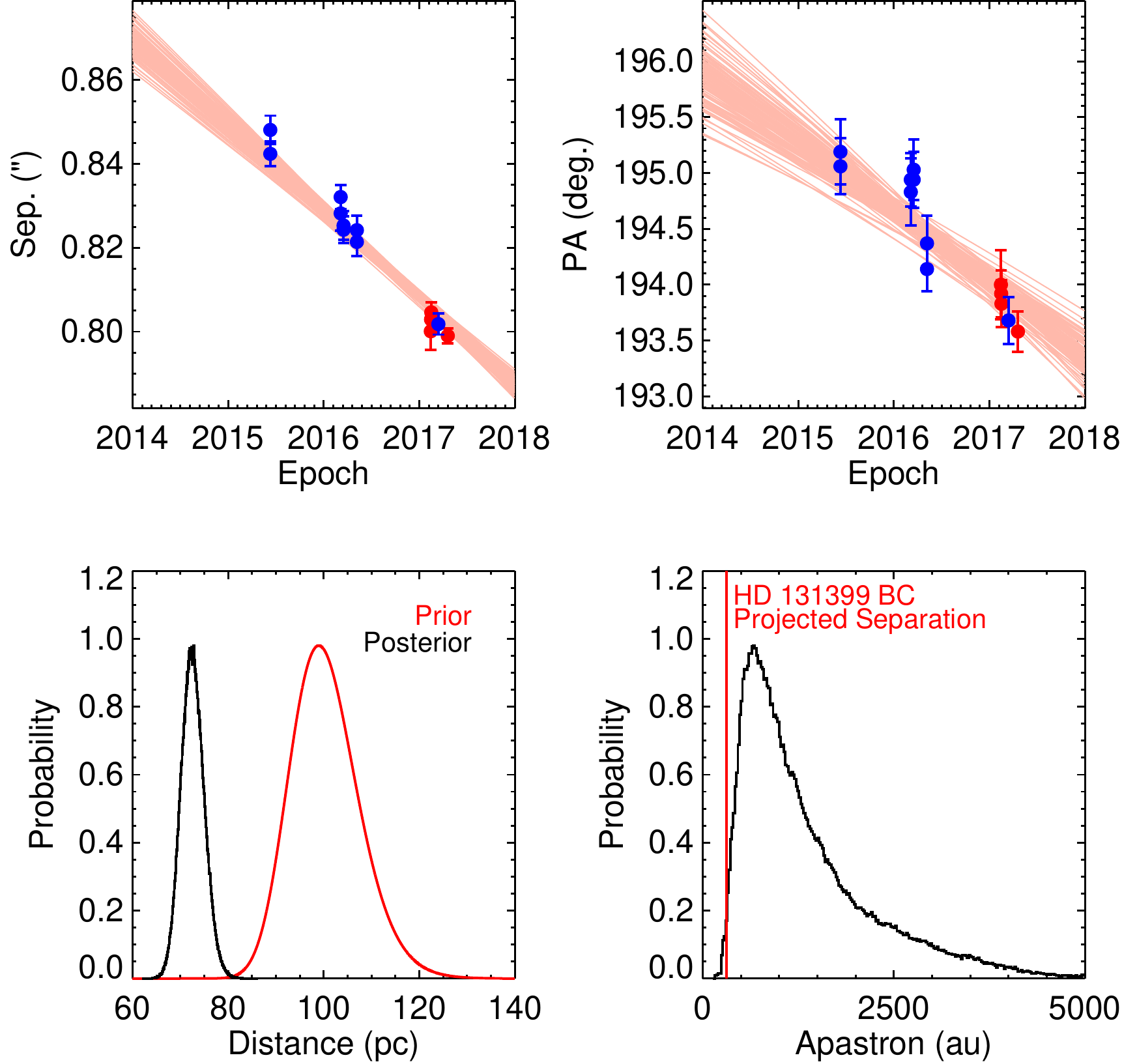}
\caption{Modified orbital fit to HD~131399~Ab where distance and mass are allowed to vary, with priors of Gaussian set by the measurements: $\pi=10.2\pm0.7$\,mas and $M=2.08\pm0.11$\,M$_\odot$.  In order to generate orbital motion fast enough to fit the data, the MCMC had to move the system distance from $\sim$98\,pc to $\sim$73\,pc, which is more than 4$\sigma$ from the \textit{Hipparcos} measurement, and inconsistent with our SED fitting.  Even with this smaller distance, the bottom right panel shows that the resulting orbits are highly eccentric, with apastron distance greater than the projected separation between HD~131399~A and HD~131399~BC for 99\% of orbits.}
\label{fig:orbit2}
\end{figure}

Allowing the distance and mass to float significantly improved the quality of the fit, reducing the minimum $\chi^2_{\nu}$ from 6.7 (Figure~\ref{fig:orbit1}) to 0.92 (Figure~\ref{fig:orbit2}). This improvement was achieved by the MCMC chains moving to a significantly smaller distance to HD~131399~A ($\sim$73\,pc) and a slightly larger total mass ($\sim$2.25\,M$_\odot$). This decreased the measured velocity of the HD~131399~Ab and increased the escape velocity of the system so that bound orbits could be fit. The posterior distribution of the parallax ($\pi = 13.79\pm0.46$\,mas) is 4.3$\sigma$ discrepant from the prior distribution (Figure~\ref{fig:orbit2}), and corresponds to a distance of $72.5\pm2.4$\,pc, consistent with the distance required in the escape velocity analysis in Section~\ref{escape_sec}. This distance is significantly discrepant from the {\it Hipparcos} measurement of $98.0\pm6.9$\,pc, the distance obtained from the SED fit of the three stars in the HD~131399 system of $107.9\pm4.0$\,pc (Section~\ref{sec:sedA}), and also the mean distance of UCL members of $140$\,pc \citep{deZeeuw:1999fe}. The posterior distribution of the mass ($M = 2.27\pm0.10$\,$M_{\odot}$) is shifted by 1.3$\sigma$ relative to the prior distribution ($M=2.08\pm0.11$\,$M_{\odot}$). As with the fit using OFTI with a fixed mass and distance, the posterior distribution of $e$ is strongly peaked at very high eccentricities. We find a median and 1$\sigma$ range of $e=0.980_{-0.017}^{+0.010}$, and the lowest eccentricity within any of the MCMC chains was $e=0.82$.

\subsection{Standard Test of Background Motion Assuming an Infinitely Distant Background Object}\label{bg_sec1}

Most stars targeted by direct imaging are relatively nearby ($\lesssim$100\,pc), and so they typically have well-measured parallaxes and proper motions (errors $\lesssim$ 1\,mas).  Thus the motion of the target star across the sky is well determined over time.  Analysis then proceeds by comparing the relative astrometry between candidate and target star over multiple epochs, and determining whether it follows the background track (which, relative to the target star, moves in the opposite direction as the parallax and proper motion of the star), or is more consistent with common proper motion (e.g. \citealt{nielsen:2013}).

\begin{figure}
\includegraphics[width=\columnwidth]{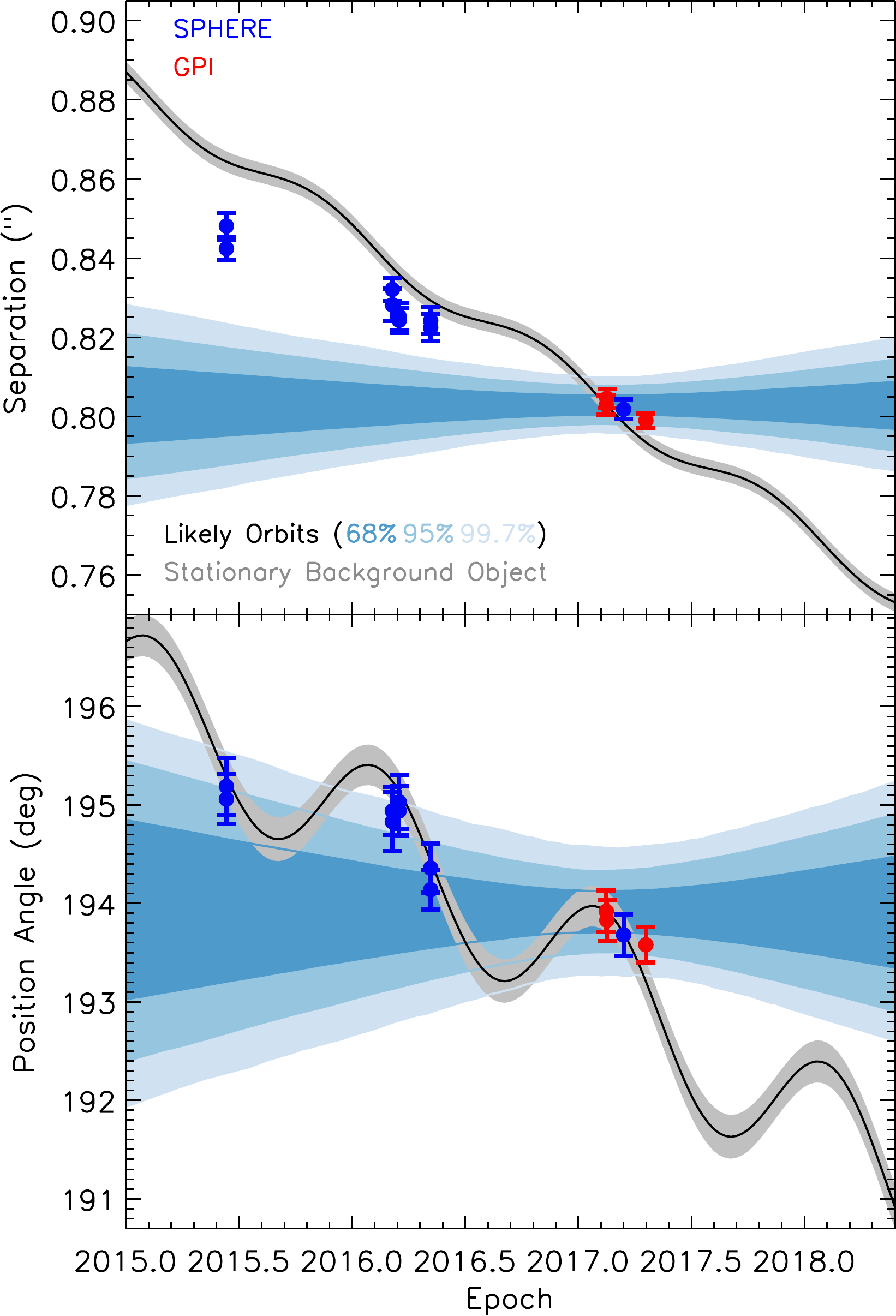}
\caption{Expected motion of the candidate companion to HD~131399~Ab relative to HD~131399~A, as measured by VLT/SPHERE (blue) and Gemini/GPI (red).  An infinitely distant background object with zero proper motion would be following the gray track, while an orbiting planet would lie within the blue cone.  Neither explanation is a good match to the data.}
\label{fig:bg1}
\end{figure}

An added complication is that candidates closer to the star may show significant orbital motion over the timeframe of the astrometric observations, as detailed in the case of the planets HR~8799~bcd \citep{Marois:2008ei}, $\beta$~Pic~b \citep{Lagrange:2010}, and 51~Eri~b \citep{DeRosa:2015jl}.  A determination of the status of the candidate can be made when it is shown to follow the background track at a projected velocity (assuming the distance of the target star) inconsistent with a bound orbit, or it does not follow the background track and has motion consistent with a bound orbit.

Figure~\ref{fig:bg1} presents this analysis for  HD~131399~Ab, and demonstrates that neither the infinitely far background object scenario nor the orbiting planet scenario is fully consistent with all the astrometric data.  The background track is tied to the GPI 2017 Feb 15 {\it H}-band point, and assumes the {\it Hipparcos} proper motion for HD~131399~A of ($-26.69\pm0.59$, $-31.52\pm0.55$)\,mas\,yr$^{-1}$, and parallax of $10.2\pm0.7$\,mas \citep{vanLeeuwen:2007dc}.  The width of the gray track corresponds to the 68\,\% confidence interval, based on a Monte Carlo error analysis that combines the errors on the reference epoch astrometry and the star's proper motion and parallax \citep{nielsen:2013}.  The orbital motion cones are calculated using the same 2017 Feb 15 GPI astrometry, using the Monte Carlo method described in \citet{DeRosa:2015jl}.  Orbital parameters are drawn from prior distributions: uniform priors in $\omega$ and $T_0$, $i$ following a $\sin i$ distribution, and $e$ following a linear fit to RV planets \citep{nielsen:2010}. $a$ and $\Omega$ are chosen to place the reference astrometry on the orbit at the reference epoch, to within a two-dimensional Gaussian centered on the measured astrometry and with standard deviation equal to the measurement errors.

The astrometry falls neither on the background track nor within the orbit cone.  The 2015 June 12 SPHERE data, in particular, is clearly between the separation values predicted for a zero proper motion background object and an orbiting planet.  We have demonstrated above that orbital motion cannot explain this offset, and so we instead re-examine the assumption that a background object has no proper motion or parallax of its own.

\subsection{A Finite-Distance Background Object with Non-Zero Proper Motion}\label{bg_sec2}

As noted above, the standard assumption when testing for common proper motion is that the background object is at infinite distance with zero proper motion.  In reality, the typical distances to background stars are $\sim$1--10\,kpc, with proper motions less than a few mas per year.  We investigate this possibility using the MCMC method described in \citet{Macintosh:2014js} and \citet{DeRosa:2015jl} to find the proper motion and parallax a background object would need to match the relative astrometry of a candidate companion.  In these previous works we assumed no errors on the astrometry at the reference epoch and no errors on the proper motion and parallax of the primary.  Here we update this method by incorporating these errors and fitting an 8-dimensional function: proper motion (in RA and Dec) and parallax of the candidate, proper motion and parallax of the primary, and separation and PA of the candidate with respect to the primary star at a reference epoch (chosen to be 2017.0 so as to be in the middle of our astrometric record).  Priors are taken to be uniform in proper motion in RA and Dec of the candidate.  We adopt the distance prior described by \citet{bailerjones2015}, which combines a uniform space density prior with an exponential drop-off in stellar density, $p(d) \propto d^2 e^{-d/L}$, where $d$ is distance and $L$ is a reference length scale, which is set to 1000\,pc.  Changing variables from distance to parallax ($\pi = 1/ d$) introduces an additional factor of $1/ \pi^2$, giving us our parallax prior of $p(\pi) \propto \pi^{-4} e^{-1/(\pi L)}$, which is truncated at 10\,kpc.  For the primary, Gaussian priors are used for the proper motion and parallax corresponding to the \citet{vanLeeuwen:2007dc} \textit{Hipparcos} measurements and errors.  Uniform priors are assumed for the separation and position angle at the reference epoch.

\begin{figure}
\includegraphics[width=\columnwidth]{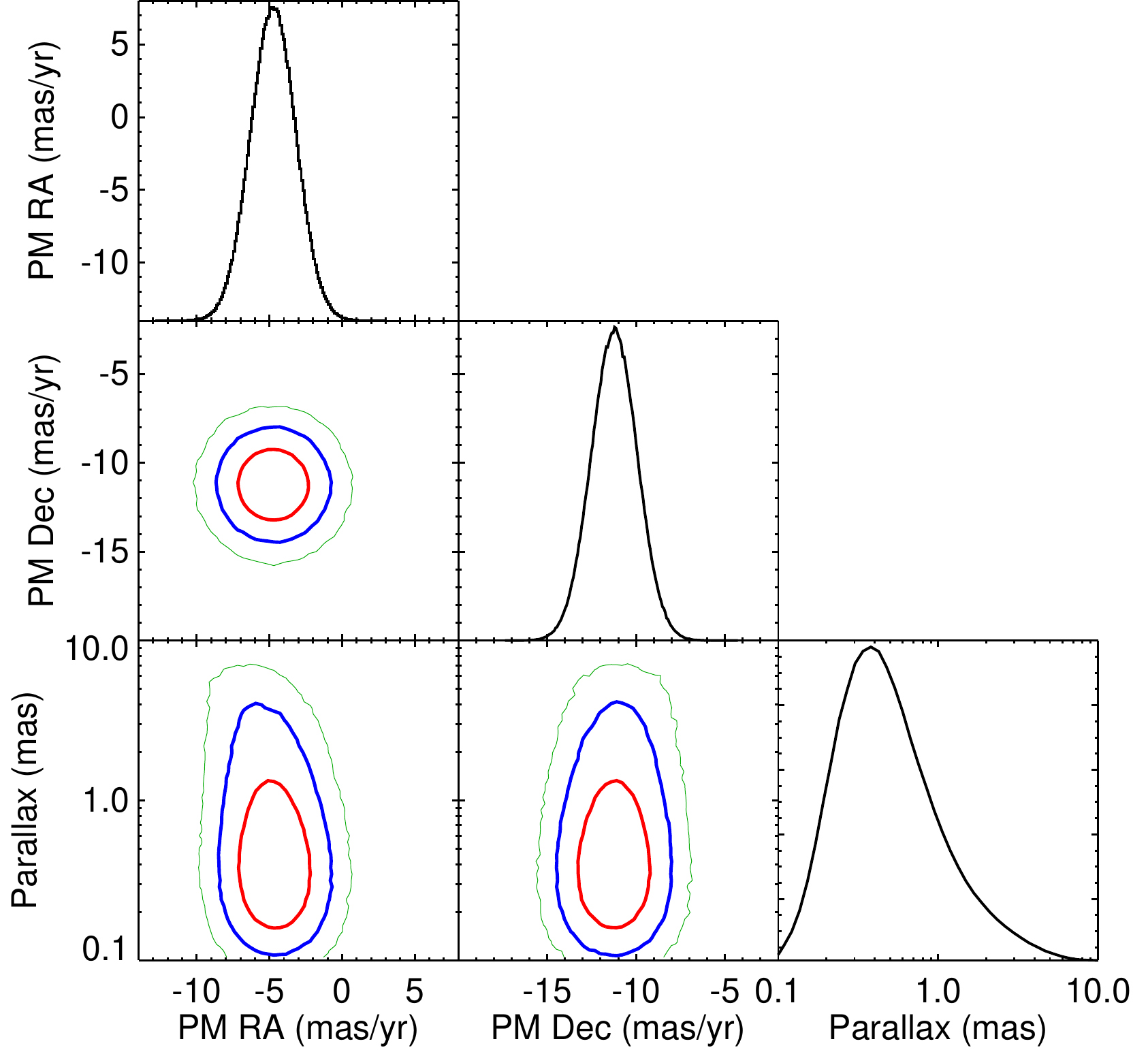}
\caption{Posterior probability distributions and covariances for the proper motion and parallax of HD~131399~Ab, given the relative astrometry of the candidate and the literature measurement of the parallax and proper motion of HD~131399~A.  Red, blue, and green contours correspond to 68\%, 95\%, and 99.7\% confidence intervals.  No appreciable parallax motion of the candidate is seen given our astrometric errors (the shape of the parallax posterior is largely set by our prior), while a significant motion in the Dec direction is observed.}
\label{fig:bg2}
\end{figure}

\begin{figure}
\includegraphics[width=\columnwidth]{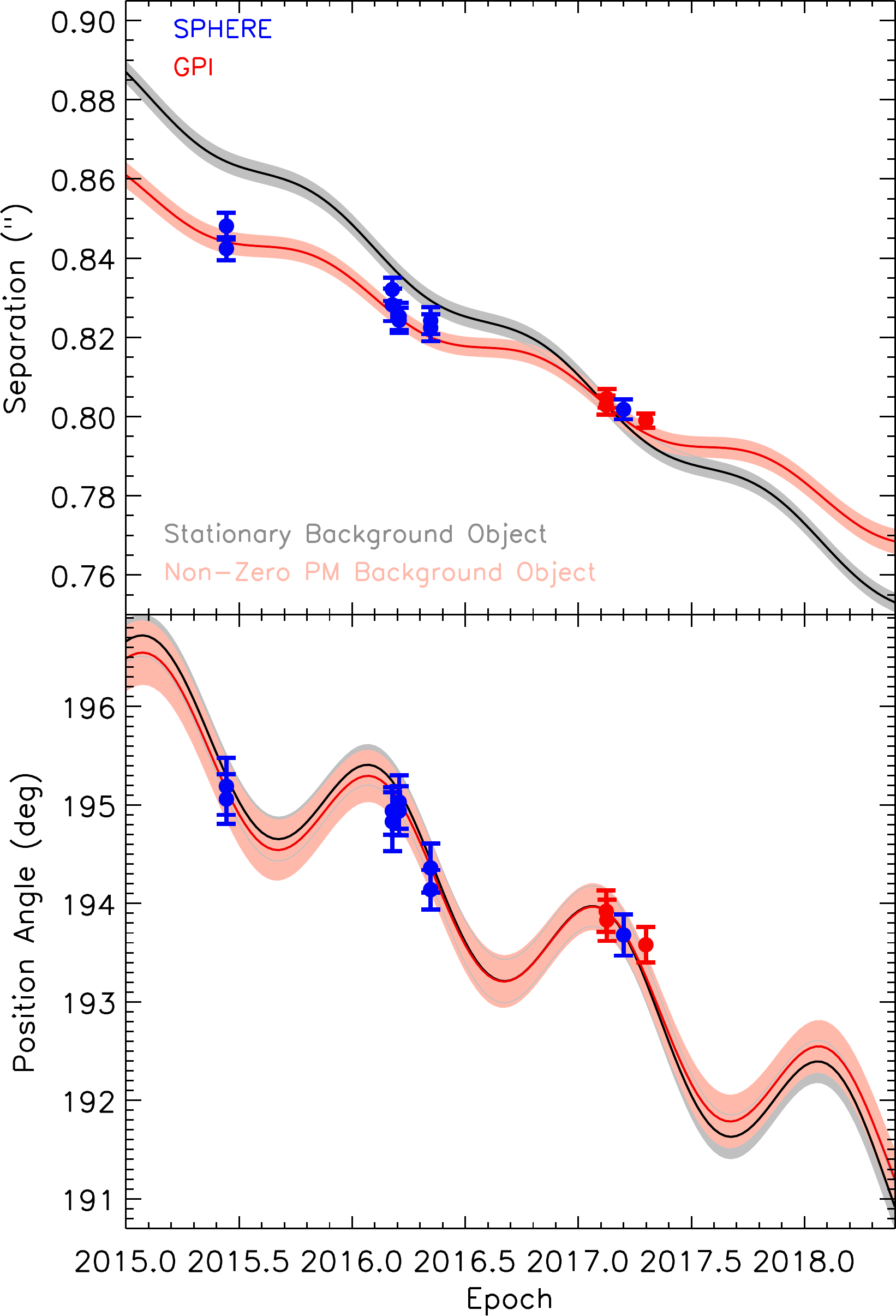}
\caption{Modified background track using our fit to the proper motion and parallax of HD~131399~Ab.  The original background track (for an infinitely distant background object with zero proper motion) is shown in gray, the modified track in red. The track gives a prediction for relative motion of HD~131399~Ab over the rest of the 2017 observing season, accelerating in PA while slowing down in separation compared to the predictions of a linear fit to the data.}
\label{fig:bg1_new}
\end{figure}

In Figure~\ref{fig:bg2} we present the posteriors on parallax and proper motion of HD~131399~Ab, assuming it is not bound to HD~131399~A.  While proper motion in the RA direction and parallax are both close to 0 ($\mu_{\alpha} = -4.7 \pm 1.6$\,mas\,yr$^{-1}$, $\pi <$ [0.64, 2.01] mas at [68\%, 95\%] confidence), Dec proper motion is significantly larger ($\mu_{\delta} = -11.2 \pm 1.3$\,mas\,yr$^{-1}$).  This motion in Declination is the departure from the background track seen in Figure~\ref{fig:bg1}, and the orbital motion examined by W16 and discussed in Section~\ref{orbit_sec}.  We display the fit proper motions along with the data in Figure~\ref{fig:bg1_new}, where the new proper motion track is the difference between the {\it Hipparcos} values of proper motion and parallax of HD~131399~A and our fit values for the proper motion and parallax of HD~131399~Ab.  Errors from the {\it Hipparcos} measurement and uncertainties in our fit are added in quadrature. While $\sim$12\,mas\,yr$^{-1}$ is a relatively large proper motion, about one-quarter the total proper motion of HD~131399~A, is this motion plausible for a star at $\sim$1\,kpc?

\begin{figure*}
\includegraphics[width=\textwidth,trim={0 0cm 0 0},clip]{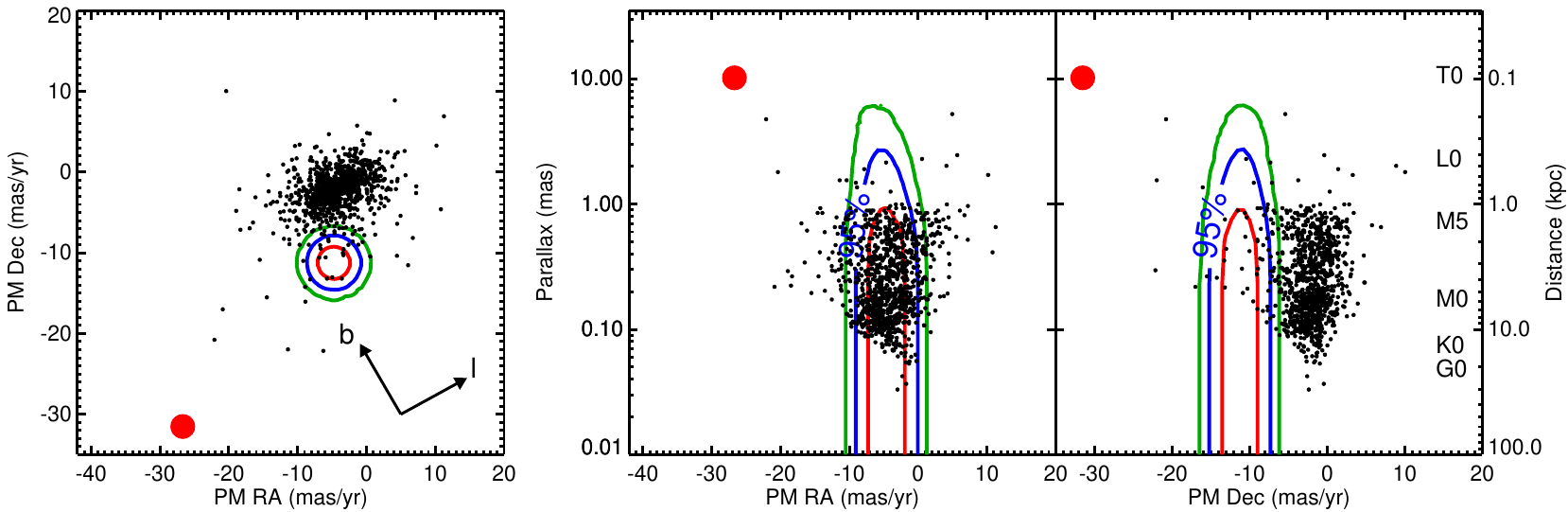}
\caption{Constraints on proper motion and parallax of HD~131399~Ab along with 1000 simulated background stars generated by the Besan\c{c}on Galaxy Model.  The red point marks the proper motion and parallax of the star HD~131399~A.  While the bulk of the Besan\c{c}on points lie outside our constraints, there are  background objects that fall within our contours (0.16\%, 0.89\%, and 1.9\% of points lie within the 1, 2, and 3$\sigma$ contours, respectively).  Thus a $\sim$1--10 kpc background object is a plausible explanation for the relative velocity observed between HD~131399~A and HD~131399~Ab.} \label{fig:bg3}
\end{figure*}

To answer this question we use the Besan\c{c}on model of stellar populations \citep{robin2003}, retrieving a set of simulated stars from the web form at \href{http://model.obs-besancon.fr}{http://model.obs-besancon.fr}.  We selected stars in the direction of HD~131399~A, with magnitudes of $19.63<H<19.87$ to match the 2$\sigma$ range of apparent magnitudes of HD~131399~Ab, with distances from 0 to 50\,kpc, and with a large solid angle of one square degree to give us a large statistical sample (6197 stars were generated).  We plot a subset of 1000 stars, along with the constraints on the proper motion and parallax of HD~131399~Ab from the relative astrometry, in Figure~\ref{fig:bg3}.  Since our parallax posterior is largely set by our choice of prior, and our data cannot distinguish between parallaxes $\lesssim$5 mas, we extend the contours from 0.1 mas to 0.  While the proper motion and parallax constraints do not encompass the majority of the simulated background stars, there is a significant subset that fall within the contours: 0.16\% fall within the 1$\sigma$ contours, 0.89\% inside 2$\sigma$, and 1.9\% within 3$\sigma$.  We note that these values do not represent the probability that HD~131399~Ab is a background object, but are instead  proportional to our constraints on proper motion and parallax.  Even if our constraints were in the middle of the cloud of background objects in Figure~\ref{fig:bg3}, as better astrometry allowed us to reduce the size of the contours fewer and fewer simulated background stars would fall inside.  Rather, this is a demonstration that the proper motion required to explain the change in relative astrometry seen in Figure~\ref{fig:bg1} is plausible for background stars in the direction of HD~131399~A with the same apparent magnitude as HD~131399~Ab.

The Besan\c{c}on points are consistent not just with the amplitude of the proper motion, but also the direction we measure for HD~131399~Ab.  These points, in the left panel of Figure~\ref{fig:bg3}, are not distributed isotropically, but instead preferentially represent stars moving south and west.  If we were simply to reverse the direction of HD~131399~Ab, multiplying the measurements of proper motion in RA and Dec by $-1$, the fraction of Besan\c{c}on stars falling into the [1, 2, 3] $\sigma$ contours drops to [0\%, 0\%, 0.048\%].  While there is no preferred direction for an orbiting planet, there is clearly a preferred direction for the proper motion of background stars at these coordinates, and HD~131399~Ab is moving in that direction.

We note the constraints at smaller distances ($\lesssim$ 1\,kpc) are largely driven by our choice of the prior on parallax, but do not strongly affect the overall conclusions.  Constraints on proper motion in RA and Dec are largely unchanged by the choice of prior.  Changing from a $p(d) \propto d^2 e^{-d/L}$ prior to a simple $p(d) \propto d^2$, cut off at 10\,kpc results in the fraction of Besan\c{c}on stars falling into [1, 2, 3] $\sigma$ contours dropping from [0.16\%, 0.89\%, 1.9\%] to [0.03\%, 0.29\%, 1.3\%], entirely due to lower values of parallax becoming less probable.  Switching to a uniform prior in parallax results in again smaller percentages falling into the contours compared to the exponentially declining prior, [0\%, 0.4\%, 1.5\%], as now more large-parallax points are accepted while more small-parallax points are rejected.  Of these three priors, both the uniform and $d^2$ prior are poor fits to the Besan\c{c}on points, while the exponentially declining prior is an excellent fit between 1 and 10\,kpc.  Since our measurement on parallax is essentially an upper limit, we adopt the values from the more realistic exponentially declining prior in our analysis, but with the small parallax constraints extended from 0.1 mas to 0 mas.

\subsection{Possible Scenarios}\label{sec:discuss}

We are presented with two unlikely scenarios for the nature of HD~131399~Ab: a planet with extreme orbital parameters (most likely currently being ejected from the system), or a background object with an unusually high proper motion.  An order of magnitude estimate for the likelihood of observing a planet being ejected just as it is detected is the orbital period divided by the system lifetime, which for a circular orbit with semi-major axis equal to the projected separation of 82\,au, is 520\,yr~/~16\,Myr, or $9 \times 10^{-6}$.  So while a background object whose proper motion is only consistent with $\sim$1\% of Besan\c{c}on simulated objected (at 95\% confidence) is unlikely, the ejected planet hypothesis is several order of magnitudes more unlikely.

\subsubsection{Bound Planet or Background Star?}

To apply a more rigorous analysis, we construct an odds ratio between the likelihood of the planet and background object scenario.  In particular, we consider three elements for each hypothesis: the overall likelihood of each object, the relative likelihood as a function of separation from the star, and relative probability as a function of projected velocity.  That is:

\begin{equation}
\frac{P_{BG} }{P_{Pl}} = \frac{P(BG) P(\rho | BG) P(v | \rho_{BG})}{P(Pl) P(\rho | Pl) P(v | \rho_{Pl})}
\label{bg_eq}
\end{equation}

\noindent where $P$ is probability, BG and Pl are Background and Planet, $\rho$ is projected separation, and $v$ is projected velocity.

For the first term, we restrict the separation range to $0\farcs1$ to $1\arcsec$ ($\approx$ 10--100\,au), and apparent magnitude $19.63<H<19.87$.  For background objects, this is the total number of Besan\c{c}on objects generated in Section~\ref{bg_sec2}, 6197 stars, multiplied by the ratio of solid angles: $6197 \times (\pi (1\arcsec)^2 - \pi (0\farcs1)^2) / ((3600\arcsec)^2) = 0.15\%$.  For planets, an apparent magnitude constraint of $19.63<H<19.87$, at 16\,Myr, corresponds to a mass range of 3.64--4.01\,$M_{\rm Jup}$, using the COND models \citep{Baraffe:2003bj}.  Early analysis of the Gemini Planet Imager Exoplanet Survey survey puts planet yield at $\sim$6\% of 2\,$M_\odot$ hosting a planet between 10--100 au, with mass between 5--13\,$M_{\rm Jup}$ (Nielsen et al. 2018, {\it in prep.}).  We convert this probability to our mass range, 3.64--4.01\,$M_{\rm Jup}$ assuming planets follow the power-law distribution in mass of \citet{cumming2008}, $dN/dM \propto M^{-1.31}$, or 0.76\% of stars having such a planet.

Next, we consider the second term, the fraction of background stars and planets at the observed projected separation, taken here to be the 2$\sigma$ range measured at the discovery epoch with the SPHERE IFS, 836.6--848.2\,mas.  Over small areas on the sky ($\lesssim10\arcsec$) background objects are uniformly distributed, so the probability of lying at this separation is just the ratio of areas between an annulus from 836.6--848.2\,mas, and one from 100--1000\,mas, 1.97\%.  For planets, the semi-major axis distribution (converted from the \citealt{cumming2008} distribution for period for solar-mass stars) is $dN/da \propto a^{-0.61}$.  So the second term for planets is then 0.85\%.  Adopting a uniform log semi-major axis distribution  ($dN/da \propto a^{-1}$) instead would have a minor effect, changing this term from 0.85\% to 0.60\%.

Finally, we consider the projected velocity measured between 2015 and 2017.  As described in Section~\ref{bg_sec2}, 0.89\% of Besan\c{c}on objects have a proper motion consistent with our 2$\sigma$ contours, and so we use this as the value for $P(v | \rho_{BG})$.  The corresponding term for planets, $P(v | \rho_{Pl})$, is then the fraction of orbits whose projected velocity is within the 2$\sigma$ range of the measured projected velocity from the relative astrometry.  Using OFTI \citep{blunt:2017}, we generate $10^7$ orbits from the first SPHERE IFS astrometric measurement, incorporating errors in separation, position angle, stellar mass, and distance.  The maximum projected velocity expected from an orbiting body is then 19.7 mas/yr.  In Section~\ref{escape_sec} we generated 10$^7$ measurements of the projected velocity from the full astrometric record of HD~131399~Ab, finding 2$\sigma$ confidence intervals between 15.3--21.5\,mas\,yr$^{-1}$ in RA and 18.5--22.6\,mas\,yr$^{-1}$ in Dec.  As expected, since we found this projected velocity to be above escape velocity, none of the $10^7$ generated orbits have a projected velocity in this 2$\sigma$ range, so we can only set an upper limit on $P(v | \rho_{Pl})$ of $10^{-7}$, or $10^{-5}$\%.

Entering these values into Equation~\ref{bg_eq} we find:

\begin{equation}
\frac{P_{BG} }{P_{Pl}} > \frac{0.15\% \times 1.97\% \times  0.94\%}{0.76\% \times 0.85\% \times 10^{-5}\%} = 43,000
\label{bg_eq2}
\end{equation}

\noindent so the probability that HD~131399~Ab is a background object is 43,000 times greater than the probability that it is a bound planet.  While this analysis depends on multiple assumptions, including the poorly known distribution of giant planets at wide separations, the dominant term in Equation~\ref{bg_eq2} is the velocity term.  If we were to consider the stability of these orbits the likelihood of being a bound planet would drop further, since the best-fitting orbits are likely to cross the orbit of HD~131399~BC.  But stability calculations (e.g., \citealt{Holman:1999}) show that the semimajor axis of the planet must actually be at least a factor of two smaller than that of HD~131399~BC, so even non-crossing orbits are not sufficient.  As we would expect from our analysis of the escape velocity, the large projected velocity of HD~131399~Ab strongly precludes the possibility that it is a bound planet.

\subsubsection{Ejected Planet?}

An unlikely scenario is that HD~131399~Ab is indeed a planetary mass object that formed around HD~131399~A, but is currently in the process of being ejected from the system, thus explaining why it is traveling faster than escape velocity.  We note this is very unlikely \textit{a priori}, since the chance of observing a 16\,Myr star just as it is ejecting its planet is very low \citep{Veras:2009}.  As noted earlier, our estimate for this probability, the ratio between the orbital period at the projected separation and the system lifetime, is $9 \times 10^{-6}$.

Planets ejected during planet-planet scattering are expected to attain a positive energy by a series of small kicks, and hence when they attain a positive energy they are traveling only slightly faster than the local escape velocity \citep{Malhotra:2002}.  If the observations are interpreted as an escaping planet with an observed speed twice the local escape velocity, the scattering must have happened much closer to the star, and hence the current motion should be radial from star A, which contradicts the observations, since the separation of Ab and A is decreasing over time.

\subsubsection{Alternative Explanations}

For the astrometric dataset to be consistent with a bound orbit one of three measurements has to change significantly: our astrometry and errors, the total mass of HD~131399~Ab, or the distance to the system.

Our SED fitting of HD~131399~A makes it difficult to significantly change either the system mass or the distance.  For the projected velocity to drop below escape velocity, either the distance to the system needs to decrease or the mass needs to increase.  But adding more mass would increase the flux in the SED fit, which would result in a larger distance for the same spectral type.  Similarly, decreasing the distance would require a later spectral type to explain the same flux, which would decrease the mass.

In order to avoid disrupting the SED fit, the extra mass in the system would have to come in the form of low-mass companions with significant mass but negligible flux.  One such configuration would be multiple low-mass companions to HD~131399~A.  Given typical magnitudes as a function of spectral type for dwarfs \citep{Pecaut:2013}, to double the mass while not raising the $V$-band flux by more than 10\% would require two G1V companions to the A1V HD~131399~A.  Such companions would be visible in the GPI $H$ images at projected separations greater than 50 mas (5 au), requiring an unlikely configuration of a very compact triple system in addition to the wide companions B and C.  Such configurations could be detectable with future RV monitoring of the system.

An even more unlikely explanation is  degenerate companions to HD~131399~A.  Given the system age of 16\,Myr, a white dwarf companion can be excluded since the time required for a $\lesssim$10\,$M_{\odot}$ star to evolve into a white dwarf has not elapsed \citep{Maeder:2009,Bertulani:2013}  Even if a high-mass star formed a white dwarf within 16\,Myr, we would be observing the white dwarf at its peak luminosity, and it would be evident in the observed SED.  The system age is plausible for a higher mass companion to go supernova and form a black hole or neutron star.  Such high-mass stars ($\gtrsim$10\,$M_\odot$) are intrinsically rare, where the progenitor would be one of the highest mass stars to form in the UCL association, and the rapid change in mass from a supernova in a binary system would likely disrupt the HD~131399~ABC system.

Systematic errors in the astrometry could be responsible for the projected velocity of HD~131399~Ab exceeding escape velocity.  We find this possibility unlikely as well, since previous analysis of GPI astrometry has shown excellent precision over time ($<1$\,mas for $\beta$~Pic~b, \citealt{Wang:2016gl}) and with respect to VLT/SPHERE and Keck/NIRC2.  The fact that the 2017 SPHERE epoch lies between the 2017 February and 2017 April GPI epochs illustrates that both instruments have well-calibrated astrometry.  A level of astrometric systematics larger than seen in any previous study would be required to explain the movement observed for this system.

A final possible explanation would be a significant error in the \textit{Hipparcos} proper motion.  Figure~\ref{fig:bg3} shows the proper motion of HD~131399~Ab in the RA direction is well-matched to the mean Besan\c{c}on points in the RA direction, but at the edge of the distribution in the Dec direction.  A smaller Dec proper motion for the star HD~131399~A (by a few mas/yr) would account for this offset.  The Tycho-2 catalog \citep{Hog:2000wk} gives a proper motion of $[-29.7 \pm 1.2, -26.2 \pm 1.3]$\,mas\,yr$^{-1}$, compared to the \textit{Hipparcos} \citep{vanLeeuwen:2007dc} values of $[-29.69 \pm 0.59, -31.52 \pm 0.55]$; the two measurements are consistent in the RA direction, but are $\sim$3$\sigma$ discrepant in the Dec direction.  The smaller Tycho-2 proper motion would translate to HD~131399~Ab needing a smaller total proper motion to explain the astrometry.  HD~131399~BC, $\sim3\arcsec$ south of the star at $\sim$220$^\circ$, could plausibly have biased the proper motion in the Dec direction, though this effect is expected to be small, given the large magnitude difference ($\Delta H_p = 4.52$) between A and BC.  Both sets of measurements are in general agreement with ground-based astrometry, with a proper motion given by \citep{bastian:1993} of $[-34.7 \pm 3.4, -29.0 \pm 3.1]$, and by \citet{platais:1998} of $[-30.6 \pm 2.9, -29.6 \pm 3.4]$\,mas\,yr$^{-1}$. HD~131399~A is not in the \textit{GAIA} first data release \citep{lindegren:2016}, likely because it is a close binary, but the final \textit{GAIA} astrometry should provide a useful check on the proper motion of this system.

\section{Conclusion}\label{sec:conc}

We present new astrometric and spectroscopic measurements of HD~131399~Ab obtained with Gemini/GPI in $JHK1$ bands, with VLT/SPHERE in {\it J} band, and with Keck/NIRC2 in {\it L}$^\prime$ combined with a re-analysis of the VLT/SPHERE datasets in $JHK1K2$ bands published by W16. Our results, derived from two independent PSF subtraction and analysis pipelines used to mitigate biases from the data reduction and analysis process, lead us to revise the status of the object. The data are inconsistent with the hypothesis that HD~131399~Ab is a bound T-type planet (W16) and instead is more consistent with it being a background star.

Using spectro-photometry, we revise the SED of HD~131399~Ab with these higher quality data and broader wavelength coverage. Empirical comparisons show that its near-IR colors and spectra preclude HD~131399~Ab from having a spectral type later than early L and and we exclude the presence of strong methane absorption. The $H$ band spectra obtained from the SPHERE data, though noisy, presents a peaky continuum characteristic of a young LT object in only one of the four datasets. The other three datasets are featureless with no sign of methane nor water absorption. We show that the peak may be caused by a speckle close to the source, which biases the extracted flux between $1.62$ and $1.64$\,\micron. Moreover, our revised SPHERE $K1K2$ magnitudes are inconsistent with the published ones (W16), while the sets of photometry are consistent for the binary B, and exclude the presence of methane. Altogether, the new GPI spectra and magnitudes of HD~131399~Ab and the revised SPHERE ones are mostly consistent with a K or a M dwarf, the most probable candidates for a background object at an apparent {\it H} band magnitude around 20.

In addition, our revised astrometry for the SPHERE datasets has a systematic offset of one degree on the position angles with respect to the values presented in W16, which was an offset not present in the position angles of HD~131399~B. The astrometry also reveals a significant change to the separation at each epoch. This reanalysis is further evidence against the planet hypothesis, with only a 0.092\% likelihood that HD~131399~Ab is moving slower than escape velocity using only the pre-2017 data.  When we include the new GPI and SPHERE astrometry from 2017 to this analysis, which we have previously shown to be self-consistent and consistent between the two instruments, this confidence increases to better than one part in $10^7$.  In order to construct bound orbits, it is necessary to reduce the system distance from 98\,pc to 73\,pc, or increase the mass of HD~131399~A by more than a factor of two, both implausible scenarios.  Even when these changes are made, though, the resulting orbit is highly eccentric, with the likely values for apastron greatly exceeding the projected separation of HD~131399~BC. Therefore all these changes represent unstable orbits.  We conclude, then, that orbital motion is an inadequate explanation for the change in relative astrometry observed in this system.  This motion is, however, consistent with a background star in this part of the sky, in both magnitude and direction, even if the proper motion required is in the top 4\% for these background objects.

Finally, we note that HD~131399~Ab is a particularly pathological background object: W16 followed all the standard steps in their analysis, but a combination of multiple unlikely factors led to the conclusion that HD~131399~Ab was a planet.  A speckle in the third SPHERE dataset, combined with the SPHERE {\it YJH} spectral coverage ending at $\sim1.64\,\micron$, mimicked a T-like spectrum in the {\it H} band.  The apparent magnitude of the background star was consistent with a planet of the inferred spectral type and age of the system.  This background object has a proper motion (in mas\,yr$^{-1}$) larger than 96\% of background objects at this magnitude range in this region of the sky, and the target star's proper motion is relatively small, only four times larger than that of the background object.  This resulted in a significant deviation from the stationary background object track, and was within a factor of two of a velocity consistent with orbital motion.  Further, the background object was one third of the way between HD~131399~Ab and HD~131399~BC, at the outer limit of predictions for stable orbits of planets in multiple systems.  In short, this background object was almost tailor-made to pass the standard tests performed in the analysis of direct imaging data.

As direct imaging technology and techniques mature, it becomes more important to consider the assumptions of the standard stationary background object test when testing for common proper motion.  As contrasts improve at smaller angular separations, planet candidates are being detected around stars at larger distances from the Sun, and therefore with smaller proper motions.  At the same time, astrometric accuracy is improving, even exceeding 1 mas for the brightest objects \citep{Wang:2016gl}.  With these trends, it is increasingly likely that the few mas/yr expected motion of background objects will become significant when differentiating between common proper motion companions and background stars, as is the case for HD~131399~Ab.  In addition, planet candidates detected closer to the target star will have a larger range of predicted orbital motions.  As a result special care should be taken when evaluating the background object scenario, in addition to spectral typing, to ensure that a sufficiently long astrometric record has been collected to definitively identify candidates as having common proper motion or being unassociated background objects.

\acknowledgments
 Based on observations obtained at the Gemini Observatory, which is operated by the Association of Universities for Research in Astronomy, Inc., under a cooperative agreement with the NSF on behalf of the Gemini partnership: the National Science Foundation (United States), the National Research Council (Canada), CONICYT (Chile), Ministerio de Ciencia, Tecnolog\'{i}a e Innovaci\'{o}n Productiva (Argentina), and Minist\'{e}rio da Ci\^{e}ncia, Tecnologia e Inova\c{c}\~{a}o (Brazil). Some of the data presented herein were obtained at the W.M. Keck Observatory, which is operated as a scientific partnership among the California Institute of Technology, the University of California and the National Aeronautics and Space Administration. The Observatory was made possible by the generous financial support of the W.M. Keck Foundation. The authors wish to recognize and acknowledge the very significant cultural role and reverence that the summit of Mauna Kea has always had within the indigenous Hawaiian community.  We are most fortunate to have the opportunity to conduct observations from this mountain. Based on observations collected at the European Organisation for Astronomical Research in the Southern Hemisphere under ESO programme 098.C-0864(A). Based on data obtained from the ESO Science Archive Facility. SPHERE is an instrument designed and built by a consortium consisting of IPAG (Grenoble, France), MPIA (Heidelberg, Germany), LAM (Marseille, France), LESIA (Paris, France), Laboratoire Lagrange (Nice, France), INAF–Osservatorio di Padova (Italy), Observatoire de Genève (Switzerland), ETH Zurich (Switzerland), NOVA (Netherlands), ONERA (France) and ASTRON (Netherlands) in collaboration with ESO. SPHERE was funded by ESO, with additional contributions from CNRS (France), MPIA (Germany), INAF (Italy), FINES (Switzerland) and NOVA (Netherlands). SPHERE also received funding from the European Commission Sixth and Seventh Framework Programmes as part of the Optical Infrared Coordination Network for Astronomy (OPTICON) under grant number RII3-Ct-2004-001566 for FP6 (2004–2008), grant number 226604 for FP7 (2009–2012) and grant number 312430 for FP7 (2013–2016).

This research has made use of the SIMBAD and VizieR databases, operated at CDS, Strasbourg, France, the Database of Ultracool Parallaxes maintained by Trent Dupuy, and the Washington Double Star Catalog maintained by the U.S. Naval Observatory at \href{http://www.usno.navy.mil/USNO/astrometry/optical-IR- prod/wds/WDS}{http://www.usno.navy.mil/USNO/astrometry/optical-IR- prod/wds/WDS}. This research has benefited from the SpeX Prism Library (and/or SpeX Prism Library Analysis Toolkit), maintained by Adam Burgasser at http://www.browndwarfs.org/spexprism, the IRTF Spectral Library, maintained by Michael Cushing, and the Montreal Brown Dwarf and Exoplanet Spectral Library, maintained by Jonathan Gagn{\'e}.

J.R., R.D. and D.L. acknowledge support from the Fonds de Recherche du Qu\'{e}bec. JRM's work was performed in part under contract with the California Institute of Technology (Caltech)/Jet Propulsion Laboratory (JPL) funded by NASA through the Sagan Fellowship Program executed by the NASA Exoplanet Science Institute. Support for MMB's work was provided by NASA through Hubble Fellowship grant \#51378.01-A awarded by the Space Telescope Science Institute, which is operated by the Association of Universities for Research in Astronomy, Inc., for NASA, under contract NAS5-26555. Supported by NSF grants AST-1411868 (K.B.F., B.M., and J.P.), AST-141378 (G.D.), and AST-1518332 (R.D.R., J.J.W., T.M.E., J.R.G., P.G.K.) . Supported by NASA grants NNX14AJ80G (E.L.N., S.C.B., B.M., F.M., and M.P.), NNX15AC89G and NNX15AD95G (B.M., J.E.W., T.M.E., R.J.D.R., G.D., J.R.G., P.G.K.) and NNX16AD44G (K.M.M). K.W.D. is supported by an NRAO Student Observing Support Award SOSPA3-007. Portions of this work were performed under the auspices of the U.S. Department of Energy by Lawrence Livermore National Laboratory under Contract DE-AC52-07NA27344. This work benefited from NASA’s Nexus for Exoplanet System Science (NExSS) research coordination network sponsored by NASA’s Science Mission Directorate.

\facility{Gemini:South (GPI), VLT:Melipal (SPHERE), Keck:II (NIRC2)}
\software{Astropy \citep{astropy}, emcee \citep{ForemanMackey:2013io}, GPI DRP \citep{Perrin:2016gm}, IDL Astronomy Library \citep{idlastro}, pyKLIP \citep{Wang:2015th}, SPHERE DRH \citep{Pavlov:2008}, SPHERE IFS \citep{Vigan:2015}}

\appendix

\section{Flux-calibrated spectrum of HD 131399 A\lowercase{b}}

\begin{deluxetable}{cccc}
\tabletypesize{\scriptsize}
\tablecaption{\label{tab:sphere_sed}SPHERE {\it YJH} spectral energy distribution of HD 131399 A and Ab, and corresponding 68\,\% confidence intervals}
\tablewidth{0pt}
\tablehead{
\colhead{Wavelength} & \colhead{$F_{\lambda}({\rm A}) \times 10^{-12}$} & \colhead{Contrast} & \colhead{$F_{\lambda}({\rm Ab}) \times 10^{-17}$}  \\
(\micron)& (W\,m$^{-2}$\,\micron$^{-1}$) & ($\times 10^{-6}$) & (W\,m$^{-2}$\,\micron$^{-1}$)
}
\startdata
$0.966$ & $11.90 \pm 0.15$ & $3.82 \pm 2.92$ & $4.53 \pm 3.46$ \\
$0.978$ & $11.67 \pm 0.15$ & $0.70 \pm 1.21$ & $0.81 \pm 1.41$ \\
$0.992$ & $11.01 \pm 0.14$ & $1.59 \pm 1.66$ & $1.75 \pm 1.83$ \\
$1.008$ & $10.22 \pm 0.14$ & $2.84 \pm 1.78$ & $2.90 \pm 1.82$ \\
$1.026$ & $10.01 \pm 0.13$ & $3.07 \pm 1.81$ & $3.07 \pm 1.82$ \\
$1.043$ & $9.66 \pm 0.13$ & $2.77 \pm 1.65$ & $2.68 \pm 1.59$ \\
$1.062$ & $9.17 \pm 0.13$ & $3.33 \pm 1.65$ & $3.04 \pm 1.51$ \\
$1.079$ & $8.39 \pm 0.12$ & $4.06 \pm 1.65$ & $3.41 \pm 1.39$ \\
$1.096$ & $7.67 \pm 0.11$ & $4.26 \pm 1.77$ & $3.26 \pm 1.36$ \\
$1.114$ & $7.53 \pm 0.10$ & $2.71 \pm 2.02$ & $2.03 \pm 1.52$ \\
$1.136$ & $7.19 \pm 0.10$ & $2.89 \pm 2.18$ & $2.08 \pm 1.57$ \\
$1.156$ & $6.77 \pm 0.10$ & $4.21 \pm 2.27$ & $2.85 \pm 1.54$ \\
$1.174$ & $6.43 \pm 0.09$ & $5.89 \pm 2.02$ & $3.79 \pm 1.30$ \\
$1.192$ & $6.14 \pm 0.09$ & $7.75 \pm 2.01$ & $4.76 \pm 1.23$ \\
$1.211$ & $5.81 \pm 0.08$ & $4.74 \pm 1.66$ & $2.75 \pm 0.96$ \\
$1.229$ & $5.51 \pm 0.08$ & $5.07 \pm 1.50$ & $2.80 \pm 0.83$ \\
$1.248$ & $5.23 \pm 0.08$ & $5.28 \pm 2.01$ & $2.75 \pm 1.06$ \\
$1.267$ & $4.85 \pm 0.07$ & $7.22 \pm 2.02$ & $3.50 \pm 0.98$ \\
$1.287$ & $4.49 \pm 0.07$ & $5.19 \pm 1.93$ & $2.33 \pm 0.87$ \\
$1.304$ & $4.44 \pm 0.07$ & $5.66 \pm 2.10$ & $2.51 \pm 0.93$ \\
$1.318$ & $4.33 \pm 0.06$ & $4.54 \pm 2.52$ & $1.97 \pm 1.09$ \\
$1.331$ & $4.20 \pm 0.06$ & $2.63 \pm 2.74$ & $1.10 \pm 1.15$ \\
$1.343$ & $4.07 \pm 0.06$ & $3.13 \pm 5.13$ & $1.27 \pm 2.09$ \\
$1.378$ & $3.72 \pm 0.06$ & $5.45 \pm 3.63$ & $2.03 \pm 1.35$ \\
$1.413$ & $3.41 \pm 0.05$ & $3.52 \pm 4.25$ & $1.20 \pm 1.45$ \\
$1.429$ & $3.27 \pm 0.05$ & $1.67 \pm 3.13$ & $0.55 \pm 1.03$ \\
$1.445$ & $3.14 \pm 0.05$ & $2.51 \pm 4.17$ & $0.80 \pm 1.31$ \\
$1.461$ & $3.01 \pm 0.05$ & $2.21 \pm 2.49$ & $0.67 \pm 0.75$ \\
$1.478$ & $2.87 \pm 0.04$ & $2.01 \pm 2.27$ & $0.57 \pm 0.65$ \\
$1.496$ & $2.73 \pm 0.04$ & $3.18 \pm 3.01$ & $0.87 \pm 0.82$ \\
$1.512$ & $2.61 \pm 0.04$ & $4.81 \pm 2.69$ & $1.25 \pm 0.70$ \\
$1.529$ & $2.51 \pm 0.04$ & $5.96 \pm 2.19$ & $1.49 \pm 0.55$ \\
$1.546$ & $2.41 \pm 0.04$ & $5.24 \pm 2.23$ & $1.26 \pm 0.54$ \\
$1.563$ & $2.32 \pm 0.04$ & $5.07 \pm 2.46$ & $1.18 \pm 0.57$ \\
$1.580$ & $2.24 \pm 0.04$ & $5.39 \pm 2.13$ & $1.21 \pm 0.48$ \\
$1.598$ & $2.17 \pm 0.03$ & $7.92 \pm 2.16$ & $1.72 \pm 0.47$ \\
$1.614$ & $2.10 \pm 0.03$ & $7.77 \pm 2.49$ & $1.63 \pm 0.52$ \\
$1.629$ & $2.04 \pm 0.03$ & $7.16 \pm 2.74$ & $1.46 \pm 0.56$ \\
$1.641$ & $1.99 \pm 0.03$ & $6.09 \pm 3.73$ & $1.22 \pm 0.74$ \\
\enddata
\end{deluxetable}

\clearpage

\startlongtable
\begin{deluxetable}{cccc}
\tabletypesize{\scriptsize}
\tablecaption{\label{tab:gpi_sed}GPI {\it JHK1} spectral energy distribution of HD 131399 A and Ab, and corresponding 68\,\% confidence intervals}
\tablewidth{0pt}
\tablehead{
\colhead{Wavelength} & \colhead{$F_{\lambda}({\rm A}) \times 10^{-12}$} & \colhead{Contrast\tablenotemark{\it a}} & \colhead{$F_{\lambda}({\rm Ab})\tablenotemark{\it a} \times 10^{-17}$}  \\
(\micron)& (W\,m$^{-2}$\,\micron$^{-1}$) & ($\times 10^{-6}$) & (W\,m$^{-2}$\,\micron$^{-1}$)
}
\startdata
$1.114$ & $7.59 \pm 0.10$ & $1.55 \pm 1.48$ & $1.17 \pm 1.13$ \\
$1.121$ & $7.49 \pm 0.10$ & $1.88 \pm 1.58$ & $1.41 \pm 1.19$ \\
$1.127$ & $7.35 \pm 0.10$ & $3.19 \pm 1.87$ & $2.35 \pm 1.38$ \\
$1.134$ & $7.23 \pm 0.10$ & $4.36 \pm 1.78$ & $3.16 \pm 1.29$ \\
$1.140$ & $7.11 \pm 0.10$ & $4.96 \pm 1.62$ & $3.53 \pm 1.15$ \\
$1.147$ & $6.99 \pm 0.10$ & $5.34 \pm 1.50$ & $3.74 \pm 1.05$ \\
$1.153$ & $6.84 \pm 0.10$ & $4.99 \pm 1.31$ & $3.41 \pm 0.90$ \\
$1.160$ & $6.69 \pm 0.10$ & $4.69 \pm 1.21$ & $3.14 \pm 0.81$ \\
$1.166$ & $6.55 \pm 0.09$ & $5.03 \pm 0.94$ & $3.30 \pm 0.62$ \\
$1.173$ & $6.43 \pm 0.09$ & $4.83 \pm 0.96$ & $3.10 \pm 0.62$ \\
$1.180$ & $6.32 \pm 0.09$ & $4.79 \pm 0.98$ & $3.03 \pm 0.62$ \\
$1.186$ & $6.22 \pm 0.09$ & $4.86 \pm 1.05$ & $3.02 \pm 0.65$ \\
$1.193$ & $6.11 \pm 0.09$ & $5.03 \pm 1.10$ & $3.07 \pm 0.67$ \\
$1.199$ & $6.00 \pm 0.09$ & $4.90 \pm 1.07$ & $2.94 \pm 0.64$ \\
$1.206$ & $5.89 \pm 0.08$ & $4.58 \pm 1.10$ & $2.70 \pm 0.65$ \\
$1.212$ & $5.78 \pm 0.08$ & $4.17 \pm 1.20$ & $2.41 \pm 0.69$ \\
$1.219$ & $5.68 \pm 0.08$ & $4.21 \pm 1.17$ & $2.39 \pm 0.67$ \\
$1.225$ & $5.58 \pm 0.08$ & $4.26 \pm 1.12$ & $2.37 \pm 0.63$ \\
$1.232$ & $5.48 \pm 0.08$ & $4.52 \pm 1.14$ & $2.47 \pm 0.62$ \\
$1.238$ & $5.38 \pm 0.08$ & $4.41 \pm 1.07$ & $2.37 \pm 0.58$ \\
$1.245$ & $5.28 \pm 0.08$ & $4.13 \pm 0.99$ & $2.18 \pm 0.52$ \\
$1.252$ & $5.18 \pm 0.07$ & $4.19 \pm 1.07$ & $2.17 \pm 0.55$ \\
$1.258$ & $5.07 \pm 0.07$ & $4.77 \pm 1.10$ & $2.41 \pm 0.56$ \\
$1.265$ & $4.92 \pm 0.07$ & $4.98 \pm 1.27$ & $2.45 \pm 0.63$ \\
$1.271$ & $4.74 \pm 0.07$ & $4.93 \pm 1.22$ & $2.34 \pm 0.58$ \\
$1.278$ & $4.55 \pm 0.07$ & $5.02 \pm 1.26$ & $2.29 \pm 0.57$ \\
$1.284$ & $4.45 \pm 0.07$ & $5.26 \pm 1.31$ & $2.34 \pm 0.58$ \\
$1.291$ & $4.44 \pm 0.07$ & $5.75 \pm 1.36$ & $2.55 \pm 0.60$ \\
$1.297$ & $4.47 \pm 0.07$ & $6.29 \pm 1.41$ & $2.81 \pm 0.63$ \\
$1.304$ & $4.47 \pm 0.07$ & $6.51 \pm 1.39$ & $2.90 \pm 0.62$ \\
$1.310$ & $4.42 \pm 0.06$ & $5.95 \pm 1.40$ & $2.63 \pm 0.62$ \\
$1.317$ & $4.35 \pm 0.06$ & $5.33 \pm 1.47$ & $2.32 \pm 0.64$ \\
$1.324$ & $4.28 \pm 0.06$ & $5.26 \pm 1.44$ & $2.25 \pm 0.62$ \\
$1.330$ & $4.21 \pm 0.06$ & $5.76 \pm 1.50$ & $2.43 \pm 0.63$ \\
$1.337$ & $4.14 \pm 0.06$ & $5.67 \pm 1.56$ & $2.35 \pm 0.65$ \\
$1.343$ & $4.07 \pm 0.06$ & $5.15 \pm 1.49$ & $2.09 \pm 0.61$ \\
$1.350$ & $4.00 \pm 0.06$ & $3.67 \pm 1.93$ & $1.47 \pm 0.78$ \\
$1.495$ & $2.73 \pm 0.04$ & $5.38 \pm 1.46$ & $1.47 \pm 0.40$ \\
$1.503$ & $2.67 \pm 0.04$ & $5.66 \pm 1.09$ & $1.51 \pm 0.29$ \\
$1.511$ & $2.61 \pm 0.04$ & $6.18 \pm 0.93$ & $1.61 \pm 0.24$ \\
$1.520$ & $2.56 \pm 0.04$ & $6.62 \pm 0.85$ & $1.69 \pm 0.22$ \\
$1.528$ & $2.51 \pm 0.04$ & $6.65 \pm 0.82$ & $1.67 \pm 0.21$ \\
$1.537$ & $2.46 \pm 0.04$ & $6.24 \pm 0.83$ & $1.53 \pm 0.21$ \\
$1.545$ & $2.42 \pm 0.04$ & $6.37 \pm 0.84$ & $1.54 \pm 0.20$ \\
$1.553$ & $2.37 \pm 0.04$ & $6.10 \pm 0.77$ & $1.45 \pm 0.19$ \\
$1.562$ & $2.33 \pm 0.04$ & $6.11 \pm 0.74$ & $1.43 \pm 0.17$ \\
$1.570$ & $2.29 \pm 0.04$ & $6.12 \pm 0.76$ & $1.40 \pm 0.18$ \\
$1.579$ & $2.25 \pm 0.04$ & $5.79 \pm 0.73$ & $1.30 \pm 0.17$ \\
$1.587$ & $2.21 \pm 0.03$ & $5.70 \pm 0.72$ & $1.26 \pm 0.16$ \\
$1.596$ & $2.18 \pm 0.03$ & $6.03 \pm 0.73$ & $1.31 \pm 0.16$ \\
$1.604$ & $2.14 \pm 0.03$ & $5.88 \pm 0.71$ & $1.26 \pm 0.15$ \\
$1.612$ & $2.11 \pm 0.03$ & $6.18 \pm 0.69$ & $1.30 \pm 0.15$ \\
$1.621$ & $2.09 \pm 0.03$ & $7.09 \pm 0.66$ & $1.48 \pm 0.14$ \\
$1.629$ & $2.05 \pm 0.03$ & $7.39 \pm 0.69$ & $1.51 \pm 0.14$ \\
$1.638$ & $1.99 \pm 0.03$ & $7.09 \pm 0.62$ & $1.41 \pm 0.12$ \\
$1.646$ & $1.97 \pm 0.03$ & $7.21 \pm 0.64$ & $1.42 \pm 0.13$ \\
$1.654$ & $1.97 \pm 0.03$ & $7.91 \pm 0.73$ & $1.56 \pm 0.15$ \\
$1.663$ & $1.94 \pm 0.03$ & $8.81 \pm 0.80$ & $1.71 \pm 0.16$ \\
$1.671$ & $1.88 \pm 0.03$ & $9.31 \pm 0.84$ & $1.75 \pm 0.16$ \\
$1.680$ & $1.82 \pm 0.03$ & $8.67 \pm 0.90$ & $1.58 \pm 0.17$ \\
$1.688$ & $1.80 \pm 0.03$ & $7.40 \pm 0.84$ & $1.34 \pm 0.15$ \\
$1.696$ & $1.82 \pm 0.03$ & $7.15 \pm 0.87$ & $1.30 \pm 0.16$ \\
$1.705$ & $1.81 \pm 0.03$ & $7.36 \pm 0.86$ & $1.33 \pm 0.16$ \\
$1.713$ & $1.78 \pm 0.03$ & $8.27 \pm 0.93$ & $1.47 \pm 0.17$ \\
$1.722$ & $1.72 \pm 0.03$ & $8.79 \pm 1.00$ & $1.51 \pm 0.17$ \\
$1.730$ & $1.65 \pm 0.03$ & $7.91 \pm 1.02$ & $1.30 \pm 0.17$ \\
$1.739$ & $1.60 \pm 0.03$ & $7.11 \pm 1.03$ & $1.14 \pm 0.17$ \\
$1.747$ & $1.60 \pm 0.03$ & $6.88 \pm 1.02$ & $1.10 \pm 0.16$ \\
$1.755$ & $1.61 \pm 0.03$ & $7.87 \pm 0.95$ & $1.27 \pm 0.16$ \\
$1.764$ & $1.61 \pm 0.03$ & $8.43 \pm 0.92$ & $1.35 \pm 0.15$ \\
$1.772$ & $1.59 \pm 0.03$ & $7.88 \pm 1.02$ & $1.25 \pm 0.16$ \\
$1.781$ & $1.56 \pm 0.03$ & $6.63 \pm 1.20$ & $1.03 \pm 0.19$ \\
$1.789$ & $1.53 \pm 0.02$ & $5.90 \pm 1.51$ & $0.90 \pm 0.23$ \\
$1.797$ & $1.49 \pm 0.02$ & $7.04 \pm 2.13$ & $1.05 \pm 0.32$ \\
$1.886$ & $1.21 \pm 0.02$ & $-6.86 \pm 10.09$ & $-0.84 \pm 1.22$ \\
$1.895$ & $1.23 \pm 0.02$ & $3.61 \pm 1.45$ & $0.44 \pm 0.18$ \\
$1.903$ & $1.22 \pm 0.02$ & $-1.15 \pm 1.88$ & $-0.14 \pm 0.23$ \\
$1.912$ & $1.20 \pm 0.02$ & $-0.61 \pm 1.44$ & $-0.07 \pm 0.17$ \\
$1.920$ & $1.18 \pm 0.02$ & $-0.82 \pm 0.36$ & $-0.10 \pm 0.04$ \\
$1.929$ & $1.14 \pm 0.02$ & $2.48 \pm 1.01$ & $0.28 \pm 0.12$ \\
$1.938$ & $1.08 \pm 0.02$ & $12.20 \pm 4.99$ & $1.33 \pm 0.54$ \\
$1.946$ & $1.05 \pm 0.02$ & $7.43 \pm 3.43$ & $0.78 \pm 0.36$ \\
$1.955$ & $1.06 \pm 0.02$ & $4.25 \pm 1.86$ & $0.45 \pm 0.20$ \\
$1.963$ & $1.07 \pm 0.02$ & $5.13 \pm 2.15$ & $0.55 \pm 0.23$ \\
$1.972$ & $1.07 \pm 0.02$ & $10.09 \pm 3.21$ & $1.08 \pm 0.34$ \\
$1.981$ & $1.06 \pm 0.02$ & $8.67 \pm 2.99$ & $0.92 \pm 0.32$ \\
$1.989$ & $1.04 \pm 0.02$ & $8.94 \pm 2.81$ & $0.93 \pm 0.29$ \\
$1.998$ & $1.03 \pm 0.02$ & $8.86 \pm 3.22$ & $0.91 \pm 0.33$ \\
$2.006$ & $1.01 \pm 0.02$ & $10.61 \pm 3.42$ & $1.07 \pm 0.35$ \\
$2.015$ & $0.99 \pm 0.02$ & $10.47 \pm 3.36$ & $1.04 \pm 0.33$ \\
$2.023$ & $0.98 \pm 0.02$ & $6.44 \pm 2.56$ & $0.63 \pm 0.25$ \\
$2.032$ & $0.96 \pm 0.02$ & $9.15 \pm 2.84$ & $0.88 \pm 0.27$ \\
$2.041$ & $0.95 \pm 0.02$ & $10.85 \pm 2.81$ & $1.03 \pm 0.27$ \\
$2.049$ & $0.93 \pm 0.02$ & $8.69 \pm 2.60$ & $0.81 \pm 0.24$ \\
$2.058$ & $0.92 \pm 0.02$ & $7.83 \pm 2.69$ & $0.72 \pm 0.25$ \\
$2.066$ & $0.91 \pm 0.02$ & $12.11 \pm 3.11$ & $1.09 \pm 0.28$ \\
$2.075$ & $0.89 \pm 0.02$ & $10.24 \pm 2.74$ & $0.91 \pm 0.25$ \\
$2.084$ & $0.88 \pm 0.01$ & $11.42 \pm 2.85$ & $1.00 \pm 0.25$ \\
$2.092$ & $0.86 \pm 0.01$ & $12.42 \pm 2.63$ & $1.07 \pm 0.23$ \\
$2.101$ & $0.85 \pm 0.01$ & $10.68 \pm 3.14$ & $0.91 \pm 0.27$ \\
$2.109$ & $0.84 \pm 0.01$ & $9.01 \pm 3.12$ & $0.75 \pm 0.26$ \\
$2.118$ & $0.82 \pm 0.01$ & $12.97 \pm 3.67$ & $1.07 \pm 0.30$ \\
$2.126$ & $0.81 \pm 0.01$ & $13.44 \pm 3.61$ & $1.09 \pm 0.30$ \\
$2.135$ & $0.80 \pm 0.01$ & $8.35 \pm 3.30$ & $0.67 \pm 0.26$ \\
$2.144$ & $0.78 \pm 0.01$ & $8.63 \pm 3.39$ & $0.68 \pm 0.26$ \\
$2.152$ & $0.76 \pm 0.01$ & $8.99 \pm 3.52$ & $0.68 \pm 0.27$ \\
$2.161$ & $0.72 \pm 0.01$ & $5.49 \pm 2.38$ & $0.40 \pm 0.17$ \\
$2.169$ & $0.71 \pm 0.01$ & $6.43 \pm 2.86$ & $0.45 \pm 0.20$ \\
$2.178$ & $0.72 \pm 0.01$ & $7.77 \pm 3.17$ & $0.56 \pm 0.23$ \\
$2.187$ & $0.72 \pm 0.01$ & $-4.05 \pm 1.87$ & $-0.29 \pm 0.14$ \\
$2.195$ & $0.72 \pm 0.01$ & $10.64 \pm 4.20$ & $0.77 \pm 0.30$ \\
\enddata
\tablenotetext{a}{The uncertainties on the contrast and flux do not take into account the uncertainty on the systematic star-to-spot flux ratio used to convert contrasts measured with respect to the satellite spots into source-to-star contrasts. These systematic uncertainties are $3\%$ at {\it J}, $5\%$ at {\it H}, and $6\%$ at $K1$ \citep{Maire:2014gs}.}
\end{deluxetable}

\end{document}